\documentclass{aa}
\usepackage{graphicx}
\usepackage{txfonts}
\usepackage{amssymb}
\usepackage[figuresright]{rotating}
\usepackage{multirow}
\usepackage[switch]{lineno}
\usepackage{calc}
\begin{document}

  \title{Multiwavelength interferometric observations and modeling of
    circumstellar disks\thanks{Based on observations made with
      telescopes of the European Organisation for Astronomical Research in the
      southern Hemisphere (ESO) at the Paranal Observatory, Chile, under the
      programs 073.A-9014, 075.C-0014, 075.C-0064, 075.C-0253,
      077.C-0750, 079.C-0101, and 079.C-0595.}}

  \author{A. A. Schegerer\inst{1,2}, T. Ratzka\inst{3},
    P. A. Schuller\inst{4}, S. Wolf\inst{5}, L. Mosoni\inst{6}, Ch. Leinert\inst{2}} 

  \offprints{A. A. Schegerer, \email{aschegerer@bfs.de}}
  
  \institute{
    Bundesamt f\"ur Strahlenschutz, Fachbereich f\"ur Strahlenschutz und Gesundheit, 
    Ingolst\"adter Landstra{\ss}e 1, 85764 Neuherberg, Germany \and 
    Max-Planck-Institut f\"ur Astronomie, K\"onigstuhl 17, 69117 Heidelberg, Germany
    \and  Ludwig-Maximilians-Universit\"at, Universit\"ats-Sternwarte
    M\"unchen, Scheinerstra{\ss}e 1, 81679 M\"unchen, Germany \and
    Universit\"at zu K\"oln, I.\,Physikalisches Institut, Z\"ulpicher Stra{\ss}e 77,
    50937 K\"oln, Germany  
    \and Universit\"at Kiel, Institut f\"ur Theoretische Physik und
    Astrophysik, Leibnizstra{\ss}e 15, 24098 Kiel, Germany 
    \and MTA Research Center for Astronomy and Earth Sciences, Konkoly Thege
    Mikl\'os Astronomical Institute, 1525 Budapest, Hungary}

  \date{}
  
  \abstract
  {}
  {We investigate the structure of the innermost region of three circumstellar
    disks around pre-main sequence stars HD\,142666, AS\,205\,N, and
    AS\,205\,S. We determine the inner radii of the 
    dust disks and, in particular, search for transition objects where dust
    has been depleted and inner disk gaps have formed at radii of a few tenths
    of AU 
    up to several AU. 
  } 
  {We performed interferometric observations with IOTA, AMBER, and MIDI in the
    infrared wavelength ranges $1.6-2.5\,\mathrm{\mu m}$ and
    $8-13\,\mathrm{\mu m}$ with projected baseline lengths between
    $25\,\mathrm{m}$ and $102\,\mathrm{m}$. The data analysis
    was based on radiative transfer simulations in 3D models of young
    stellar objects (YSOs) to reproduce the spectral energy distribution and the
    interferometric visibilities simultaneously. Accretion effects and disk
    gaps could be considered in the modeling approach. Results from previous
    studies restricted the parameter space.}
  {The objects of this study were spatially resolved in the infrared
    wavelength range using the interferometers. Based on these 
    observations, a disk gap could be found for the source HD\,142666 that
    classifies it as transition object. There is a disk hole up to a
    radius of $R_\mathrm{in}=0.30\,\mathrm{AU}$ and a (dust-free) ring between
    $0.35\,\mathrm{AU}$ and $0.80\,\mathrm{AU}$ in the disk of
    HD\,142666.
    The classification of AS\,205 as a system of classical T\,Tauri stars
    could be confirmed using the canonical model approach, i.\/e., there are
    no hints of disk gaps in our observations. 
  }
  {}  
  \keywords{Infrared: stars -- Accretion disks -- Planetary systems:
    protoplanetary disks -- 
    Instrumentation: interferometer -- Radiative transfer -- Stars:
    individual:  HD\,142666, AS\,205\,N, AS\,205\,S} 
  
  \authorrunning{Schegerer et al.}
  \titlerunning{Multiwavelength interferometric observations and modeling of
    YSOs}
  \maketitle
 
  \section{Introduction}\label{section:introduction}
  Observations of the innermost region of circumstellar disks, i.\/e., at
  a distance of a few tenths of AU 
  from the central star, are motivated by
  the outcomes of theoretical investigations that predict planet
  formation there. The ambient conditions in this
  region, such as high direct stellar radiation, high dust particle densities, 
  and long-lived vortices, can favor planet formation (Klahr \& Bodenheimer~\cite{klahr};  
  W\"unsch et al.~\cite{wuensch}; Johansen et al.~\cite{johansen}). Close to
  the star, the
  dust temperature reaches values up to the sublimation temperature of
  $\sim$$1500\,\mathrm{K}$ resulting in near-infrared (NIR) and mid-infrared
  (MIR) radiation excess. 
  First, highly resolved, i.\/e., interferometric
  observations of YSOs in the NIR wavelength range (Millan-Gabet et al.~\cite{millan-gabet})
  and the subsequent successful modeling of these measurements using simple 
  rings have motivated the idea of a compact, potentially puffed-up rim at the
  inner dust disk (Dullemond et al.~\cite{dullemondIII}) where dust at the
  inner edge of the disk is strongly heated by direct stellar
  irradiation. However, further 
  interferometric measurements have suggested a more complex disk model where
  accretion and/or an optically thin envelope close to the inner edge are
  additionally implemented (Schegerer et al.~\cite{schegerer}). 

  The viscous evolution of the disk, i.\/e., accretion and spreading
  (Lynden-Bell \& Pringle~\cite{lynden}; Hartmann
  et al.~\cite{hartmann}; Armitage et al.~\cite{armitage}), photoevaporation (Hollenbach et
  al.~\cite{hollenbach}; Clarke et al.~\cite{clarke}; Alexander et
  al.~\cite{alexanderI}; Alexander et al.~\cite{alexanderII}), dust evolution
  (Monnier \& Millan-Gabet~\cite{monnier}), and planet 
  formation (Youdin \& Shu~\cite{youdin}) strongly affect the inner
  disk structure. The inner dust disk radius $R_\mathrm{in}$ is initially 
  determined by the sublimation temperature of dust in the case of disks with
  ages up to several million years. For
  T\,Tauri stars, the primary inner disk radius $R_\mathrm{in}$ is in the range
  of $0.1\mathrm{AU}$. But if the viscous accretion inflow rate falls below
  the photoevaporation rate, the dust density strongly reduces in the inner
  disk regions where the inner
  disk radius can increase or a nearly dust-free inner ring can form on a time
  scale of $10^{5}\,$years  
  (Clarke et al.~\cite{clarke}; see Dullemond et al.~\cite{dullemondII} for a
  review). Tidal 
  forces, which can be caused by a stellar companion or a newly formed planet,
  can accelerate the formation of disk gaps 
  (Bouwman et al.~\cite{bouwmanIV}; Bryden et
  al.~\cite{bryden}). Spectrophotometric measurements of 
  YSOs show a strong decrease in infrared radiative excess with age (Haisch et
  al.~\cite{haisch}; Carpenter et al.~\cite{carpenter}; 
  Sicilia-Aguilar et al.~\cite{sic06}; Sicilia-Aguilar et al.~\cite{sic09})
  which is assumed to be caused
  by the depletion of {\it small} dust particles (Weinberger et
  al.~\cite{weinberger}; Grady et al.~\cite{grady}). Classical
  T\,Tauri stars (CTTS) 
  usually fulfill the photometric conditions $K-N>2\,\mathrm{mag}$ and
  $K-L>0.4\,\mathrm{mag}$ (Kenyon \&
  Hartmann~\cite{kenyon}). But 
  there are some YSOs where 
  $K-N>2\,\mathrm{mag}$ and $K-L<0.4\,\mathrm{mag}$. McCabe et
  al.~(\cite{cabe}) assume that the latter color relation could be a hint
  of a transition disk object (TO) where an inner disk gap has been
  formed. 
  To derive further indicators for the classification of
  YSOs as CTTS and TO, various
  infrared color relations of huge  
  samples of pre-main sequence stars from different star-forming regions have been
  studied (e.\/g., Sicilia-Aguilar et al.~\cite{sic06}; Sicilia-Aguilar et 
  al.~\cite{sic09}). 
  These studies conclude that objects with no significant infrared
  excess at $\lambda < 6\,\mathrm{\mu m}$ (but with excess at
  longer wavelengths) can be classified as TO. However, 
  as pointed out by Ercolano et al.~(\cite{ercolano}), the missing excess in
  the NIR wavelength range is a necessary but not sufficient condition for a
  TO, for late-type stars in particular. 

  Observations that
  spatially resolve the inner disk region can directly identify a TO. In fact, Brown et
  al.~(\cite{brown}) spatially 
  resolved a $40\,\mathrm{AU}$ radius inner gap in the
  circumstellar disk around \object{LkH$\alpha\,$330} in the far-infrared
  (FIR) wavelength range with the Submillimeter
  Array. There are more spatially resolved observations of correspondingly
  {\it large} gaps in circumstellar disks (e.\/g., Guilloteau et
  al.~\cite{guilloteau}; Sauter et al.~\cite{sauter}). However, the spatial
  resolution of the innermost disk region, where the formation
  of disk gaps is assumed to start, is possible by interferometric
  observations in the MIR and NIR wavelength range. Considering the spectral
  energy distribution (SED) {\it and}
  interferometric observations in the 
  MIR wavelength range, Ratzka et al.~(\cite{ratzkaIII}) found a
  depletion of the innermost disk regions 
  from warm dust grains, i.\/e., an 
  increase in the inner disk radius of the TO \object{TW\,Hya}, which 
  supplements the model of Calvet et al.~(\cite{calvet}). Akeson et
  al.~(\cite{akeson}) have recently presented a three-component model of this
  transition object.  

  In this study, we present high-resolution interferometric, multiwavelength
  observations of three YSOs using the Infrared Optical Telescope Array (IOTA),
  the Astronomical 
  Multi-BEam combineR (AMBER), and the MID-infrared Interferometer instrument 
  (MIDI). The interferometric visibilities $V$ that were obtained from these
  observations allow study of the spatial distribution of dust with
  temperatures between several $100\,\mathrm{K}$ 
  (MIDI) and $<$$1500\,\mathrm{K}$ (IOTA, AMBER) supposedly at the
  innermost regions of protoplanetary disks. 
  In Sect.~\ref{section:observation}, we
  describe our interferometric observations and the subsequent data
  reduction. Computer models of YSOs where the
  Monte Carlo method is used can explain the observations
  (Sect.~\ref{section:modeling}). We additionally consider previous studies
  concerning stellar and accretion 
  parameters, as well as photometric measurements, to get a coherent model of
  the YSOs. The
  results of our interferometric observations and modeling efforts are
  presented in Sect.~\ref{section:models}. The uniqueness of our results is
  discussed in more detail in Sect.~\ref{section:discussion}. There,
  we also discuss a more advanced modeling approach. We conclude the study in
  Sect.~\ref{section:conclusion}. 
  
  \section{Interferometric observations and data
    reduction}\label{section:observation} 
  
  \begin{table*}
    \centering
    \begin{minipage}{0.83\textwidth} 
      \caption{Object properties derived in previous measurements
        (flags refer to the references). The table lists the
        coordinates (RA and DEC in J2000.0), distances ($d$), visual extinctions
        ($A_\mathrm{V}$), spectral types (SpTyp), stellar masses ($M_\star$),
        effective stellar temperatures ($T_\star$), stellar luminosities
        ($L_\star$), and ages of the objects. See Appendix~\ref{appendix}
        for further details.} 
      \begin{tabular}{lllllllllc}\hline\hline
        Object &  $RA$ & $DEC$ &
        $d$ & $A_\mathrm{V}$ & SpTyp & $M_\star$ & $T_\star$ & $L_\star$ & Age \\
         &  (h\ m\ s) & ($^{\circ}$\ \arcmin\ \arcsec) &
        (pc) & (mag) & & (M$_\mathrm{\odot}$)
        & (K) & (L$_\mathrm{\odot}$) & (Myr) \\
        \hline
        \object{HD142666} & $15\ 56\ 40.0_\mathrm{ (1)}$ & $-22\ 01\ 40_\mathrm{ (1)}$
        & $114_\mathrm{ (8)}$ & $0.8_\mathrm{ (9)}$ & 
        A$7$\,III$_\mathrm{ (10)}$ & $1.6_\mathrm{ (9)}$ & $7600_\mathrm{ (9)}$ & $8_\mathrm{
          (9)}$ & $6.3_\mathrm{ (11)}$\\[1.0ex]
        \object{AS\,205\,N} & $16\ 11\ 31.3_\mathrm{ (1)}$ & $-18\ 38\ 25_\mathrm{
          (1)}$ & $160_\mathrm{ (2)}$ & $3.6_\mathrm{ (3)}$ & K\,$5_\mathrm{
          (2)}$ & $1.5_\mathrm{ (2)}$ & $4400_\mathrm{ (2)}$ &  
        $7.1_\mathrm{ (2)}$ & $1_\mathrm{ (12)}$\\[1.0ex]
        \object{AS\,205\,S} & $16\ 11\ 31.3_\mathrm{ (1)}$ & $-18\ 38\ 25_\mathrm{
          (1)}$ & $160_\mathrm{ (2)}$ & $3.6_\mathrm{ (3)}$ & M$3_\mathrm{
          (2)}$ & $0.30_\mathrm{ (2)}$ 
        & $3400_\mathrm{ (2)}$ & $2.2_\mathrm{ (2)}$ & 
        $1_\mathrm{ (12)}$ \\[1.0ex]
        \hline  
      \end{tabular} 
      \label{table:properties-midisurvey}  
      {\newline \scriptsize {References:} {\bf 1}:  
        2\,MASS catalog~(Cutri et al.~\cite{cutri}); {\bf 2}:~Prato
        et al.~(\cite{prato}); {\bf 3}:  
        Eisner et al.~(\cite{eisnerII}); {\bf 4}:  
        FEPS data base; {\bf 5}:  
        Pottasch \& Parthasarathy~(\cite{pottasch}); {\bf 6}: Houk et
        al.~(\cite{houk}); {\bf 7}: Valenti \& Fischer~(\cite{valenti}); {\bf
          8}: Sylvester et al.~(\cite{sylvester}); {\bf 9}: Garcia-Lopez et
        al.~(\cite{garcia-lopez}); {\bf 10}: Blondel et al.~(\cite{blondel});
        {\bf 11}: van Boekel et al.~(\cite{boekelII}); {\bf 12}: Liu et
      al.~(\cite{liu})}
    \end{minipage}
  \end{table*} 

  \subsection{Object sample}
  The
  objects of our study are HD\,142666 and AS\,205 which belong 
  to the 
  $\rho$ Ophiuchi star formation region. Selected object parameters
  known from literature are
  compiled in Table~\ref{table:properties-midisurvey} and in Appendix~\ref{appendix}. Considering the
  spectral type, the A$7$\,III source HD\,142666 is a Herbig Ae/Be
  object while the other two sources are T\,Tauri
  objects. HD\,142666 is assumed not to have any
  stellar companion (e.\/g., Fukagawa et al.~\cite{fukagawa}). Considering our
  interferometric measurements, a sinusoidal
  visibility spectrum could not be found that would be characteristic of a close (stellar)
  companion in both systems.  
  AS\,205 is a triple
  system (Eisner et al.~\cite{eisnerII}; Koresko et al.~\cite{koresko}). In
  this study, we also present the interferometric 
  observation of the main southern (S) companion of AS\,205.\footnote{Note: For the 2\,MASS
    catalog, AS\,205\,S is the main component.} 

  \subsection{Observations with IOTA}
  IOTA at the Smithsonian
  F.~L.~Whipple Observatory on Mount Hopkins, Arizona, belonged to the first
  Michelson interferometers observing in the NIR wavelength range (Carleton et
  al.~\cite{carleton}; Traub~\cite{traub}; Traub~\cite{traubII}) and was operated
  until 2006 (Schloerb et al.~\cite{schloerb}). IOTA consisted of three 
  $0.45\,\mathrm{m}$-Cassegrain telescopes that could be shifted
  on an L-shaped 
  $15\,\mathrm{m} \times 35\,\mathrm{m}$-track, allowing a flexible
  modification of the baseline lengths. An angular resolution of
  five milli-arcseconds (mas) at a
  maximum  
  baseline length of $B=38\,\mathrm{m}$ could theoretically be reached with
  IOTA. Not only the visibilities and the corresponding differential
  phases, but also the closure phase could be 
  measured with IOTA. The closure phase, unaffected by atmopheric effects, is a
  prerequisite for determining asymmetries in the measured brightness
  distribution, so essential for image reconstruction (e.g.,
  Jennison~\cite{jennison}; Leinert~\cite{leinertII}). However, numerous
  interferometric observations are necessary for an acceptable image (Kraus et
  al.~\cite{kraus}). Therefore, the closure phase is not discussed any further
  in this study. 

  Atmospheric effects were compensated by a piezo-actuated tip-tilt
  tertiary mirror. After the compensation of 
  the optical path difference by a sequence of light reflections in an
  evacuated tank, two beams were combined using the IONIC\,3 Beam Combiner
  (Berger et al.~\cite{berger}). A single
  exposure consisted of $200$ to $300$ readouts in 
  four minutes. Photometric measurements and exposures of dark current
  concluded the observation sequence. The observation of a main-sequence star
  with known diameter (calibrator) before and after the observation sequence
  allowed the determination of the modulation transfer function ($MTF$) of the
  interferometer by linear
  interpolation. Also, atmospheric and instrumental background were
  eliminated by 
  the calibration measurements.  

  In June 2006, the bright YSO
  HD\,142666 could be observed with 
  IOTA at a maximum projected baseline length of $\sim$$27\,\mathrm{m}$. The H
  band 
  visibility within the wavelength interval of $(1.65 \pm 0.15)\,\mathrm{\mu m}$
  was measured. For data calibration, we observed the main-sequence 
  stars \object{HD\,134758}, \object{HD\,143033}, and \object{HD\,139663} with
  the uniform disk (UD) diameters $(1.42\pm0.02)\,\mathrm{mas}$,
  $(1.36\pm0.02)\,\mathrm{mas}$, and $(1.99\pm0.02)\,\mathrm{mas}$, 
  respectively, in H band (Merand et al.~\cite{merand}; Bord\'e et 
  al.~\cite{borde}). The projected baseline lengths as well as the position
  angles during our interferometric observations with IOTA are 
  listed in Table~\ref{table:journal_iota}.  
  The visibilities are shown in Fig.~\ref{figure:hd142666}.
  \begin{table}[!tb]
    \centering
    \caption{Journal of IOTA observations of HD\,142666.} 
    \label{table:journal_iota}
    \begin{tabular}{cccc}
      \hline\hline
      Date of     &  \multicolumn{2}{c}{proj.~Baseline} & Flags\\
      observation &  \	  (m) & (deg)		          & \\
      \hline
      \noalign{\smallskip}
      05-06-2006 & 25/20/11 & 60/34/-66  &r/u/u\\
      \noalign{\smallskip}
      05-06-2006 & 25/20/11 & 61/34/-65  &r/u/u\\
      \noalign{\smallskip}
      05-06-2006 & 24/19/12 & 64/34/-63  &r/u/u\\
      \noalign{\smallskip}
      08-06-2006 & 29/23/10 & 54/35/-80  &u/u/u\\
      \noalign{\smallskip}
      08-06-2006 & 29/23/10 & 54/35/-79  &u/u/u\\
      \noalign{\smallskip}
      08-06-2006 & 28/23/10 & 54/35/-78  &u/u/u\\
      \noalign{\smallskip}
      08-06-2006 & 28/22/10 & 55/35/-76  &u/u/u\\
      \noalign{\smallskip}
      09-06-2006 & 27/22/11 & 57/34/-72  &r/r/u\\
      \noalign{\smallskip}
      09-06-2006 & 26/21/11 & 57/34/-71  &r/r/u\\
      \noalign{\smallskip}
      09-06-2006 & 26/21/11 & 58/34/-70  &r/u/u\\
      \noalign{\smallskip}
      \hline
    \end{tabular}
    {\newline \scriptsize Observations that are specified with the flags r and u are
      spatially resolved/unresolved, respectively. Measurements are assumed to be
      unresolved if the visibility plus standard error is greater than 1.}
  \end{table}

  The IOTA Data Reduction Software (IDRS) was used. It offers two alternative
  methods of determining the visibility $V$ ({\tt IDRS v0.7} and {\tt v0.8};
  Kraus et al.~\cite{kraus}). A first
  approach uses the linear relation between the measured power spectrum and
  the visibility (Jackson~\cite{jackson}). In this standard 
  approach, data sets with low signal-to-noise ratios can be
  added easily. Gaussian-shaped, narrow signal profiles in the power spectrum
  result 
  from data sets with low noise. Fitting linear functions to the tails
  of the Gaussian-shaped profiles allows background
  noise to be subtracted. Furthermore, a lower and an upper limit are defined to eliminate noise
  with low and high spatial frequencies. The signal profiles that were 
  measured in this study are broad as a result of a low signal-to-noise
  ratio. Therefore, the error bars of the visibilities that we obtained
  for HD\,142666 with IOTA are high.

  Another approach is based on the continous
  wavelet transform (CWT). It 
  results from a convolution of the interference signal with a Morlet
  function from which the power spectrum can be
  determined. Atmospheric perturbances, instrumental vibrations, as well as
  background noise, can effectively be eliminated with the CWT method (Kraus et
  al.~\cite{kraus}), in particular for data sets of faint objects. The
  CWT was used to confirm the results obtained from the standard method 
  described above.
 
  \subsection{Observations with AMBER}
  \label{amber_observations}

  The system AS\,205 was observed with AMBER (Petrov et al.~\cite{petrov}; see
  Table~\ref{table:journal_amber}), and it operates at the Very Large Telescope Interferometer (VLTI). 
  The combination of three telescopes with 
  AMBER allows the closure phase to be determined. Analogous to our measurements
  with IOTA, the closure phase is not discussed further in this study. 
  In
  contrast to observations with IOTA, AMBER allows the observations to be
  spectrally resolved in H and K band ($1.6\,\mathrm{\mu m} - 2.5\,\mu$m). 
  
  For our observations with AMBER, we used the Unit
  Telescopes UT\,2, UT\,3, and UT\,4. This telescope combination 
  not only includes two almost perpendicular baselines, but also offers an
  appropriate angular resolution. The baseline UT\,2-UT\,4 reached a baseline
  length of $89\,\mathrm{m}$, which corresponds to a resolution of 1.9\,mas at
  a wavelength of 1.6\,$\mu$m corresponding to $0.22\,\mathrm{AU}$ and
  $0.30\,\mathrm{AU}$ at the distance of the sources HD\,142666 and
  AS\,205, respectively. Correspondingly, the 
  angular resolution of the shortest baseline UT2-UT3 with a maximum baseline
  length of 47\,m reaches 3.5\,mas, which corresponds to
  $0.40\,\mathrm{AU}$ and $0.56\,\mathrm{AU}$. The resolution of AMBER 
  on this baseline is thus a factor 2-3 higher than the resolution that
  can be reached by MIDI in the MIR wavelength range with the longest baseline
  used (see 
  Sect.~\ref{midi_observations}). This allowed us to continuously trace the
  circumstellar disks from the hot inner regions to the warm parts at distances
  of several AU from the central star. We chose the
  low-resolution prism mode with a spectral
  resolving power of $\lambda / \Delta \lambda\approx 35$, which is similar to the
  resolving power provided by MIDI in prism mode, and it allows good spectral 
  coverage. For the observation of the weak source AS\,205\,S, only measurements in 
  K band were possible, since IRIS
  (VLTI InfraRed Image Sensor; Gitton et al.~\cite{gitton}) was operated in
  the H band to stabilize the beams.  
  
  Observations with AMBER start with exposures that are used for the wavelength
  calibration and the calibration of the offsets between the interferometric
  and the photometric 
  channels. In a second step, the so-called pixel-to-visibility
  matrix ${\bf M_\mathrm{V2P}}$ is determined. This matrix is the $MTF$ between
  visibility  
  $\vec{V}$ and measured interferogram $\vec{I}$, i.\/e., 
  \begin{eqnarray}
    \vec{I} = {\bf M_\mathrm{V2P}} \cdot \vec{V}{\rm .}
    \label{eq:visibility}
  \end{eqnarray}
  The quantity $\vec{I}$ represents here the interferogram measured on the
  detector after the total flux is subtracted. The total flux is obtained from
  the photometric channels. 
  The scientific measurements 
  with AMBER consist of several thousands of exposures. A single exposure takes
  $25$\,ms. For a calibrated 
  visibility, a standard star with 
  known diameter, i.e., with known visibility is observed. Standard stars were
  chosen from the CalVin list\footnote{\tt 
    www.eso.org/instruments/amber/tools/}. 
  We used the standard stars HD\,143033 and HD\,143900
  with the UD diameters $(1.36\pm0.02)\,\mathrm{mas}$ and 
  $(1.25\pm0.02)\,\mathrm{mas}$,
  respectively. Additional
  measurements are necessary for determining of the detector background
  and the sky background. 
  
  The AMBER Data Reduction Software {\tt amdlib v3.0.3} was used to reduce the
  data.\footnote{\tt www.jmmc.fr/data\_processing\_amber.htm} This
  recent release of the software is based on a new algorithm with an improved
  data and 
  noise model~(Chelli et al.~\cite{chelli}). The standard procedure
  described in the user manual was followed. For the frame selection, the following
  criteria were chosen. 
  The
  piston, i.\/e., the mean optical path length difference of the incoming
  wavefronts,  was limited 
  to $8.0\,\mathrm{\mu m}$ (Hummel~\cite{hummel}). Furthermore, only 20\% of
  the frames with the 
  highest signal-to-noise ratio were taken into account to avoid degrading 
  the fringe signal. This percentage is slightly more constraining than the
  value used in different previous studies using AMBER (e.g., Weigelt et
  al.~\cite{weigelt}).
  
  Subsamples of the data from several thousand 
  single exposure frames were created by comparing the resulting visibilities with 
  those derived from the complete data sets. Obvious outliers in
  the subsample set were discarded. If the overall data quality is 
  low, the selection of the highest signal-to-noise ratios could lead to
  significantly different results for the complete data set and the
  subsamples were finally selected  
  as different frames. 
  During our observations, a second calibrator 
  was available. We used this calibrator for a
  consistency test, i.\/e., to cross-check the results. 
  
  The visibilities presented in
  this paper were derived by averaging the 
  calibrated subsamples. The error bars that are presented in
  Sect.~\ref{section:models}
  represent the corresponding standard deviation. 
  The journal of
  our observations with AMBER is provided in Table~\ref{table:journal_amber}.
  \begin{table}[!tb]
    \centering
    \caption{Journal of AMBER observations.} 
    \label{table:journal_amber}
    \begin{tabular}{ccccc}
      \hline\hline
      Date of     &  Object	 &  \multicolumn{2}{c}{proj.~Baseline, U2-U3-U4} & Flags\\
      observation & 		 &  \	  (m) & (deg)		          & \\
      \hline
      \noalign{\smallskip}
      13-04-2006 & \object{AS\,205\,S}  & 44/53/75 & 49/128/93  &$^{x}$\\
      \noalign{\smallskip}
      20-03-2008 & \object{AS\,205\,N}	& 46/62/89 & 38/109/80  &\\
      \noalign{\smallskip}
      \hline
    \end{tabular}
    {\newline \scriptsize $^{x}$ measurement is not considered in data analysis
      because of technical failures and/or bad weather conditions}
  \end{table}
  
  \subsection{Observations with MIDI}
  \label{midi_observations}

  The studied YSOs have also been observed with MIDI at the VLTI (Leinert et
  al.~\cite{leinert03a}), see Tables~\ref{table:journal_hd142666} and 
  \ref{table:journal_as205}. 
  Operating in N-band (8-13\,$\mu$m), MIDI combines 
  two telescopes and offers -- as in our case -- a spectral resolving power of
  $\lambda / \Delta \lambda \approx 30$ using the prism mode. 
  
  To determine the $MTF$ of the instrument, different calibrators
  that were taken from the ``MIDI Calibrator Catalogue''\footnote{can be found on
    webpage \tt www.eso.org/} were observed
  before or after the observation of the scientific object. Five of 
  these calibrators, HD\,95272, HD\,102461, HD\,133774, HD\,139127,
  and HD\,178345, were used for absolute flux
  calibration.\footnote{\tt www.eso.org/sci/facilities/paranal/
    instruments/midi/tools/} Observations of both scientifc target and calibrator
  consist of $8000$ 
  exposures. A single exposure takes $18$\,ms. For 
  descriptions of MIDI and its operation, see Leinert et
  al.~(\cite{leinert03a}), Leinert et al.~(\cite{leinert03}), and Morel et
  al.~(\cite{morel}). The results of our MIDI
  observations of the objects HD\,142666, AS\,205\,N, and AS\,205\,S,
  as well as a basic data analysis, have already been published in Schegerer~(\cite{buch}).
  
  For data reduction, we used a custom software called {\tt
    MIA+EWS}\footnote{\tt
    www.strw.leidenuniv.nl/$\sim$nevec/MIDI/, 
    www.mpia-hd.mpg.de/MIDISOFT/}. The data reduction steps are
  described in a tutorial on the cited web page, as well as in Leinert et
  al.~(\cite{leinert}) and Ratzka et al.~(\cite{ratzka05}). The visibilities
  shown in this paper were derived by 
  using all calibrators that were observed during one night with the same
  baseline and instrument setup as the scientific targets, which show no
  peculiarities. The correction
  of the calibration that is required due to the diameters of the calibrators 
  is considered in the software package. But this correction has a
  negligible effect on the calibration. The error bars in the visibility
  curves represent the standard deviation when using the ensemble of
  calibrator measurements. We discarded a measurement whenever the result that
  was obtained from the MIA software package differs from the results
  obtained from the {\tt EWS} software package.
  
  The journals of our observations with MIDI are provided in
  Tables~\ref{table:journal_hd142666}~and~\ref{table:journal_as205}. Depending on the
  baseline that was used, an angular resolution between 8.6\,mas and 17\,mas
  corresponding to 0.98\,AU - 1.9\,AU and 1.4\,AU - 2.7\,AU at the distances
  of the sources HD\,142666 and AS\,205,
  respectively, was reached in our observations with MIDI.  
  \begin{table}[!tb]
    \centering
    \caption{Journal of MIDI observations of HD~142666.}
    \label{table:journal_hd142666}
    \begin{tabular}{cccc}
      \hline\hline
      Date of      &	  \multicolumn{2}{c}{proj.~Baseline} & Flags\\
      observation  &  \     (m) & (deg)                  & \\
      \hline
      \multicolumn{4}{c}{U1-U3}\\
      \hline
      \noalign{\smallskip}
      08-06-2004  &\ 102 &\ 37 & $^{f}$\\
      08-06-2004 &\ 91 &\ 44 & $^{x}$\\
      08-06-2004 &\ 90 &\ 44 & \\
      \noalign{\smallskip}
      \hline
    \end{tabular}
    {\newline \scriptsize $^{f}$ fringe track repeated; 
      $^{x}$ measurement is not considered in data analysis
      because of technical failures and/or bad weather} 
  \end{table}
  
  \begin{table}[!tb]
    \centering
    \caption{Journal of MIDI observations of AS\,205\,N and AS\,205\,S. The length and
      position angle of the projected baseline has been determined from the fringe-tracking sequence.} 
    \label{table:journal_as205}
    \begin{tabular}{ccccc}
      \hline\hline
      Date of     &  Object	 & \multicolumn{2}{c}{proj.~Baseline} & Flags\\
      observation & 		 & \     (m) & (deg)                  & \\
      \hline
      \multicolumn{5}{c}{U1-U2}\\
      \hline
      \noalign{\smallskip}
      29-05-2005 & \object{AS\,205\,N}   & 56 &\ 26 & \\
      29-05-2005 & \object{AS\,205\,S}   & 57 &\ 29 & \\
      \noalign{\smallskip}
      \hline
      \multicolumn{5}{c}{U1-U3}\\
      \hline
      \noalign{\smallskip}
      29-05-2005 & \object{AS\,205\,N}   & 90 &\ 42 & \\
      29-05-2005 & \object{AS\,205\,S}   & 87 &\ 42 & $^{pfx}$\\
      \noalign{\smallskip}
      \hline
      \multicolumn{5}{c}{U3-U4}\\
      \hline
      \noalign{\smallskip}
      30-05-2005 & \object{AS\,205\,N}   & 55 & 125 & \\
      30-05-2005 & \object{AS\,205\,S}   & 52 & 129 & \\
      \noalign{\smallskip}
      \hline
    \end{tabular}
    {\newline \scriptsize $^{f}$ fringe track repeated; 
      $^{p}$ photometry of one or both beams repeated;
      $^{x}$ measurement is not considered in data analysis because of
      technical failures and/or bad weather}
  \end{table}

  For the determination of spatially unresolved spectra, 
  the dedicated software package {\tt MIA+EWS} was used. 
  It extracts and calibrates raw spectra. 
  The calibration was done by comparing the measured
  spectrum with the measured spectrum of spectrophotometric calibrators
  (Table~\ref{table:photometry_as205}). 
  \begin{table}[!tb]
    \centering
    \caption{Photometric datasets used to derive the spectra.}
    \label{table:photometry_as205}
    \begin{tabular}{cccc}
      \hline\hline
      Date of     &  Object   & Airmass & Flags\\
      observation &	    &	      & \\
      \hline
      \noalign{\smallskip}
      29-05-2005  & AS\,205\,N  & 1.01/1.01 & \\
      \noalign{\smallskip}
      29-05-2005  & AS\,205\,S  & 1.02/1.02 & \\
      \noalign{\smallskip}
      30-05-2005  & AS\,205\,N  & 1.17/1.18 & $^{x}$\\
      \noalign{\smallskip}
      30-05-2005  & AS\,205\,S  & 1.24/1.25 & $^{x}$\\
      \hline
    \end{tabular}
    {\newline \scriptsize $^{s}$ spectrophotometric calibrator; $^{x}$ measurement
      is not considered in data analysis because of 
      technical failures and/or bad weather}\\
  \end{table}
  
  \section{Modeling approach}\label{section:modeling}
  \subsection{Canonical disk model}\label{section:canonical}
  For a detailed analysis, we used the radiative-transfer
  code MC3D which is 
  based on a Monte Carlo method (Wolf et al.~\cite{wolfI}) and which was
  developed by Schegerer et
  al.~(\cite{schegerer}) to the form used here .  
  A predefined, parameterized model of a passive disk is the basic element
  of our modeling approach. 
  The approach of Shakura \& Sunyaev~(\cite{shakura}) was used where the
  density distribution 
  is given as
  \begin{eqnarray}
    \label{eq:shakura}
    \hfill{}
    \rho(r,z)=\rho_{0} \left( \frac{R_{\star}}{r} \right)^{\alpha} 
    \exp \left[ - \frac{1}{2} \left( \frac{z}{h(r)} \right)^{2} \right].
    \hfill{}
  \end{eqnarray} 
  The quantity $\rho_{0}$ allows scaling the total disk mass
  $M_\mathrm{disk}$. 
  The quantities $r$ and $z$ are the radii measured from the disk center and
  the vertical distance from the disk midplane, respectively, while the
  quantity $R_\mathrm{\star}$ is the stellar radius. 
  The function $h(r)$ represents the scaleheight\footnote
  {
    The scaleheight is defined as the vertical distance from the midplane
    where the density has  
    decreased by a factor of $e^{-1}$.
  } 
  \begin{eqnarray}
    \label{eq:scaleheight}
    \hfill{}
    h(r)=h_\mathrm{100} \left( \frac{r}{100\,\mathrm{AU}} \right)^{\beta}
    \hfill{}
  \end{eqnarray} 
  with $h_\mathrm{100}=h(r$=$100\,\mathrm{AU})$. Determining the shape of the
  disk profile, the parameters $\beta$ and
  $h_\mathrm{100}$ are called profile parameters in this study. 
  In our approach the exponents
  $\alpha$ and $\beta$ satisfy the relation
  \begin{eqnarray}
    \label{eq:alpha-beta}
    \hfill{}
    \alpha=3 \left( \beta-\frac{1}{2} \right).
    \hfill{}
  \end{eqnarray}   
  The latter relation results from the coupling of surface density
  and temperature of the disk based on the assumption of a disk in
  hydrostatic equilibrium (Shakura \& Sunyaev~\cite{shakura}). We note
  that this approach is only correct for
  geometrically thin 
  disks, where $z \ll r$. However, since the scaleheight
  $h(r)$ is at least one order of magnitude smaller than the corresponding
  radius $r$, 
  this approximation should be justified for those parts of the disk
  where infrared emission has its maximum. The
  error in the density of regions $z \gtrapprox r$ is $z^5/r^7$ for Taylor's series. 
  A rounded puffed-up inner rim was not considered as an analytical supplement
  to our modeling approach (Natta et al.~\cite{nattaV}, Isella \& Natta
  \cite{isella}).
  
  The infrared excess that is emitted from YSOs originates in heated
    dust and scattered light in the circumstellar 
  environment (Vinkovic~\cite{vinkovic}). Assuming compact, homogeneous,
  spherical dust grains, their 
  optical properties such as  
  scattering and extinction cross sections were determined by Mie
  scattering theory from the measured  complex refractive index of the specific material (Wolf \&
  Voshchinnikov~\cite{wolfV}). A dust mixture of a ``astronomical
  silicate'' and graphite was assumed with relative abundances of $62.5$\% and $37.5$\%,
  respectively (Draine \& Malhotra~\cite{draine}). We used a two-layer disk
  model to take dust settling into account. The disk interior, which is
  optically thick in N band,
  contains dust grains with a
  maximum particle size of $a_\mathrm{max}=1\,\mathrm{mm}$. 
  A size of
  $a_\mathrm{max}=1\,\mathrm{mm}$ was  
  found for several T\,Tauri stars with millimeter observations using
  the Very Large Array  
  (Rodmann et al.~\cite{rodmann}). The optical depth was
  measured for constant radii, from the disk  
  atmosphere vertically to the disk midplane. 
  The 
  surface regions consisted of interstellar, unevolved dust with
  $a_\mathrm{max}=0.25\,\mathrm{\mu m}$  
  as found in the interstellar medium (Mathis et al.~\cite{mrn}; MRN dust
  composition). 

  Accretion effects were taken into account. As extensively described in
  Schegerer et al.~(\cite{schegerer}), the potential energy of a particle on
  its way towards the star is partly released in the disk midplane
  (Lynden-Bell \& Pringle~\cite{lynden}, and
  Pringle~\cite{pringle}). Considering the ``magnetically mediated'' modeling
  approach (e.\/g., Uchida \& Shibata~\cite{uchida}; Bertout et
  al.~\cite{bertout}; Calvet \& Gullbring~\cite{calvet+gullbring}), more than
  half of the potential 
  energy of the accreting particles is released in a boundary region above the
  stellar surface. 
  The accretion rate $\dot{M}$, which determines the
  total accretion luminosity $L_\mathrm{acc}$, was not an independent model
  parameter, but its value was constrained by the results of
  previous 
  measurements. Correspondingly, the stellar mass $M_\mathrm{\star}$, the
  effective stellar temperature $T_\mathrm{\star}$, and the stellar luminosity
  $L_\mathrm{\star}$, which were derived in previous studies, were fixed
  parameters in the modeling. 
  Because the SED and the infrared emission, in particular, only marginally depend on the
    disk outer radius $R_\mathrm{out}$, this disk parameter was fixed to
    $R_\mathrm{out}=100\,\mathrm{AU}$.
  Fitted model parameters of our approach were the scaleheight
  $h_\mathrm{100}$, the flaring parameter $\beta$,
  and the inclination $i$ of the circumstellar disk.\footnote{An
    inclination of {$i=0$\degr} corresponds to a face-on disk.} 
  By considering a stellar
  temperature $T_\star$, the quadratic distance law, a mean dust sublimation
  temperature of $1500\,\mathrm{K}$, and the specific absorption coefficients
  $\kappa$ of the adopted dust set, the temperature of single dust grains at
  any distance from the central star can be determined. The resulting radius 
  $R_\mathrm{sub}$ where dust sublimates was the initial value of
  $R_\mathrm{in}$. However, this approach is an approximation for optically thick
  media since the buildup factor is not considered. 
  
  By means of manual modifications of the 
  parameters, we searched for the model that most successfully reproduces the
  measured SED, NIR, and MIR visibilities simultaneously. A manual
  modification was necessary because a disk model with a specific SED and
  images\footnote{used for calculating the
    spectrally dispersed visibility} at different wavelengths takes several hours to
  compute. Therefore, 
  in our search for the best-fit model, the uniqueness of our final models
  cannot be proven.\footnote{There is a faster and parallelized version of the
    MC3D code available, now, where the computation could be strongly
    accelerated, allowing computation of
    up to $10^6$ models in several days.} We verified whether the modeling
  results can be improved by varying the model parameters using the following
  changes: \medskip\\ 
  \begin{tabular}{lll}
    $\Delta M_\mathrm{disk} = 0.5\,\mathrm{M_\mathrm{disk}}$, & 
    $\Delta R_\mathrm{in} = 0.05\,\mathrm{AU}$, & 
    $\Delta \beta = 0.1$, and \medskip\\ 
    $\Delta h_\mathrm{100} =  1\,\mathrm{AU}$. && \\ 
  \end{tabular}\medskip\\
  The step widths can be considered as
  the precision to which the local 
  minimum in the $\chi^2$-surface can be determined. The determination of the
  modeling errors would require a fit of a polynomial to the $\chi^2$-surface
  of the
  simulation grid of the independent parameters. For inclination $i$, only an
  upper value could be determined. Depending on the flaring of the disk,
  the sight of the observer onto the star and the innermost disk regions get
  worse for an increasing inclination, resulting in a decrease in the visible
  (and infrared) flux from the source.

  \subsection{Extended disk model for TOs}\label{section:extended}

  As a result of photoevaporation, as well as planet
  formation and planet motion, the mass
  density at the inner disk regions decreases and disk gaps
  can be formed. To reproduce 
  the SED that was obtained from the 8\,Myr-T\,Tauri object \object{RECX\,5}
  using the InfraRed Spectrograph (IRS) onboard the Spitzer Space Telescope,
  Bouwman et al.~(\cite{bouwmanIII}) decreased the mass density in the disk
  model by a constant factor of $100$ up to a disk radius of
  $r<33\,\mathrm{AU}$. This clearing of the disk around the TO RECX\,5
  was assumed to be a consequence of the formation of a planet. In our approach, we generally
  assumed the canonical disk model, even for more evolved T\,Tauri stars and
  TOs in particular. We found, however, that this conventional approach cannot
  reproduce all the observations obtained from the source
  HD\,142666. Therefore, a modification of the conventional approach was
  subsequently necessary for this source. 

  Depletion of small dust grains at smaller disk radii is characteristic of
  evolved T\,Tauri stars, and TOs in particular. To
  simulate dust depletion in disks, the primary disk density
  $\rho(r,z)$ (Eq.~\ref{eq:shakura}) was multiplied with an exponential
  function, 
  \begin{eqnarray}
    (1+f_{\rho})-exp\left(\frac{R_\mathrm{in}-r}{R_\mathrm{in}-R_\mathrm{out}}*log(10^{-6}+f_{\rho})\right),
  \end{eqnarray}
  that decreases
  the density at the inner disk edge $R_\mathrm{in}$ by a factor $f_{\rho}$
  with $f_{\rho}<1$. 
  The density is not decreased for $f_{\rho}=1$. 
  This exponential function, which is strictly increasing, is
  unity at $r=R_\mathrm{out}$, and it guarantees a smooth transition to the
  primary disk model in contrast to the approach of Bouwman et
  al.~(\cite{bouwmanIII}). 
  For the density reduction factor, only the two different values
  $f_{\rho} = 0.1$ and $f_{\rho} = 0.01$ were used.  

  Considering the theoretical effects of photoevaporation or planet formation,
  another characteristic of TOs is the depletion of small dust grains within
  a {\it defined} disk region, resulting in the formation of an inner hole or ring with
  (negligibly) low dust density. A dust-free
  inner hole was implemented by increasing the inner disk radius $R_\mathrm{in}$. A dust-free
  ring with sharp boundaries, $R_\mathrm{gapin}$ and $R_\mathrm{gapout}$, could
  also be cut from the model. The remaining disk structure was not
  affected by the gap. 

  The gap parameters were varied using the following step widths:\medskip\\ 
  \begin{tabular}{ll}
    $\Delta R_\mathrm{gapin} = 0.05\,\mathrm{AU}$, & $\Delta R_\mathrm{gapout} =
    0.05\,\mathrm{AU}$. \\
  \end{tabular}\medskip\\

  \begin{table*}
    \centering
    \begin{minipage}{0.95\textwidth}  
    \caption{Parameters of the models whose SED, NIR, 
      and MIR visibilities are shown in Figs.~\ref{figure:hd142666},~\ref{figure:as205n}, and~\ref{figure:as205s}. The parameters of model vi of HD\,142666 and
      the parameters of AS\,205\,N and AS\,205\,S simulate the observations
      best.  The last column lists the ratio between
      the stellar and disk model flux in H band at $1.6\,\mathrm{\mu m}$.} 
    \begin{tabular}{lllrrrrrrrrr}\hline\hline
      object & model & $M_\mathrm{disk}$ &  $R_\mathrm{in}$ & $\beta$ &
      $h_\mathrm{100}$ & $i$ & $\dot{M}$ & $f_\mathrm{\rho}$ &
      $R_\mathrm{gapin}$ & $R_\mathrm{gapout}$ &
      $\frac{F_\mathrm{\star}(\lambda=1.6\,\mathrm{\mu m})}{F_\mathrm{disk}(\lambda=1.6\,\mathrm{\mu m})} $ \\   
      & no. & (M$_\mathrm{\odot}$) & (AU) & &
      (AU) & ($^\circ$) & (M$_\mathrm{\odot}$/a) & & (AU) & (AU) & \\ \hline  

      \multirow{7}{*}{HD\,142666} & i & $0.4$ & $0.10$ & $1.0$ & $11$ & $0-50$ & $1 \cdot
      10^{-8}_\mathrm{ (1)}$ & 1 & -- & -- & $1$ \\
      & ii & $0.4$ & $0.50$ & $1.0$ & $11$ & $0-50$ & $1 \cdot
      10^{-8}_\mathrm{ (1)}$ & 1 & -- & -- & $5$ \\
      & iii & $0.4$ & $0.10$ & $1.1$ & $11$ & $0-50$ & $1 \cdot
      10^{-8}_\mathrm{ (1)}$ & 1 & -- & -- & $3$ \\
      & iv & $0.4$ & $0.50$ & $1.0$ & $11$ & $0-50$ & $7 \cdot
      10^{-8}$ & 1 & -- & -- & $1$ \\
      & v & $0.4$ & $0.10$ & $1.0$ & $11$ & $0-50$ & $1 \cdot
      10^{-8}_\mathrm{ (1)}$ & 0.01 & -- & -- & $2$ \\
      & vi & $0.4$ & $0.30$ & $1.0$ & $11$ & $0-50$ & $1 \cdot
      10^{-8}_\mathrm{ (1)}$ & 0.01 & 0.35 & 0.80 & $2$ \\[1.0ex]
      & vii & $0.4$ & $0.30$ & $1.0$ & $11$ & $0-50$ & $1 \cdot
      10^{-8}_\mathrm{ (1)}$ & 1 & 0.35 & 0.80 & $3$ \\[1.0ex]

      AS\,205\,N & & $0.08$ & $0.10$ & $1.0-1.1$ & $19-21$ 
      & $40<i<45$ & $7.0 \cdot 10^{-7}_\mathrm{ (2)}$ & -- & -- & -- & $-8$ \\[1.0ex]
      AS\,205\,S & & $0.05$ & $0.10$ & $1.0$ & $21$ &
      $<45$ & $\sim$$2.0 \cdot 10^{-7}_\mathrm{ (2)}$ & -- & -- & -- & $5$ \\[1.0ex]
     \hline
   \end{tabular}
        {\newline \scriptsize {References: } {\bf 1}: Garcia-Lopez et 
          al.~(\cite{garcia-lopez}); {\bf 2}: Eisner et
          al.~(\cite{eisnerII}); }
        \label{table:properties-midisurveyII}
      \end{minipage}
  \end{table*}

  \section{Observations and modeling results}\label{section:models}
  All scientific targets could be resolved spatially
  with IOTA or AMBER in the NIR range and MIDI in the MIR range. Because of
  technical failures and/or bad weather conditions during the measurements,
  the observations of AS\,205\,S with AMBER failed. 
  Furthermore, some single observations with
  AMBER or MIDI at
  specific baselines and PAs, which are flagged in
  Tables~\ref{table:journal_hd142666} and 
  \ref{table:journal_as205}, 
  are also not considered in the data analysis for the same reason.
 
  Table~\ref{table:properties-midisurveyII} presents the
  parameters resulting from our effort to model the
  YSOs. The ratio between the stellar and disk model flux at
  $1.6\,\mathrm{\mu m}$ is also listed. The H-band
  visibilities depend on the ratio between the stellar and disk flux, as well as
  on the disk geometry. The excess at $1.6\,\mathrm{\mu m}$ was calculated by subtracting
  the observed flux from the value for the assumed
  stellar photosphere. A negative ratio results from a high inclination of
  the disk. Flared, outer disk regions screen the infrared flux from
  the center. 
  Along with the observations, we present and discuss the modeling
  results for each object of our sample in the following. Previous
  observations, including photometric measurements, are mentioned in
  Appendix~\ref{appendix}. 
  The measured and modeled SED, as well as NIR and MIR visibilities of the
  objects in this study, are presented in
  Figs.~\ref{figure:hd142666}-\ref{figure:as205s}.

  \subsection{HD\,142666}\label{section:hd142666}
\begin{figure*}[!tb]
    \center
    \parbox{0.04\linewidth}{\raisebox{1.em}{ii}}
    \resizebox{0.23\textwidth}{!}{\includegraphics{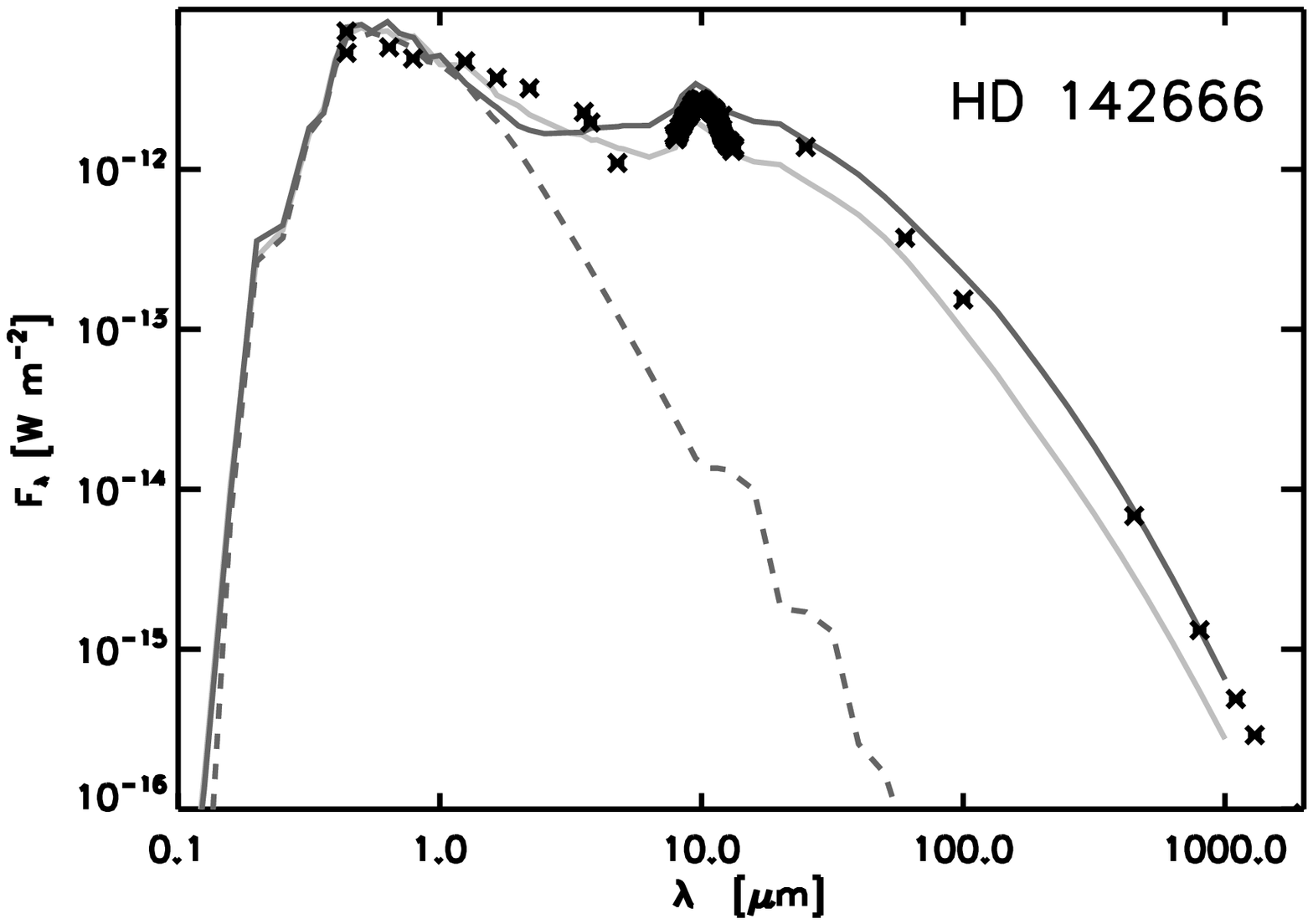}}
    \resizebox{0.23\textwidth}{!}{\includegraphics{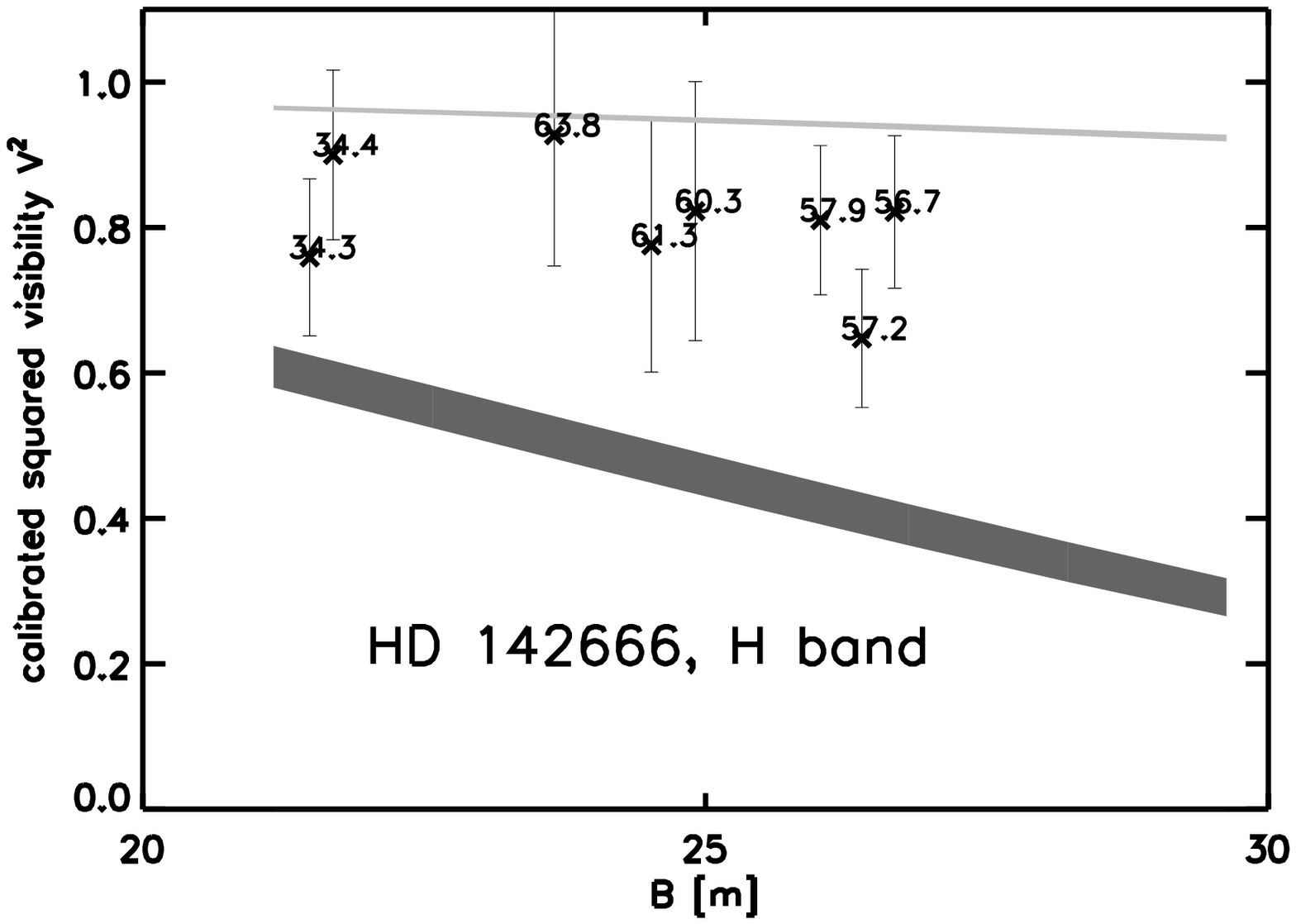}}
    \resizebox{0.23\textwidth}{!}{\includegraphics{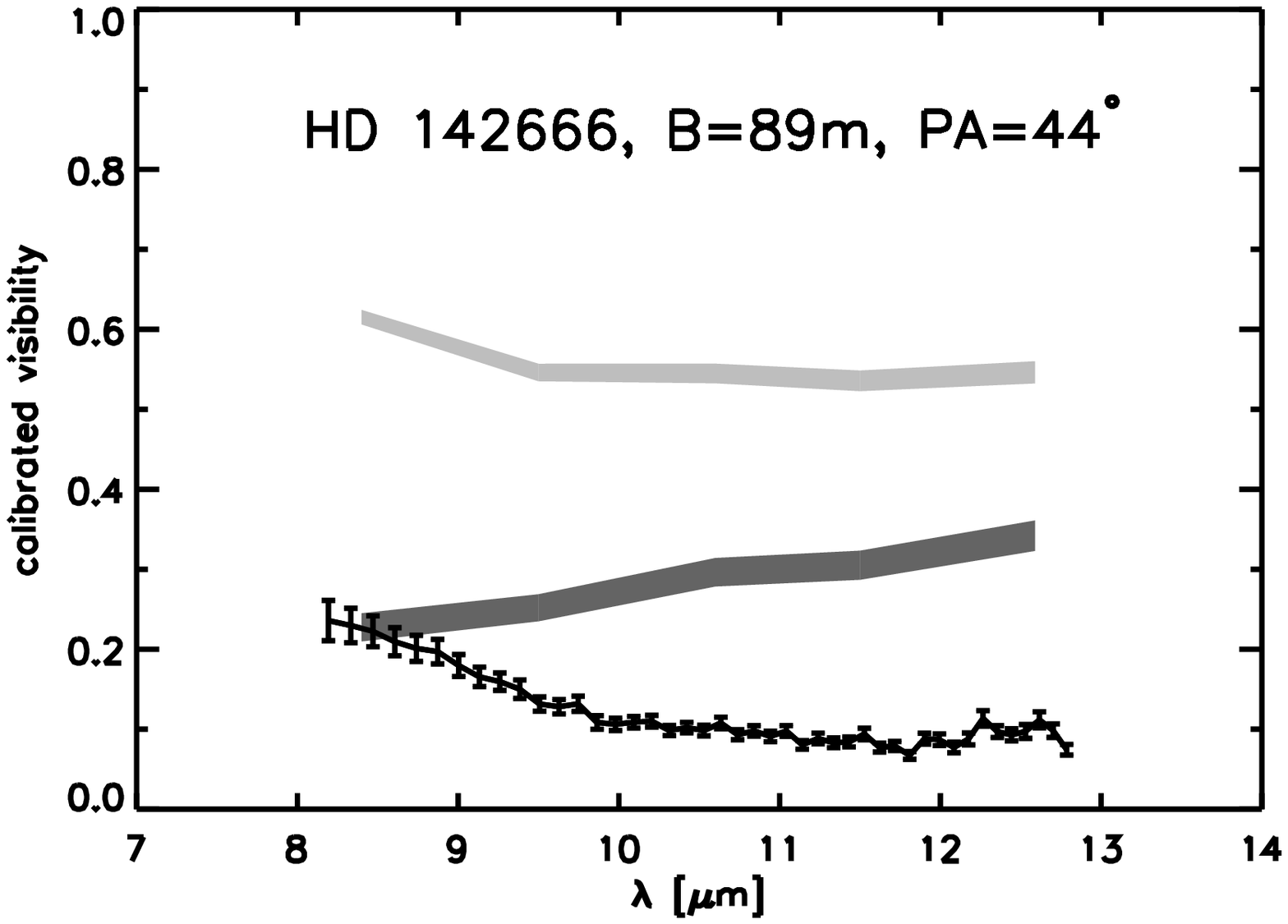}}
    \resizebox{0.23\textwidth}{!}{\includegraphics{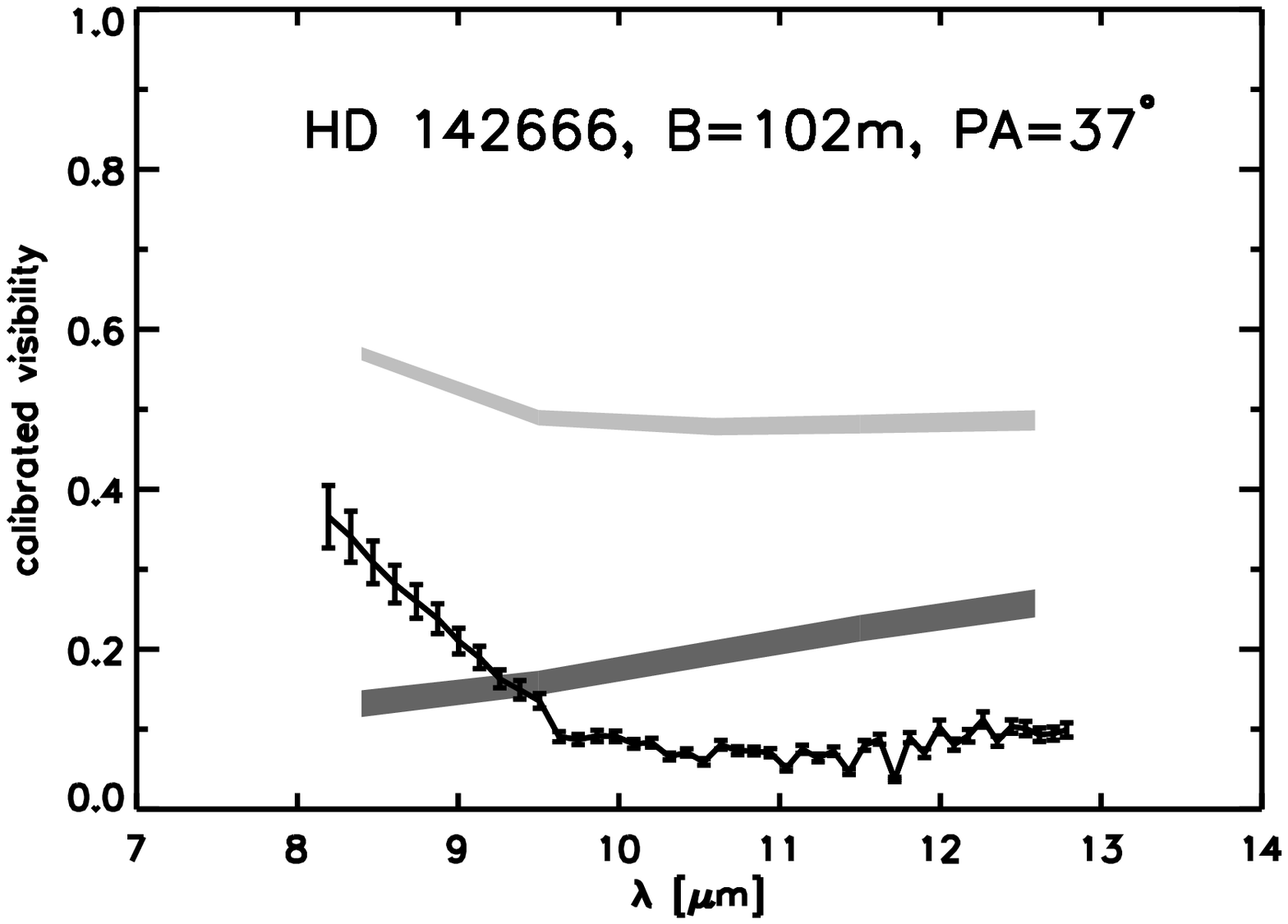}}\newline
    \parbox{0.04\linewidth}{\raisebox{1.em}{iii}}
    \resizebox{0.23\textwidth}{!}{\includegraphics{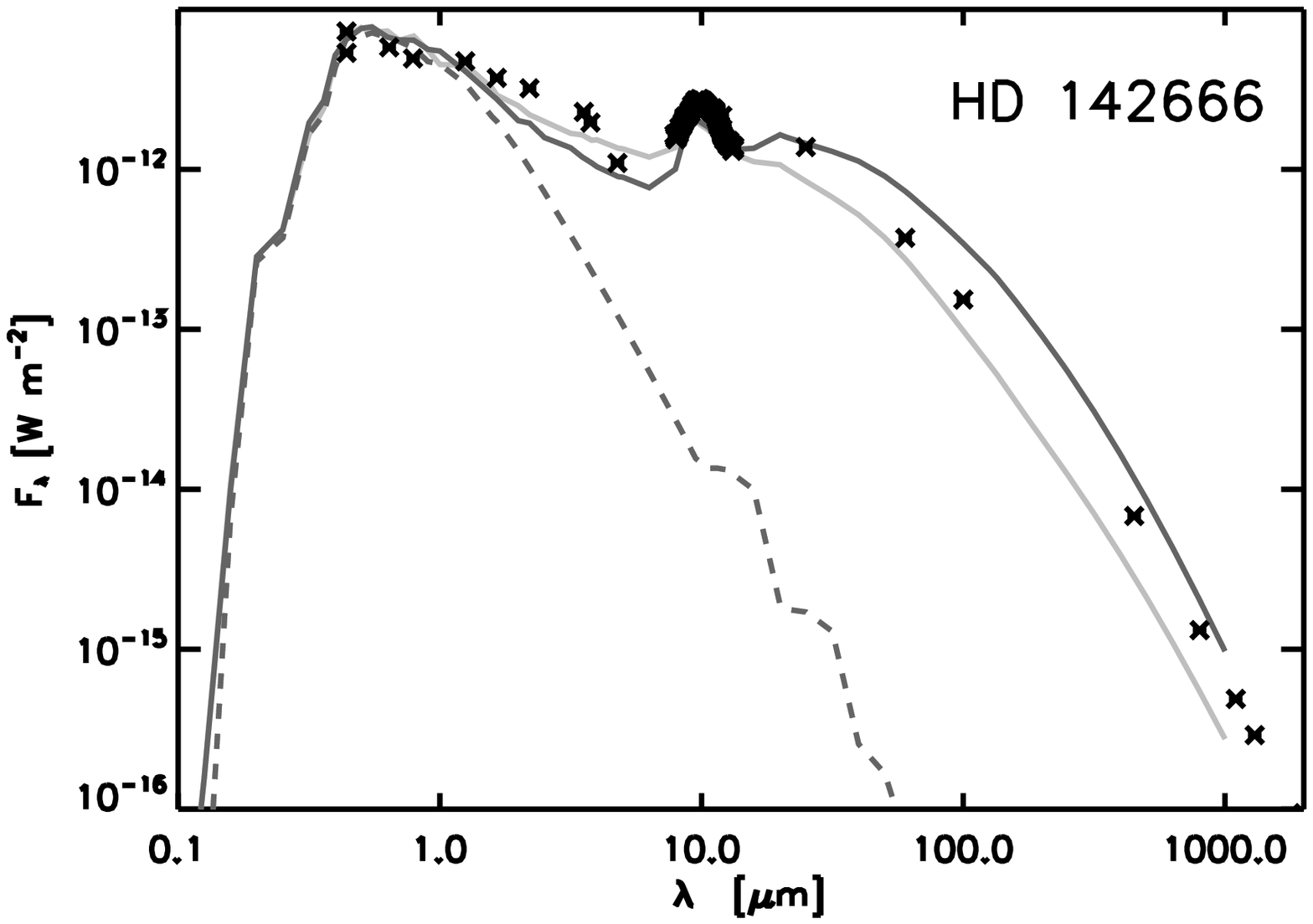}}
    \resizebox{0.23\textwidth}{!}{\includegraphics{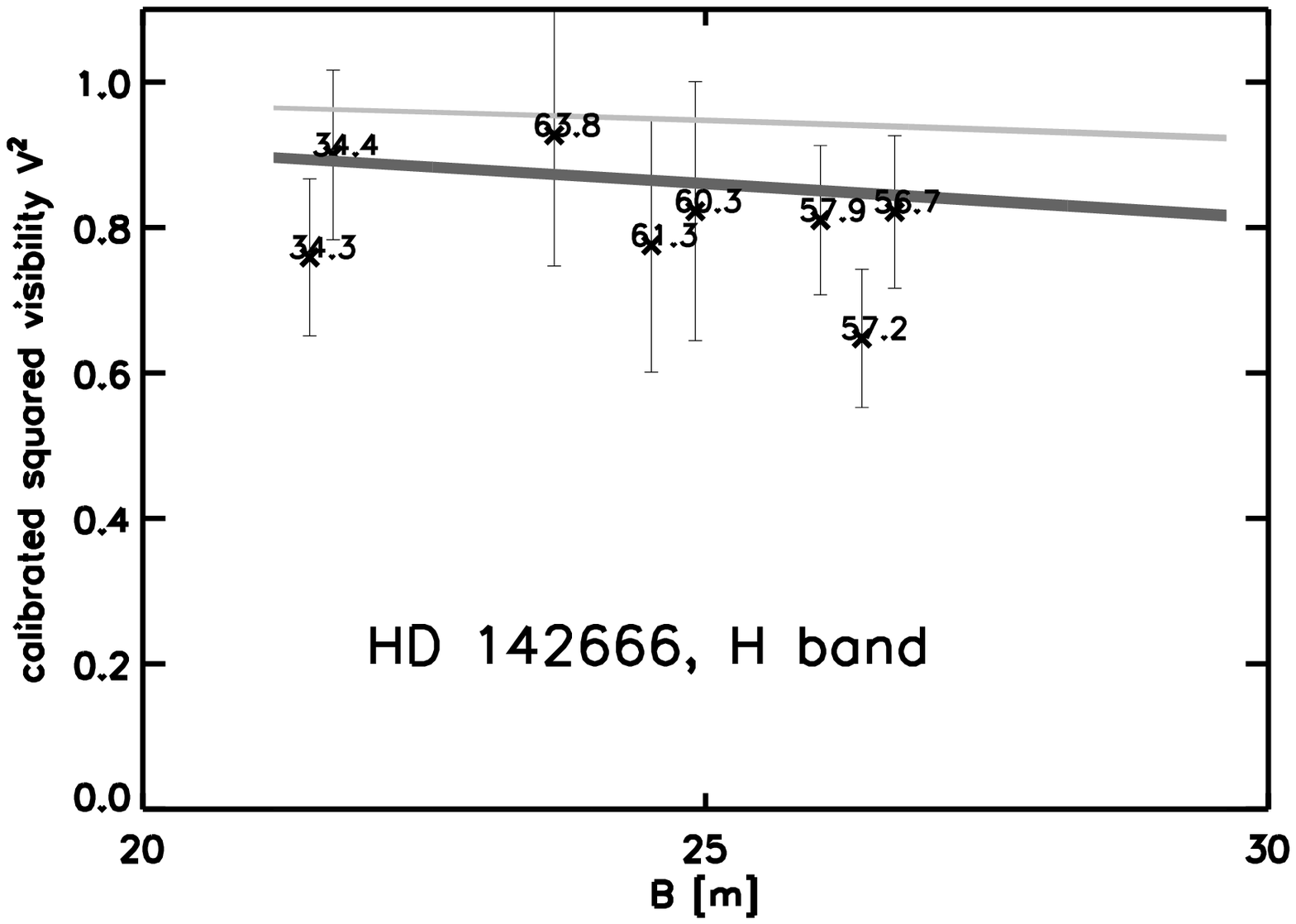}}
    \resizebox{0.23\textwidth}{!}{\includegraphics{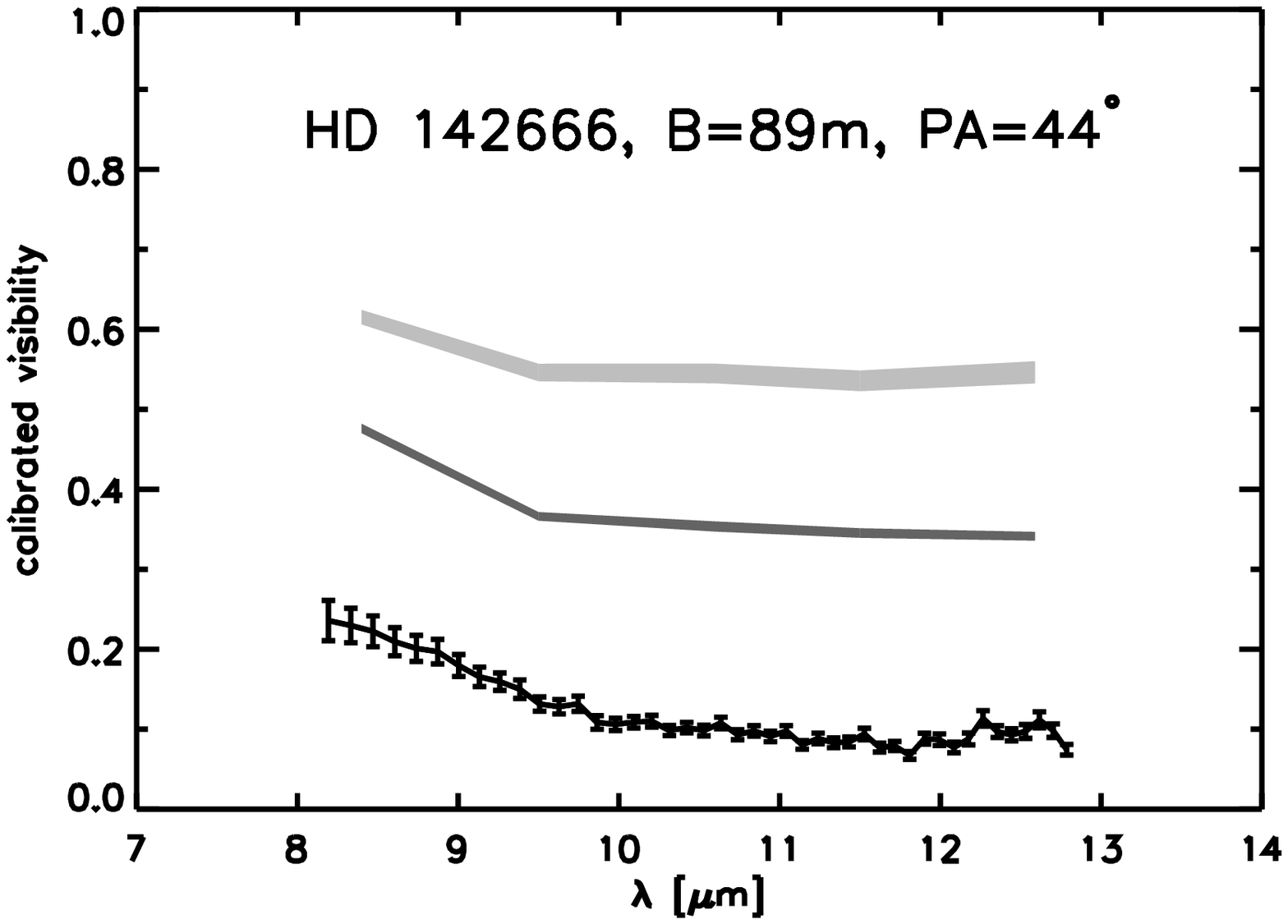}}
    \resizebox{0.23\textwidth}{!}{\includegraphics{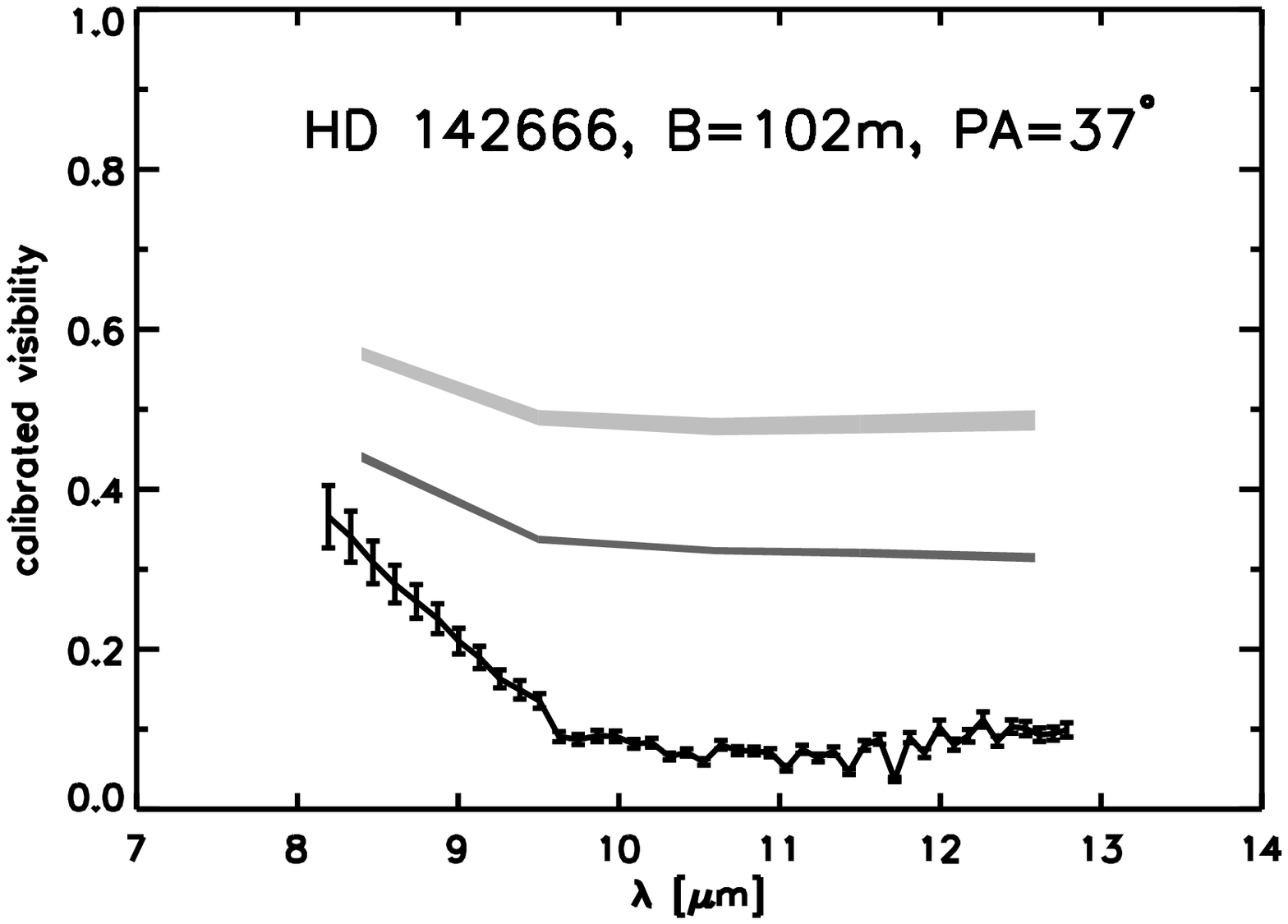}}\newline
    \parbox{0.04\linewidth}{\raisebox{1.em}{iv}}
    \resizebox{0.23\textwidth}{!}{\includegraphics{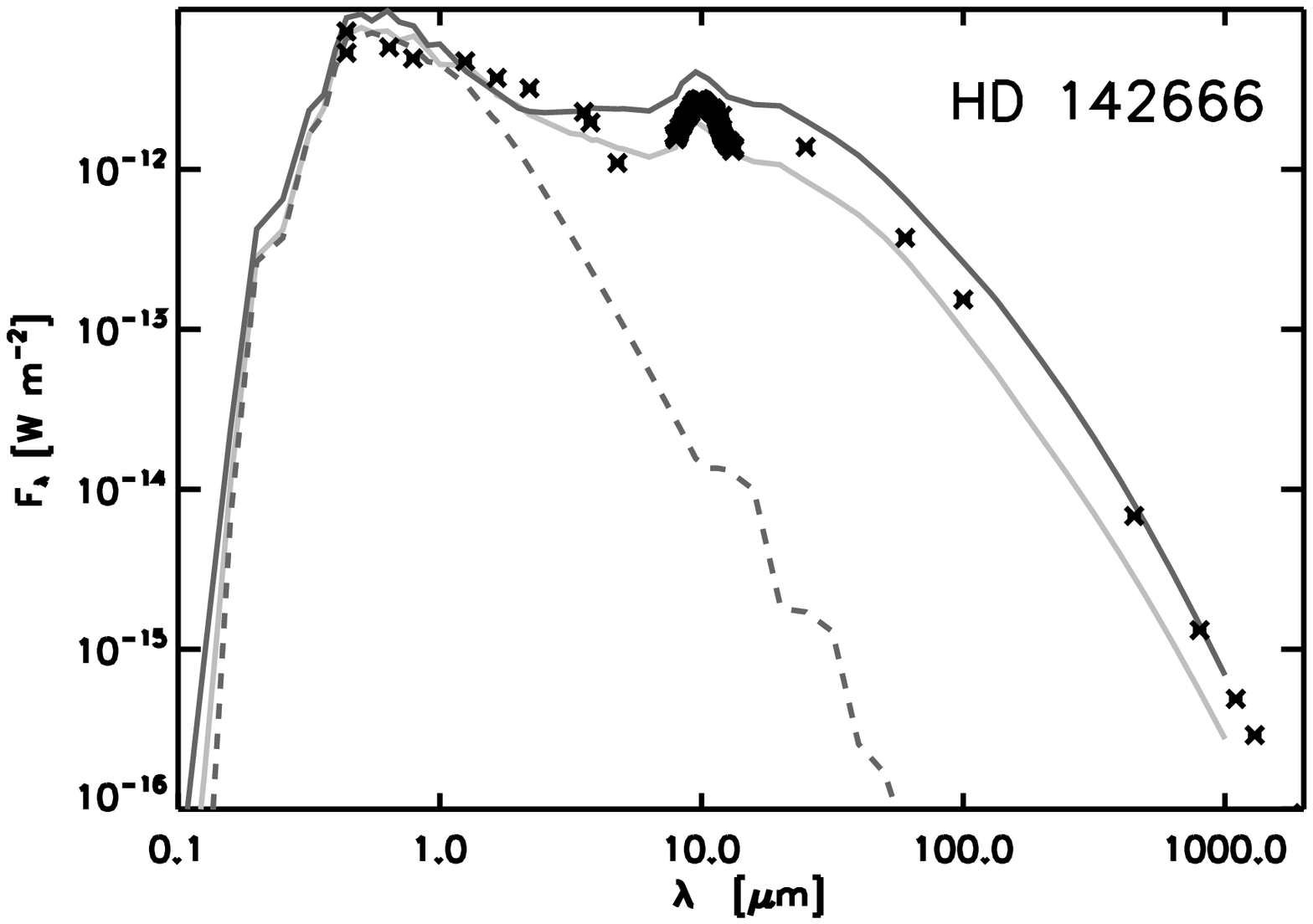}}
    \resizebox{0.23\textwidth}{!}{\includegraphics{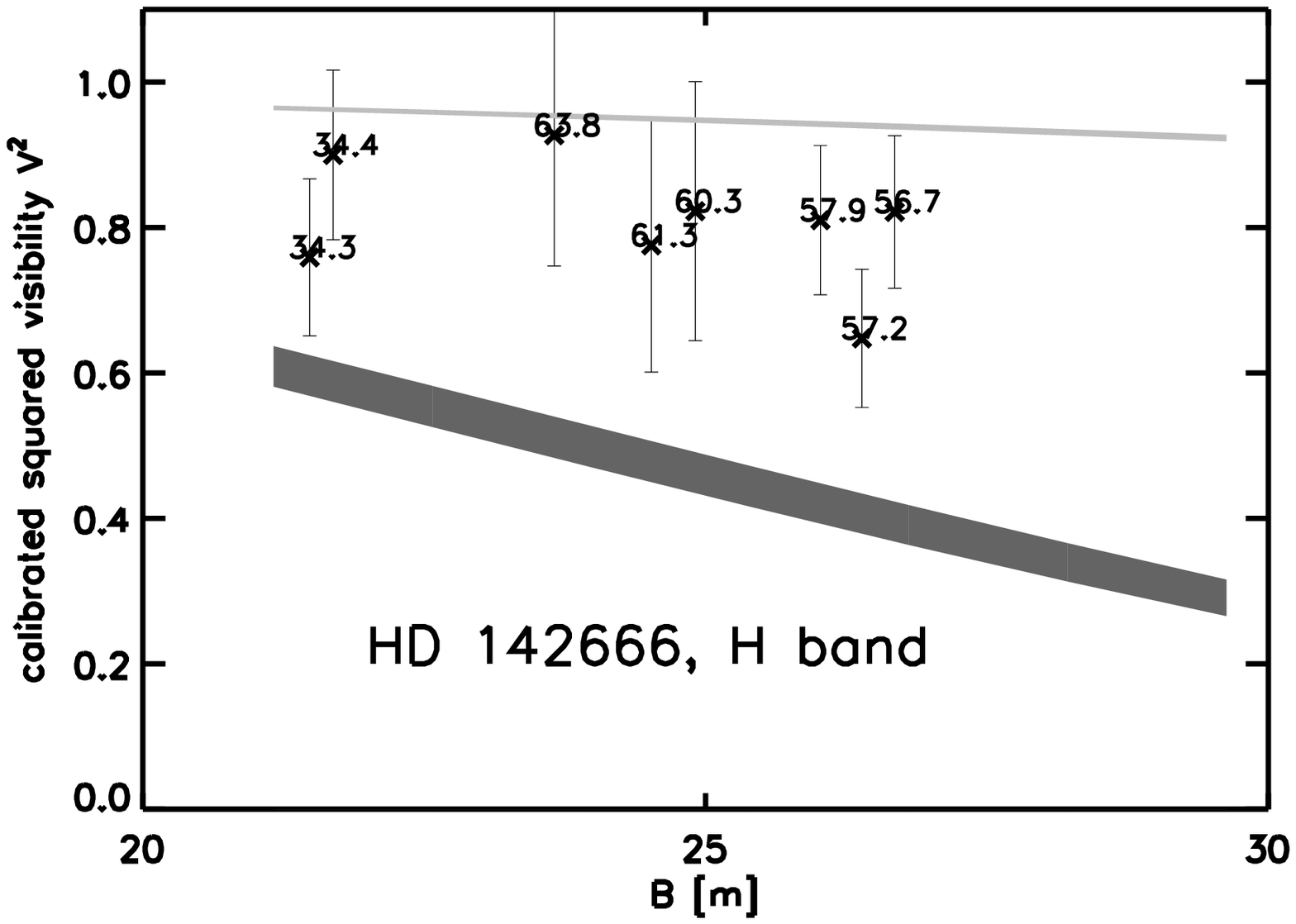}}
    \resizebox{0.23\textwidth}{!}{\includegraphics{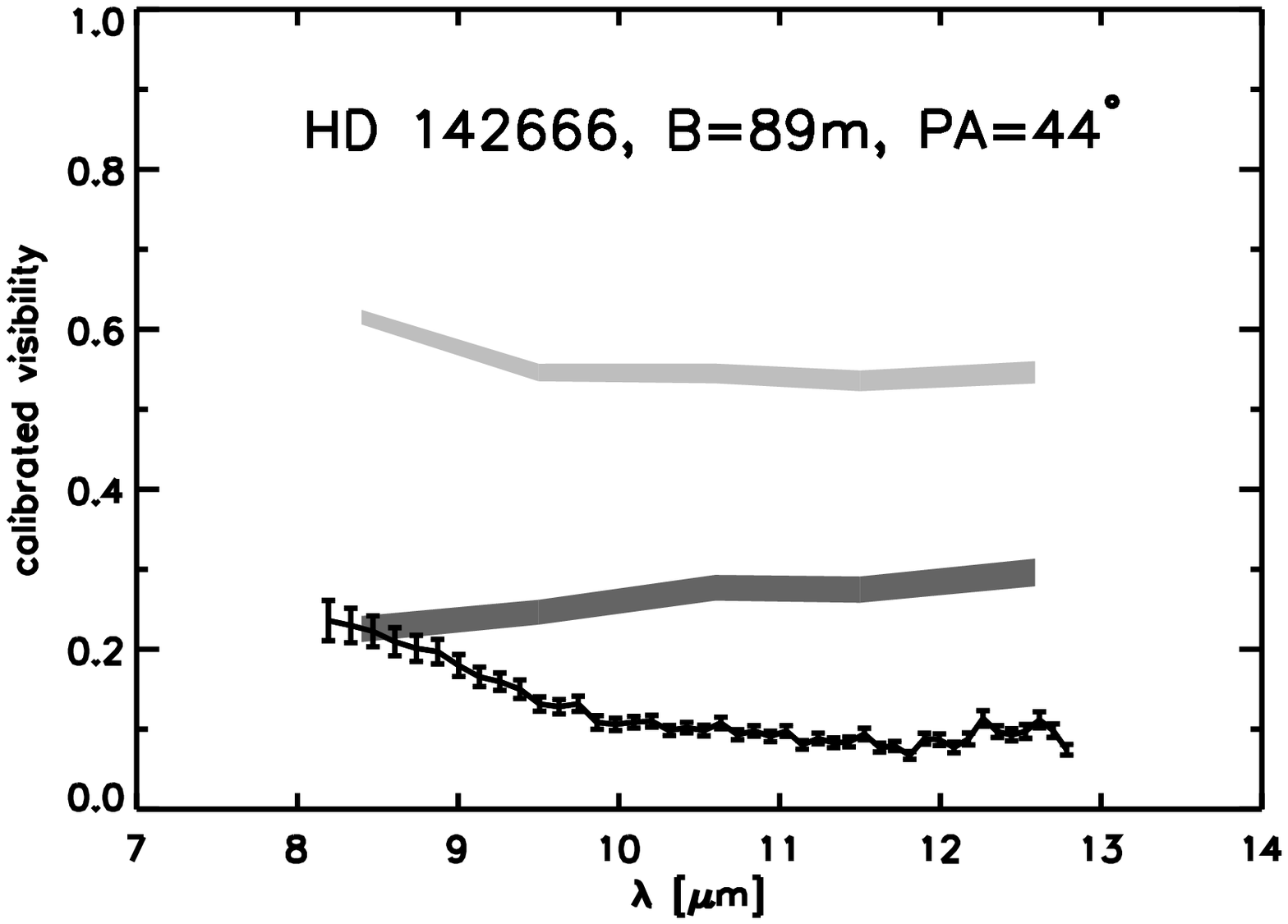}}
    \resizebox{0.23\textwidth}{!}{\includegraphics{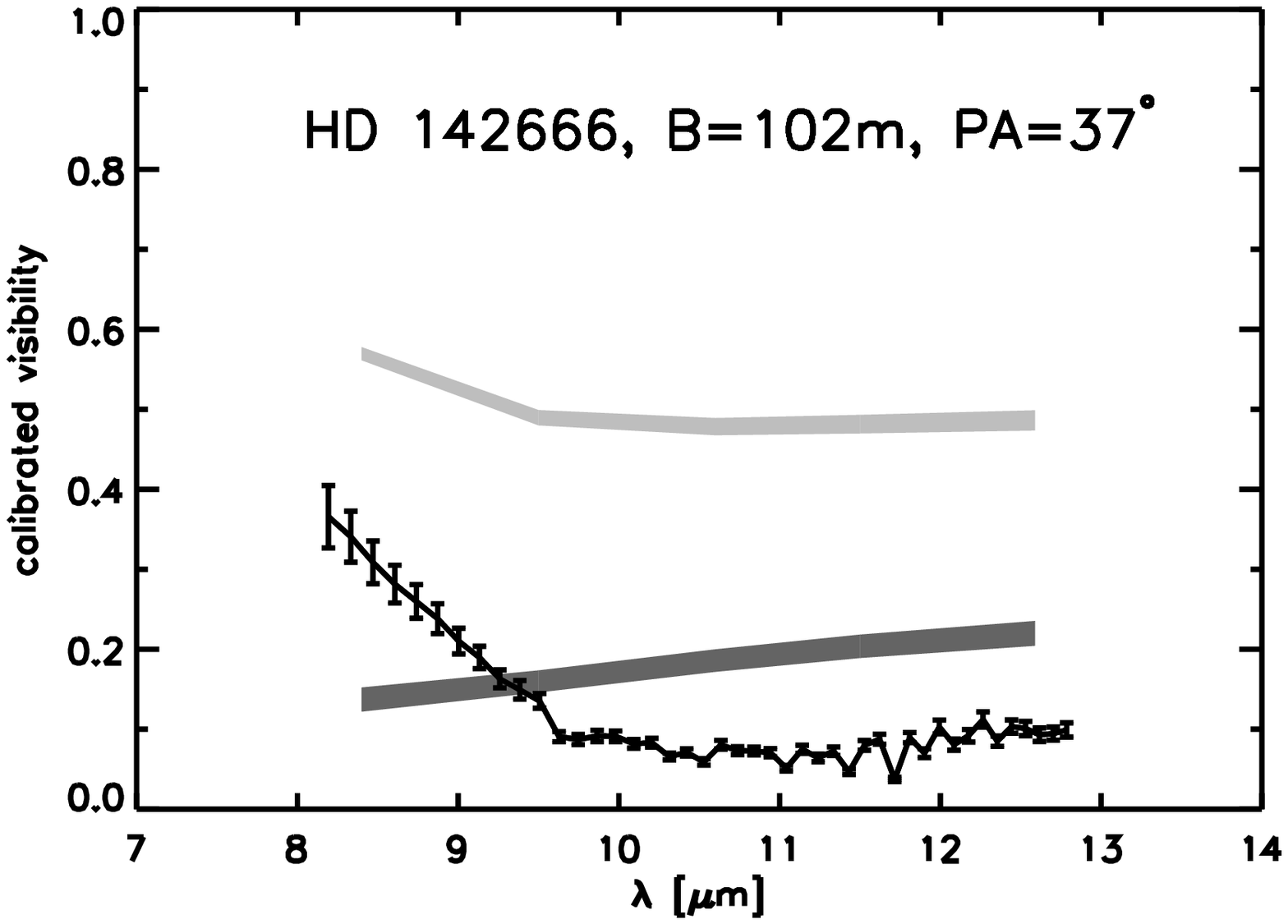}}\newline
    \parbox{0.04\linewidth}{\raisebox{1.em}{v}}
    \resizebox{0.23\textwidth}{!}{\includegraphics{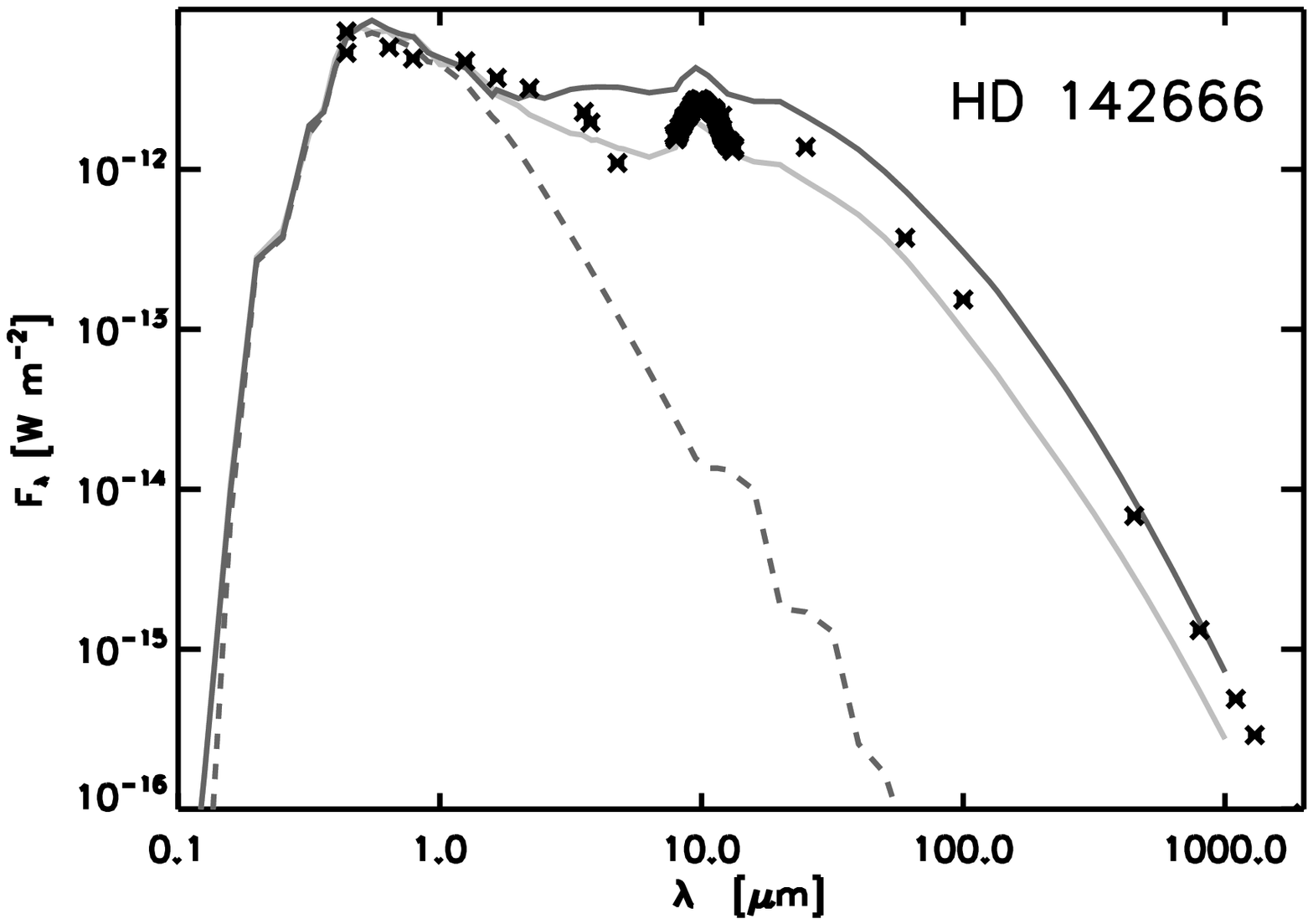}}
    \resizebox{0.23\textwidth}{!}{\includegraphics{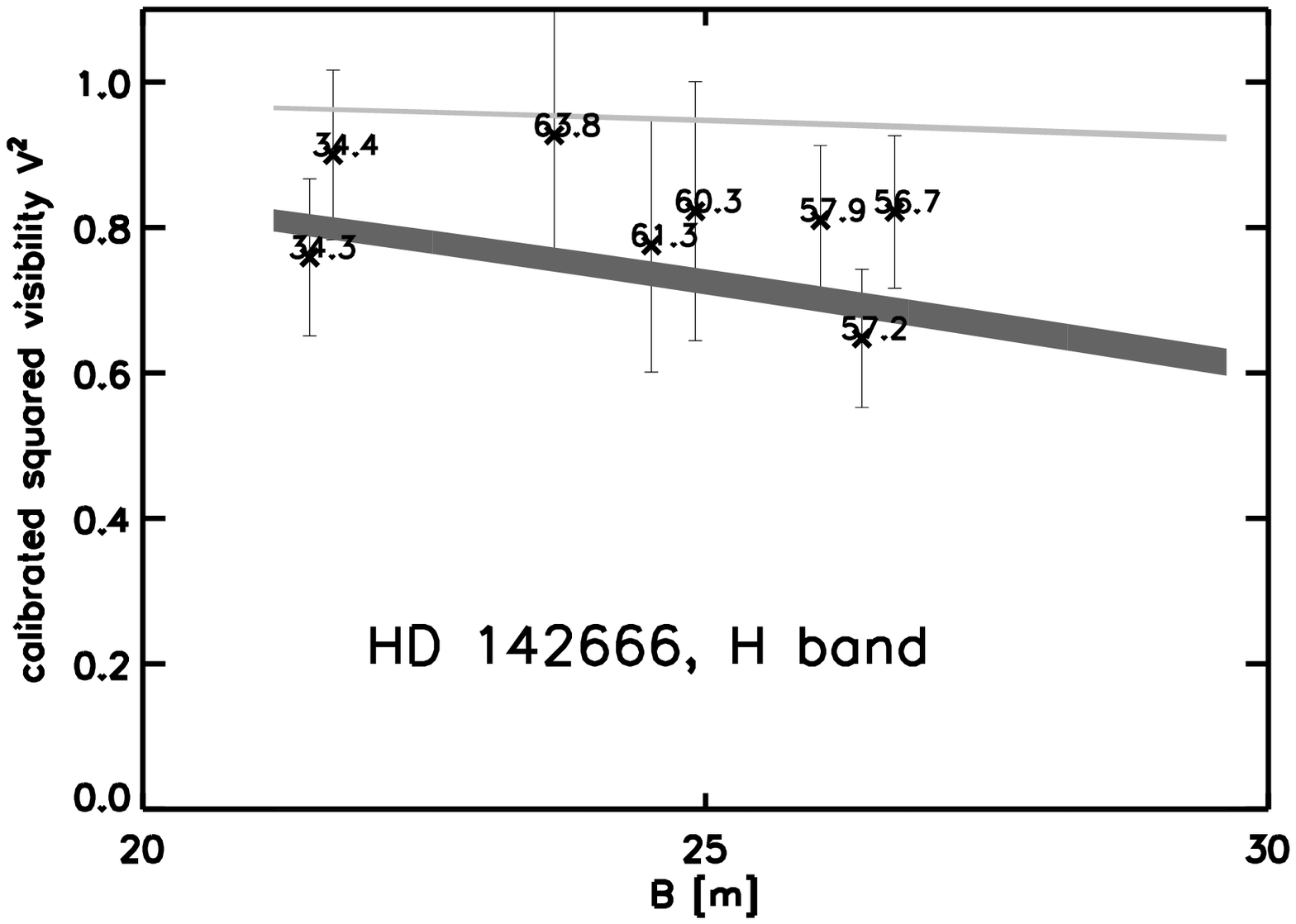}}
    \resizebox{0.23\textwidth}{!}{\includegraphics{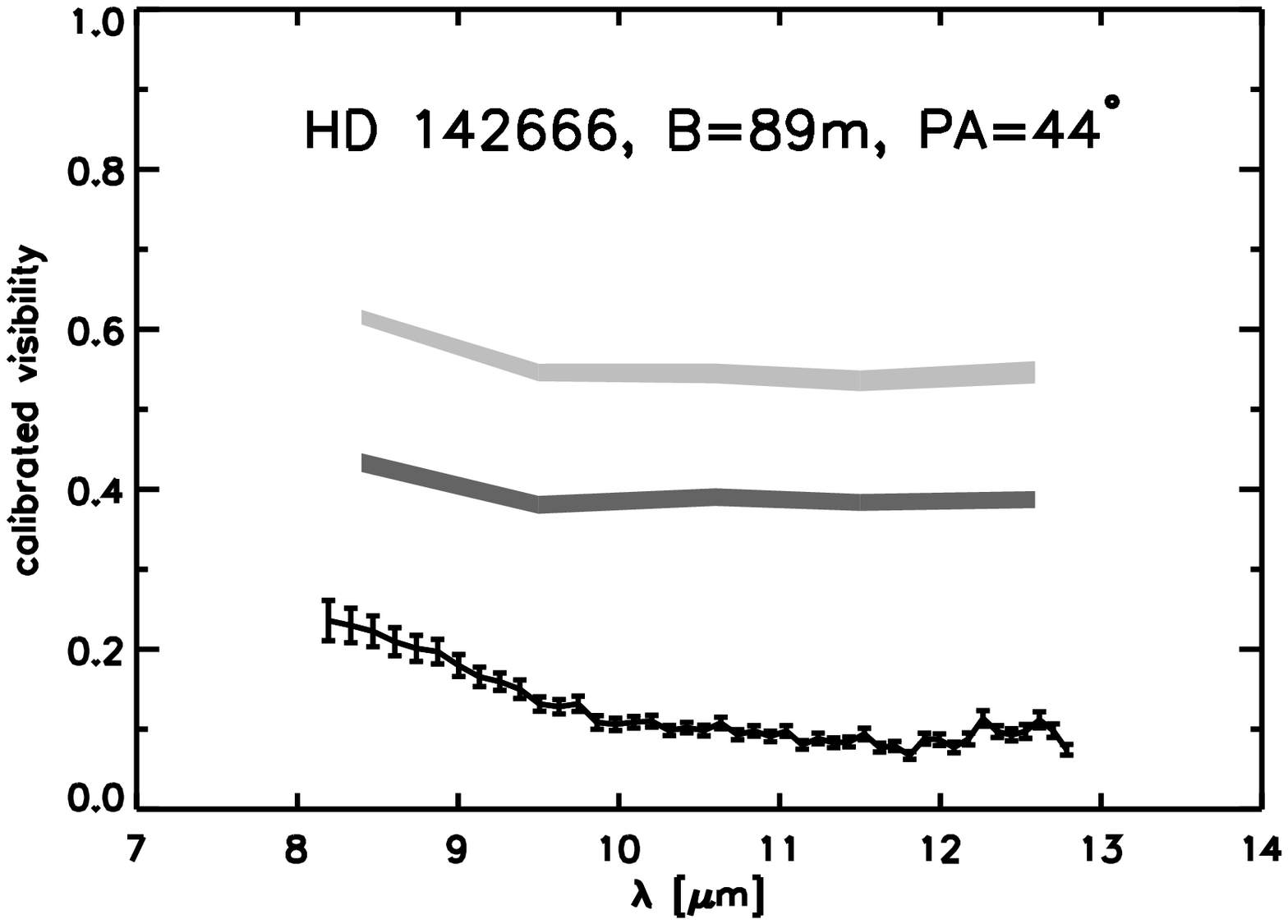}}
    \resizebox{0.23\textwidth}{!}{\includegraphics{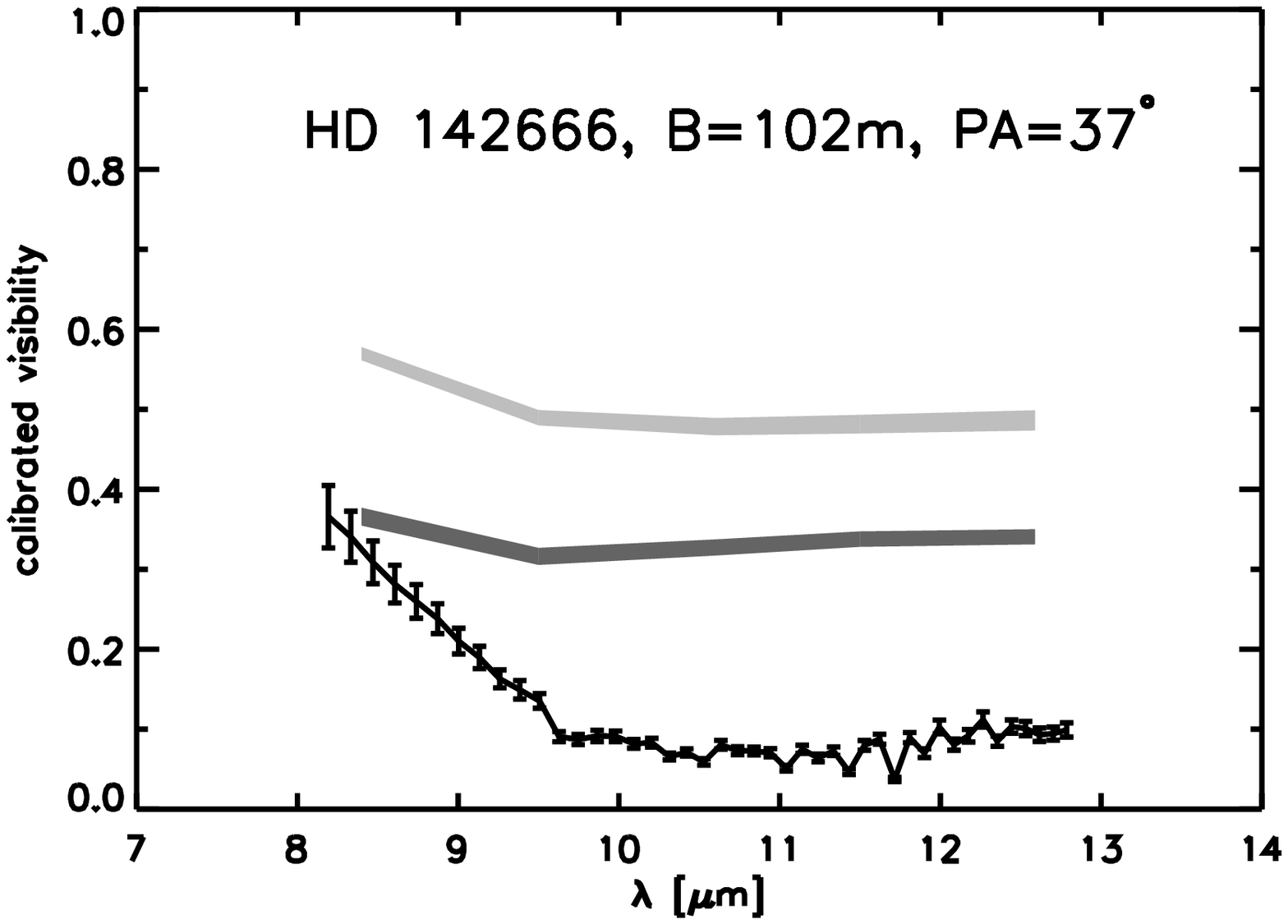}}\newline
    \parbox{0.04\linewidth}{\raisebox{1.em}{vi}}
    \resizebox{0.23\textwidth}{!}{\includegraphics{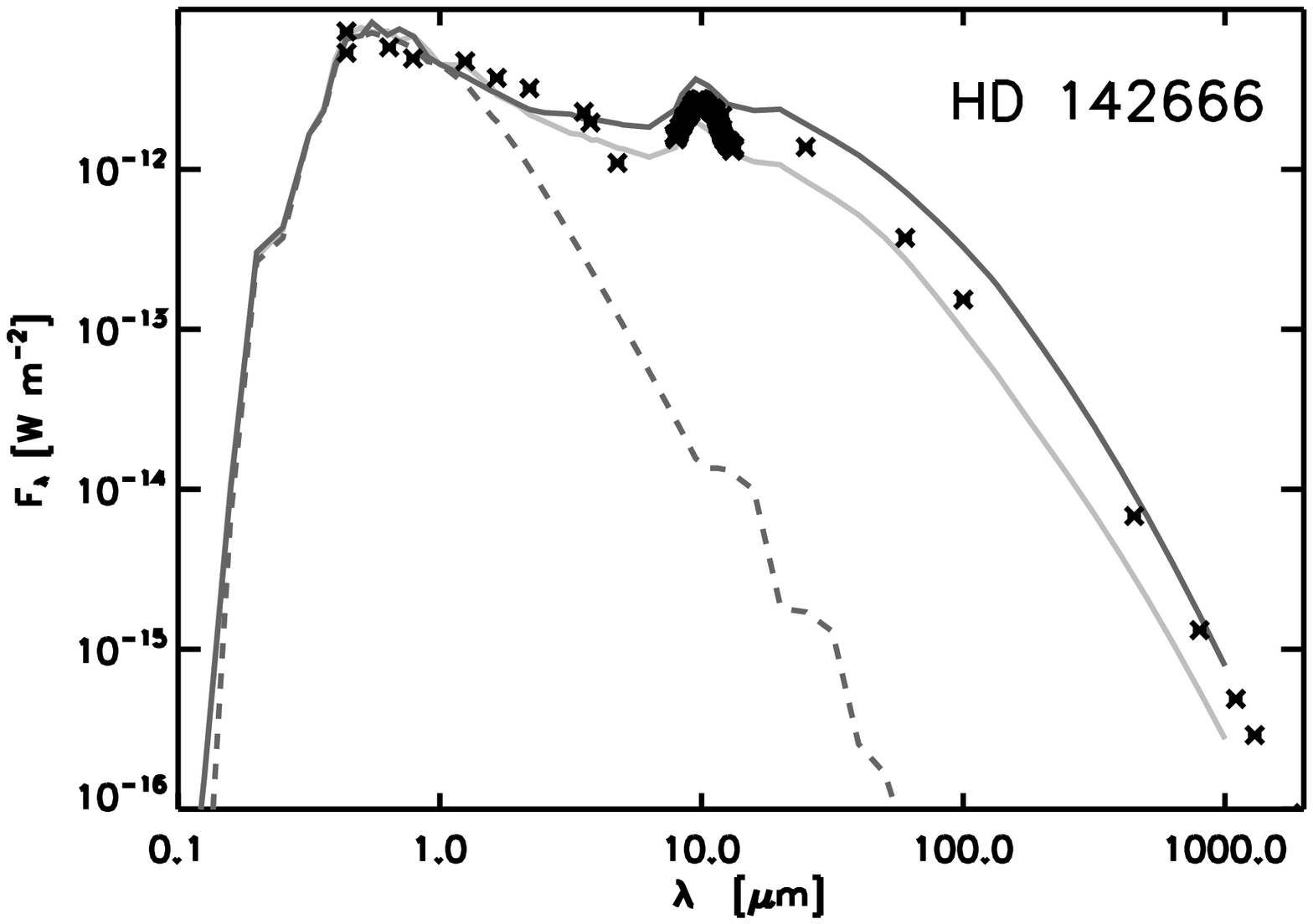}}
    \resizebox{0.23\textwidth}{!}{\includegraphics{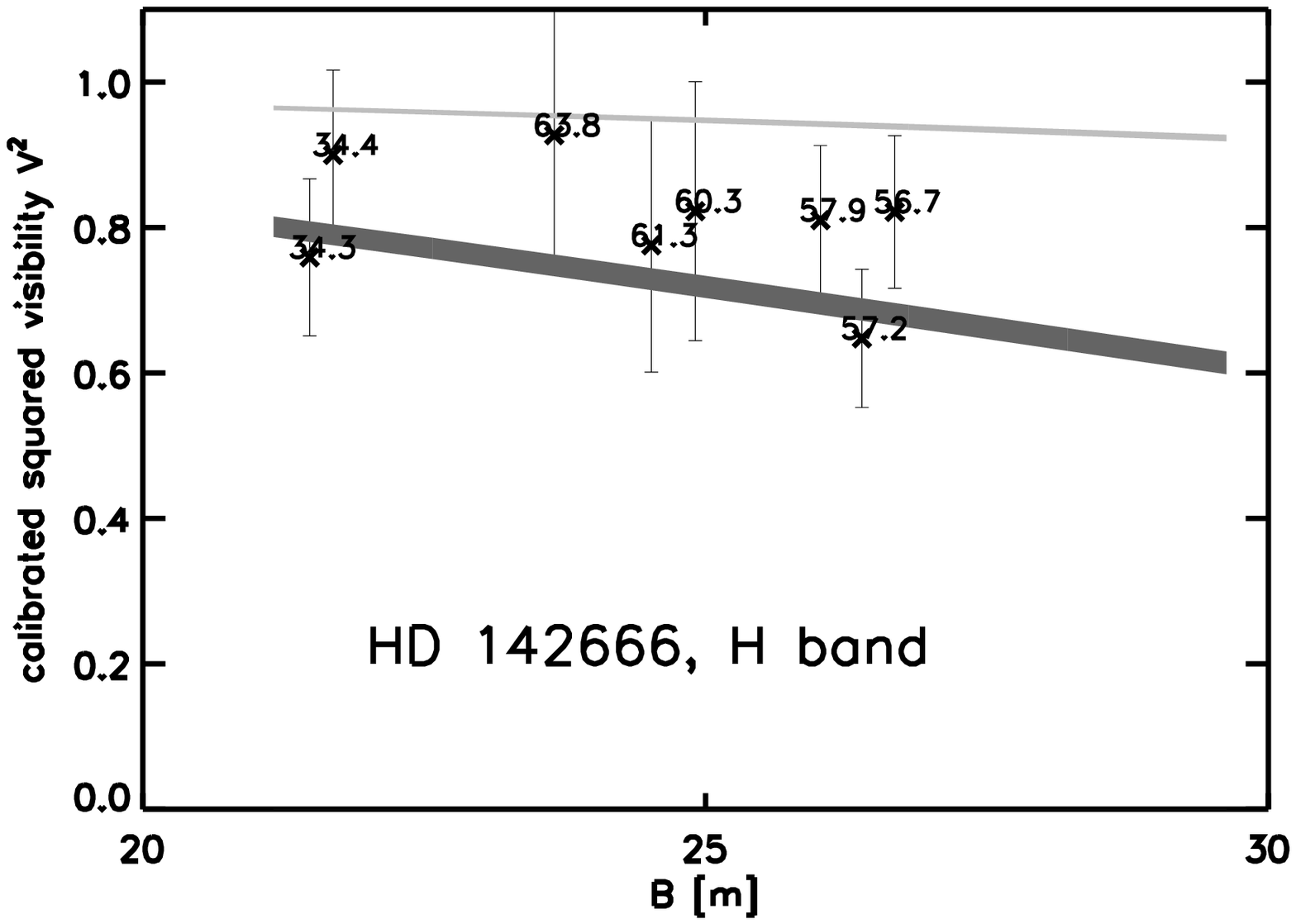}}
    \resizebox{0.23\textwidth}{!}{\includegraphics{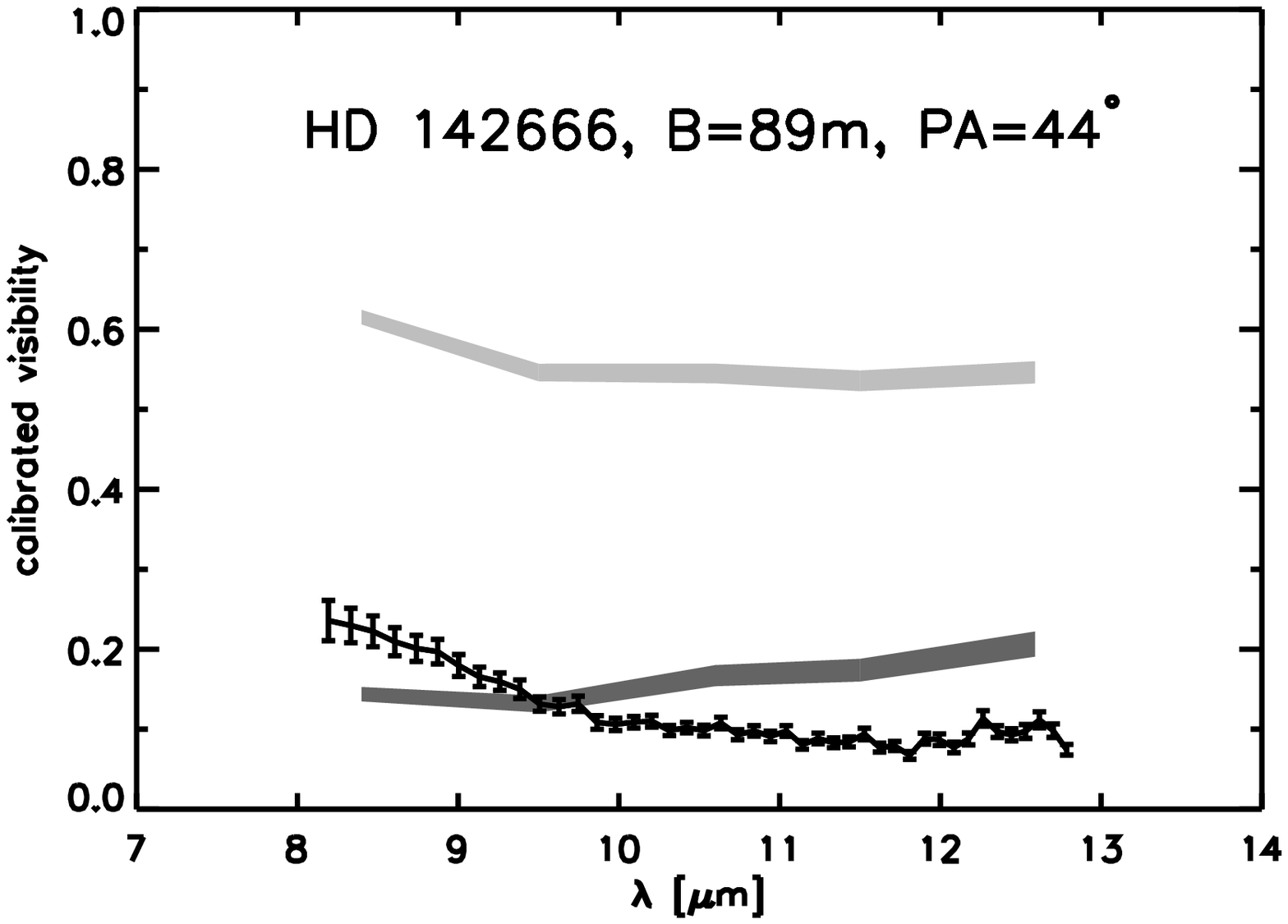}}
    \resizebox{0.23\textwidth}{!}{\includegraphics{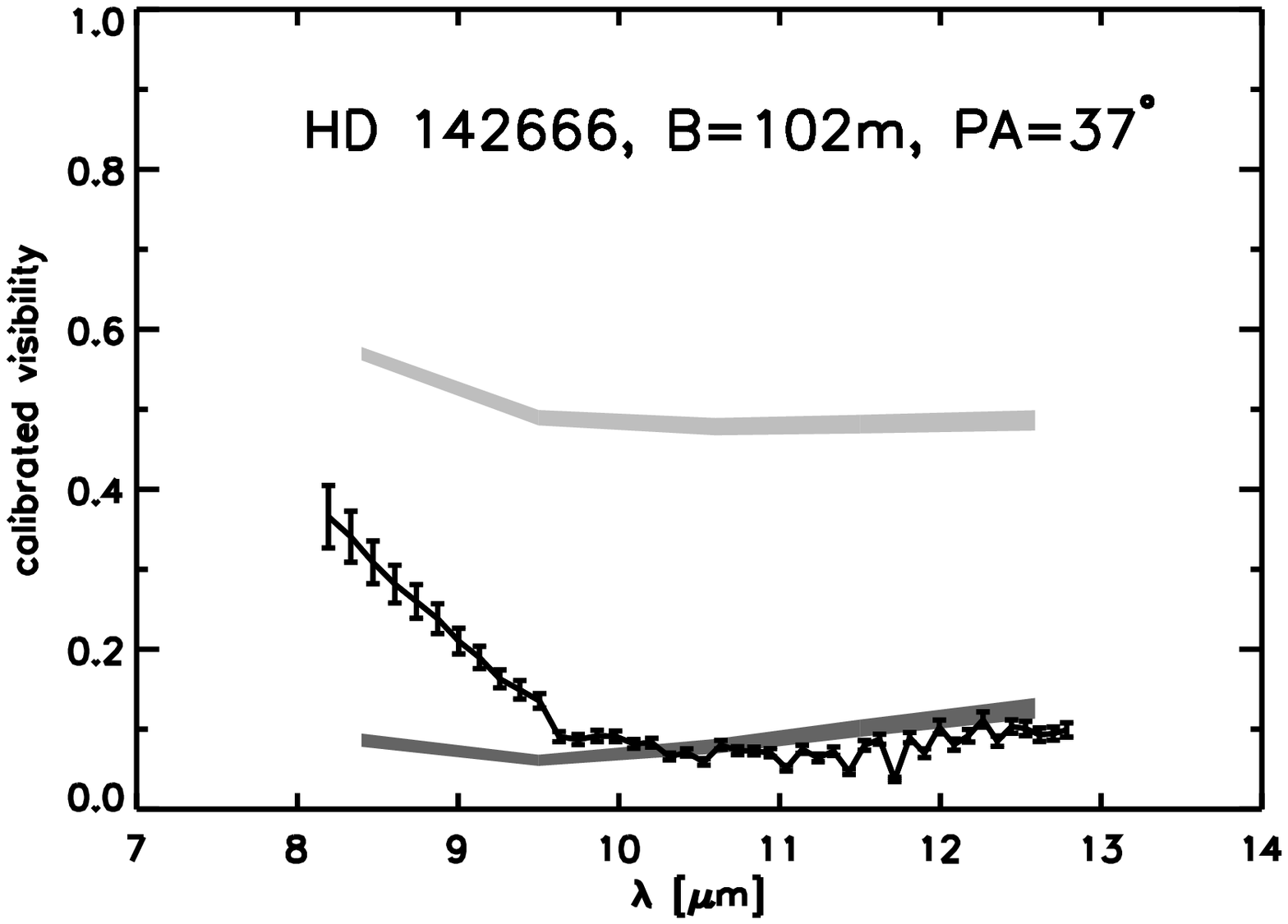}}\newline
    \parbox{0.04\linewidth}{\raisebox{1.em}{vii}}
    \resizebox{0.23\textwidth}{!}{\includegraphics{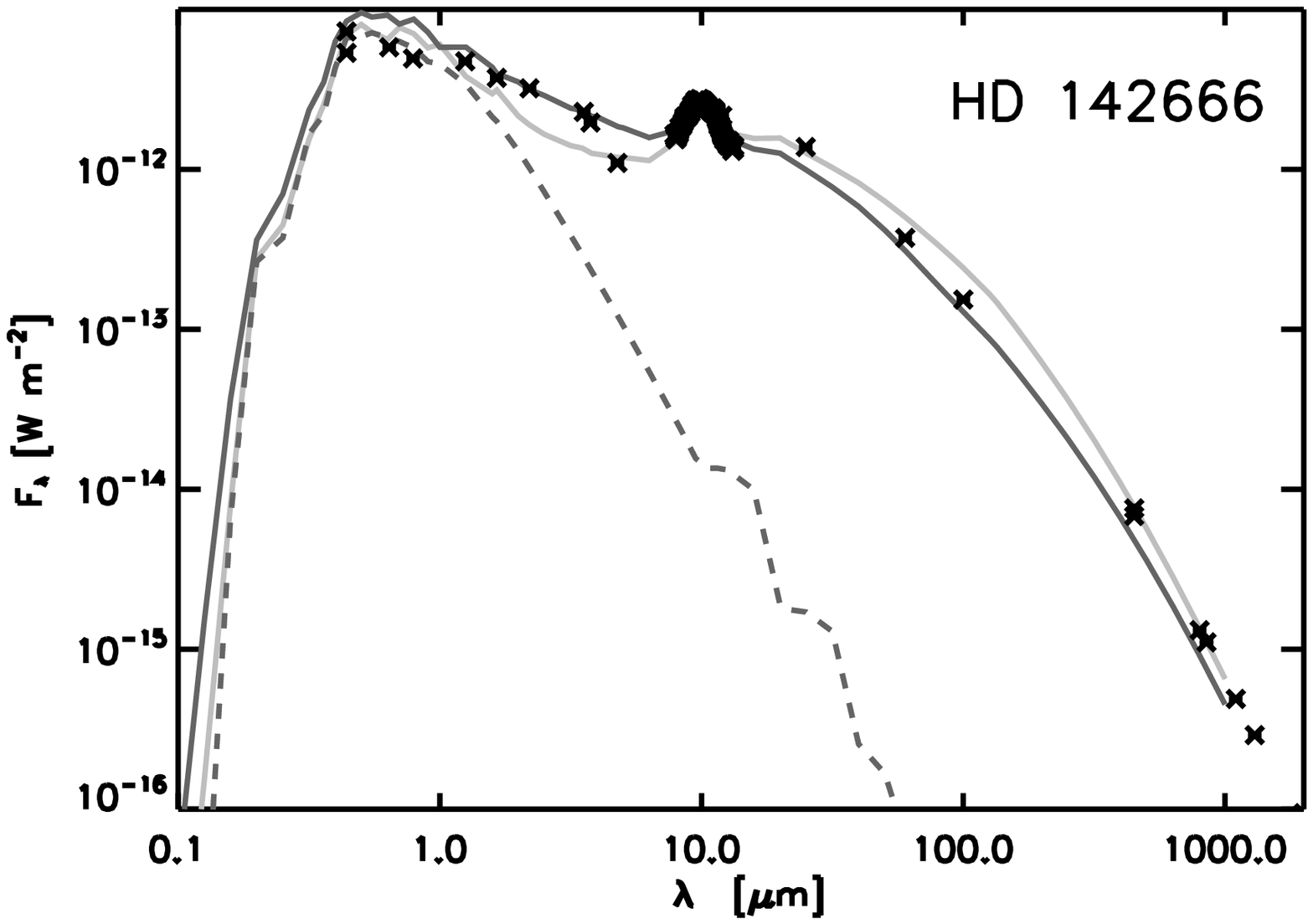}}
    \resizebox{0.23\textwidth}{!}{\includegraphics{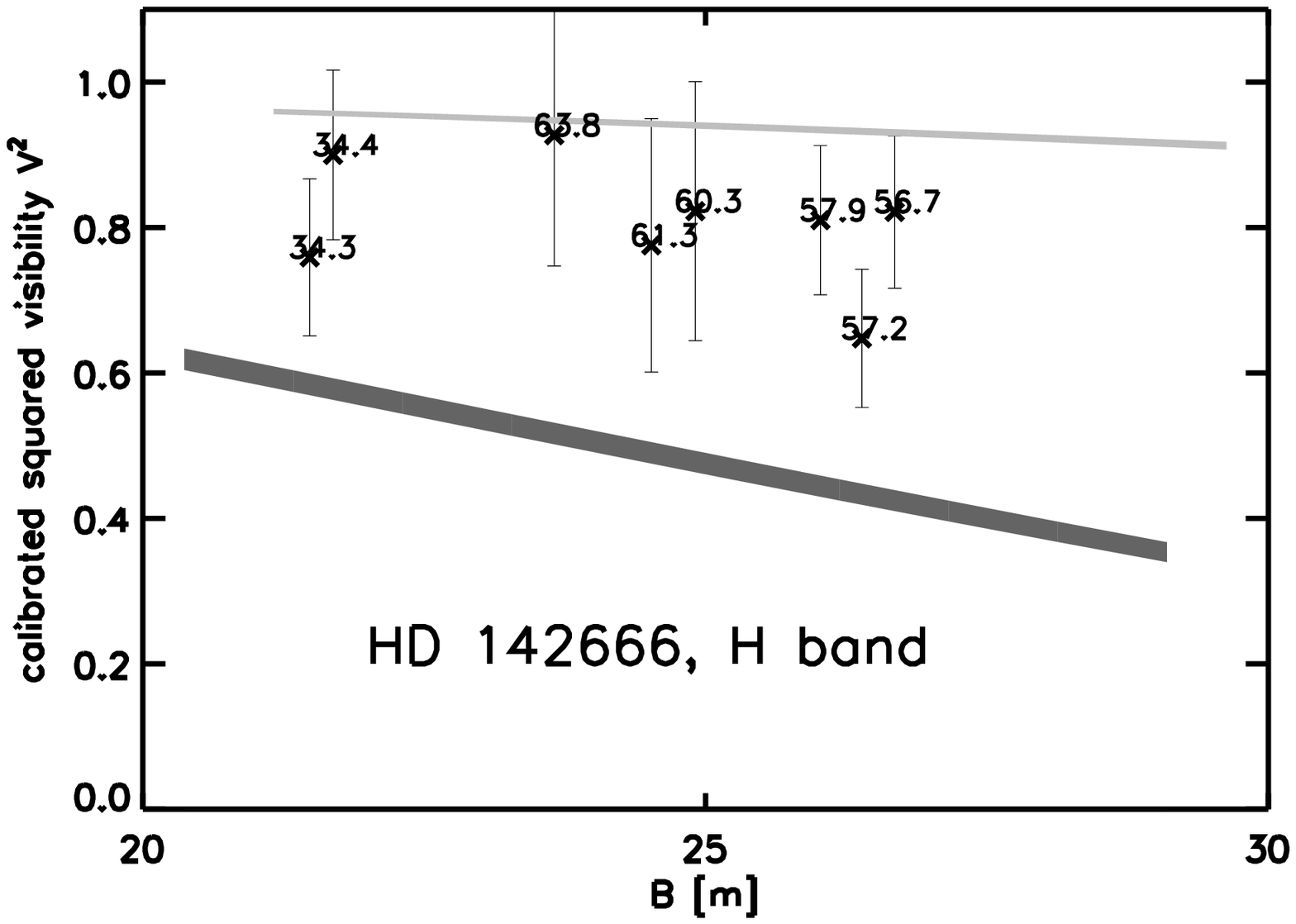}}
    \resizebox{0.23\textwidth}{!}{\includegraphics{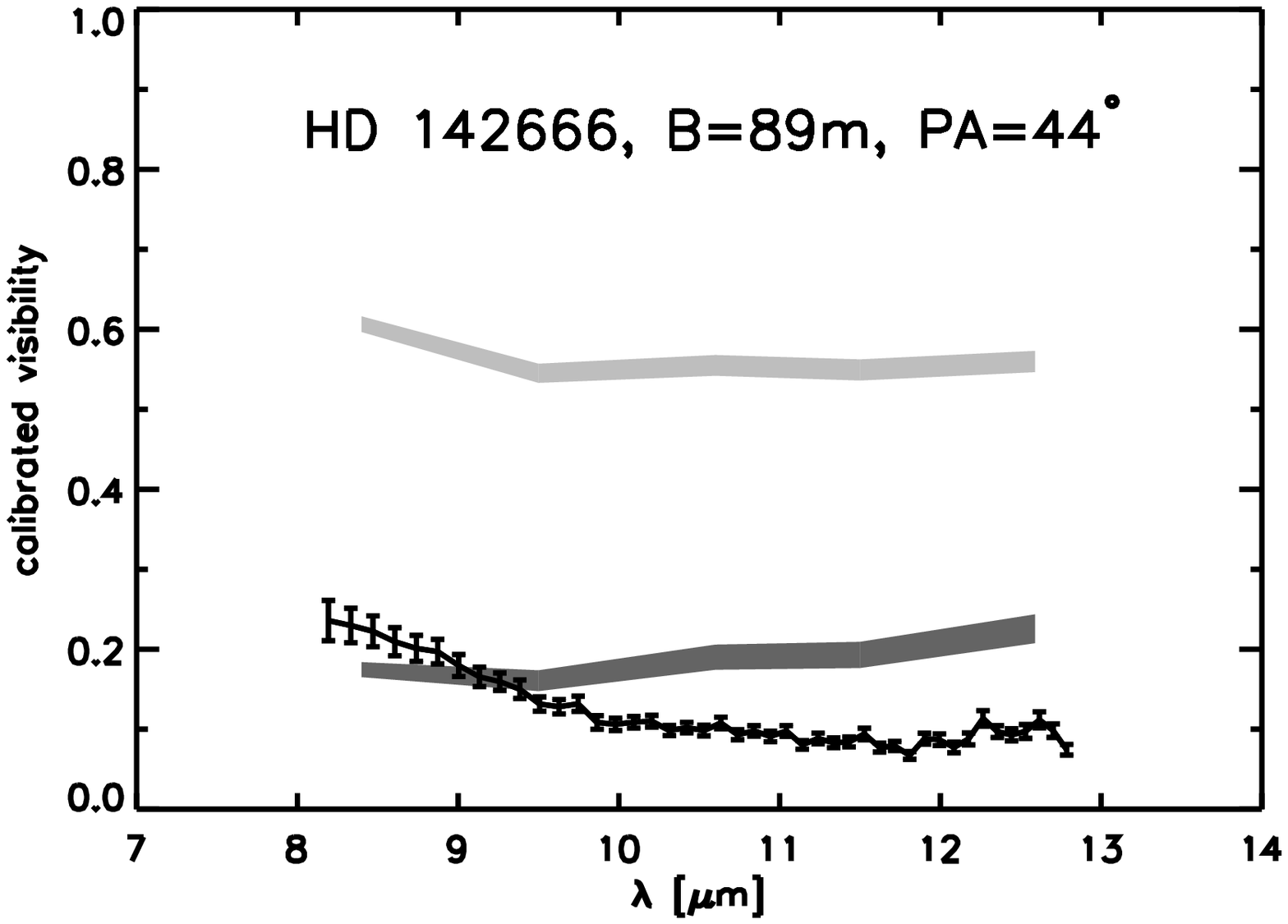}}
    \resizebox{0.23\textwidth}{!}{\includegraphics{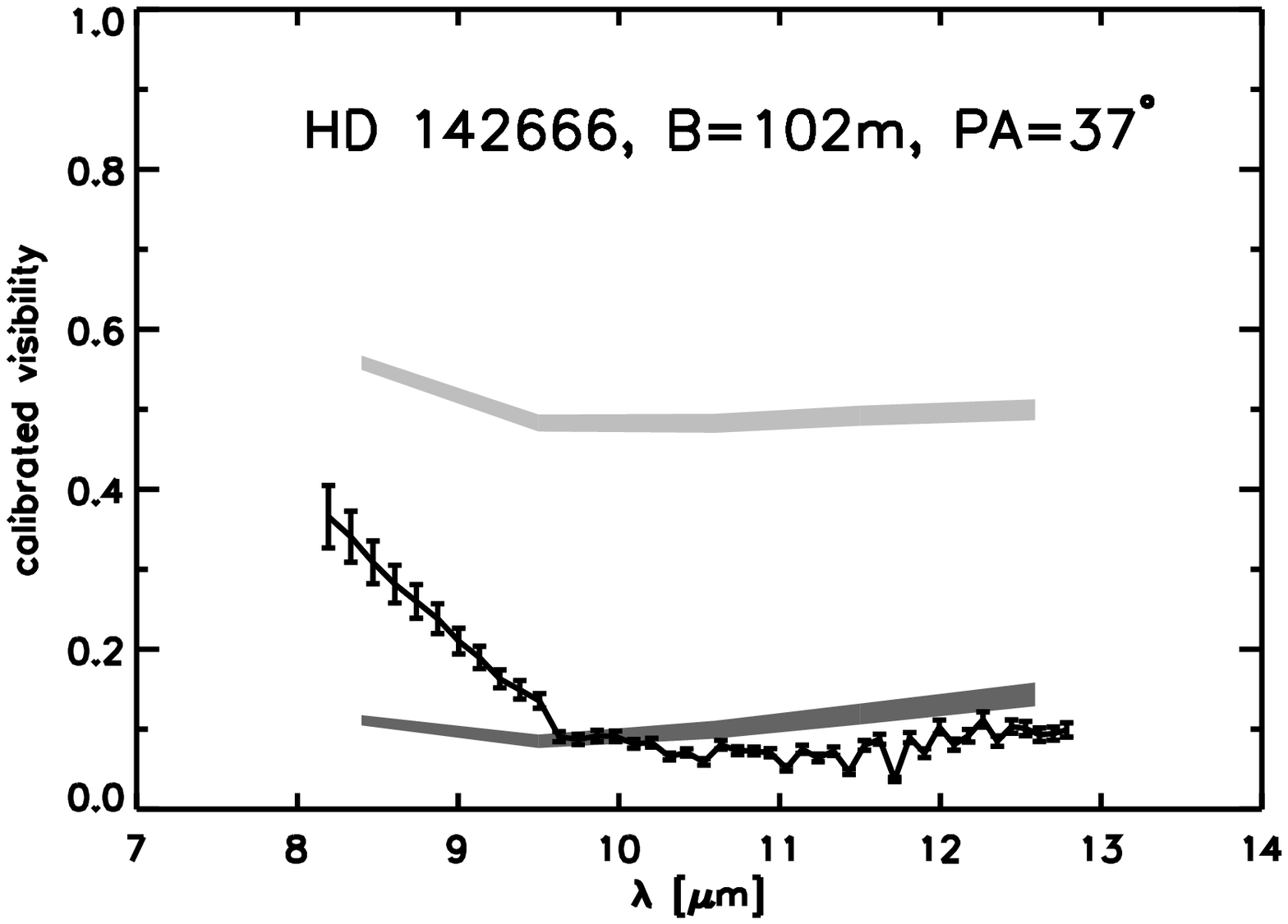}}\newline
    \caption{{Crosses in the SEDs ({\it first column}) represent
      measured photometric data obtained from the literature
      (Appendix~\ref{appendix}). The solid and the dashed curves represent the SED 
      obtained from computer models, as well as its initial stellar flux,
      respectively. The dark and light gray bars in the figures of the {\it second column} result from
      H band images at $\lambda=1.65\,\mathrm{\mu m}$ of the models of
      HD\,142666, while the crosses with the vertical error bars represent the
      measurements with IOTA. The bar width
      represents the interval that constrains the NIR visibilities $V(\lambda)$ for
      different PAs but the same inclination of the model. Numbers added to
      the crosses are the position  
      angles of the corresponding interferometric baseline. The measured MIR, spectrally
      dispersed visibilities ({\it $3^{rd}$ 
        and $4^{th}$ columns}) are  
      represented by solid lines with error bars. The modeled visibilities
      ({\it gray bars}) were derived from model images at wavelengths of
      $8.5\,\mathrm{\mu m}$, $9.5\,\mathrm{\mu m}$, $10.6\,\mathrm{\mu m}$,
      $11.5\,\mathrm{\mu m}$, and $12.5\,\mathrm{\mu m}$. The bar width
      represents the interval that constrains the MIR visibilities $V(\lambda)$ for
      different PAs but the same inclination of the model. 
      The {\it upper row} shows the effect after increasing the inner disk
      radius by a factor of five, from
      $R_\mathrm{in}=0.1\,\mathrm{AU}$ (light gray; initial model i) to
      $R_\mathrm{in}=0.5\,\mathrm{AU}$ (dark gray, model ii). The second row shows the
      result after increasing the profile parameters from $\beta=1.0$ to
      $\beta=1.1$ (model iii). The mass accretion rate is increased
      to $\dot{M} = 7 \times 10^{-8}\,\mathrm{M_{\odot}a^{-1}}$ in
      model (iv) displayed in the third row while the mass density at the inner edge is decreased
      by a factor of $f_\mathrm{\rho}=0.01$ in the model (v) shown in the fourth
      row. In the sixth row, a dust-free disk gap is also cut in model
      (vi) after the density has decreased (model v). Model of row (vii) is
      identical to model (vi) but without decreasing the density as in model
      (v). The light gray bars
      also presented in all rows show the initial model (i) for comparison.}} 
    \label{figure:hd142666}
  \end{figure*}

  Figure~\ref{figure:hd142666}
  shows the SED of HD\,142666, the spectrally
  unresolved but spatially resolved NIR visibilities from IOTA, as well as the
  spectrally and spatially resolved MIR visibility spectra from MIDI. With
  IOTA, an angular resolution of 6.3\,mas ($0.72\,\mathrm{AU}$ at
  $114\,\mathrm{pc}$) could be reached. The HAeBe object  
  was observed with MIDI in the context of guaranteed time
  observations where a spatial resolution of $1.1\,\mathrm{AU}$ and
  $0.98\,\mathrm{AU}$ for $\lambda = 8-13\,\mathrm{\mu m}$ could be reached
  using the projected baselines $B=89\,\mathrm{m}$ 
  and $B=102\,\mathrm{m}$, respectively. The low MIR 
  visibility spectra correspond to a strongly spatially resolved circumstellar
  disk in the MIR range.
 
  We used different disk parameter sets and different approaches for disk models
  until we could finally fit all the data obtained. The models (i-iv;
  Table~\ref{table:properties-midisurveyII}) show the modeling results
  that consider the classical approach of Eq.~\ref{eq:shakura} and accretion,
  while the disk densities were decreased in models (v-vi) using the
  modifications described in Sect.~\ref{section:extended}. 
  \renewcommand{\labelenumi}{\roman{enumi}}
  \begin{enumerate}
  \item Considering the stellar properties, we initially assumed an inner disk radius
    of $R_\mathrm{in}=0.1\,\mathrm{AU}$. After scanning the parameter space of
    the canonical approach of Shakura \& 
    Sunyaev~(\cite{shakura}), no model could be found that reproduces
    the SED, MIR visibility, and NIR visibilities of this source,
    simultaneously. While the NIR visibilities
    are consistent with the dust sublimation radius, the $10\,\mathrm{\mu m}$
    emission appears to arise from a region that is more extended than 
    predicted by assuming a smooth radial profile of the dust density and
    temperature distribution. 
  \item The low MIR visibility spectra favor an inner radius $R_\mathrm{in} >
    0.1\,\mathrm{AU}$. For an increasing inner radius, the NIR and MIR flux
    distribution shift to larger radii resulting in a decrease in the
    corresponding visibilities. However, a decrease in the NIR visibilities does
    not solely reproduce the IOTA data. The NIR photometric flux, which has its origin
    mainly at the inner disk edge (Fig.~\ref{figure:rad-int}), generally
    decreases for an increasing inner disk radius $R_\mathrm{in}$, and the fit 
    of the photometric measurements in the NIR range gets worse. As the
    disk mass is kept constant, there are more particles that emit in the
    MIR wavelength range at $>0.1\,\mathrm{AU}$ and the MIR photometric flux
    increases. The upper row of Fig.~\ref{figure:hd142666} shows the modifications
    of the simulation results if the inner disk radius is increased by a
    factor of $5$ from $R_\mathrm{in}=0.1\,\mathrm{AU}$ in model (i) to
    $R_\mathrm{in}=0.5\,\mathrm{AU}$ in model (ii). 
  \item To shift the MIR flux distribution to larger disk radii and, thus, to
    decrease the MIR visibility spectra in the model, we increased the profile
    parameters $h_\mathrm{100}$ and $\beta$ resulting in disk flaring where
    outer disk regions are more strongly illuminated and heated by the central
    star. Another effect of disk flaring is the increase in the far-infared
    (FIR) flux by orders of magnitude. Outer disk regions are still too cold
    to contibute sufficiently to the NIR flux, so the NIR visibility only  
    slightly decreases. Considering
    Eqs.~\ref{eq:shakura}~and~\ref{eq:alpha-beta}, the mass density of the
    inner disk regions also 
    decreases, resulting in too low NIR flux with respect to the photometric
    measurements. The second row of
    Fig.~\ref{figure:hd142666} shows the model (iii) where 
    the profile parameter $\beta$ has been increased from $\beta=1.0$ to
    $\beta=1.1$, while all the other parameters were kept constant. 
  \item Assuming an inner disk radius $R_\mathrm{in} > 0.1\,\mathrm{AU}$
    to reproduce the MIR visibilities, an increase in the NIR flux and
    NIR visibility can result from an increase in the accretion rate
    $\dot{M}$ in model (ii). But in contrast to the effect on the SED, a
    modification of the accretion rate only slightly increases the NIR
    visibility in the model with $R_\mathrm{in} =
    0.5\,\mathrm{AU}$. Furthermore, considering the high age of the object of 
    $6.3\,$million years and the measurements of Garcia-Lopez et
    al.~(\cite{garcia-lopez}) where an accretion rate of $\dot{M} = 1 \times
    10^{-8}\,\mathrm{M_{\odot}yr^{-1}}$ was determined by analyzing the FWHM of
    the Br$\gamma$ line, a large accretion rate can be
    excluded. Apart from the accretion rate, any further (infrared) source
    in the disk region that cannot be resolved with IOTA, such as a close 
    stellar/planetary companion or hypothetical inner
    rim wall that is puffed up by the strong irradiation of the central star,
    could increase the NIR flux and NIR visibility in disks with larger
    inner radii. A puffed up inner rim wall, however, also results in a higher
    concentration of warm dust. Consequently, the MIR visibility does not
    decrease since it is necessary for a better fit. Furthermore, the
    stellar temperature and luminosity of HD\,142666 are similar to the 
    corresponding properties of the YSO RY\,Tau. In an extensive
    study using a self-consistent disk model, Schegerer et
    al.~(\cite{schegerer}) have shown that there are
    not hints of a puffed upper rim wall for this YSO.
    Finally, we also implemented an optical and
    geometrically thin disk for $R_\mathrm{sub} < r < R_\mathrm{in}$ with a
    dust grain radii $a \lessapprox 0.01\,\mathrm{\mu m}$. Eisner et
    al.~(\cite{eisnerIII}) succeed in reproducing the SED and NIR
    visibilities of the TO TW\,Hya, assuming such an optically thin structure in
    the inner hole of the disk.  However, not all of the photometric and interferometric data
    could be reproduced by assuming such an inner thin disk. 
  \item To shift the MIR flux distribution to larger disk radii and decrease
    the MIR visibility, the disk density was exponentially reduced according to
    Sect.~\ref{section:extended} using $f_\mathrm{\rho}=0.01$. Another effect
    is the increase in the infrared flux. But such a modification in 
    model (v) is not sufficient to fit the MIR visibilities (fourth row in
    Fig.~\ref{figure:hd142666}).  
  \item Therefore, in addition to the latter modification in model (v), we
    increased the inner disk 
    radius to $R_\mathrm{in}=0.3$ and cut a dust-free ring from the disk between
    $R_\mathrm{gapin}=0.35\mathrm{AU}$ and
    $R_\mathrm{gapout}=0.8\,\mathrm{AU}$ in model (vi). The remaining dust ring between
    $0.3\,\mathrm{AU}$ and $0.35\,\mathrm{AU}$ is large enough to 
    provide enough NIR flux and to reproduce the NIR visibility. In
    contrast, the dust-free ring causes a shift in the MIR flux
    distribution towards larger disk radii, resulting in a decrease in the
    corresponding visibilities. The model flux at $1.6\,\mathrm{\mu m}$
    deviates by $19\%$ from the measured H band flux.
  \item In disk model (vii), there is an identical disk gap to the one in model
    (vi). But the density was not decreased. Here, the MIR visibility is
    decreased, but not enough to reproduce the observations. Furthermore,
    the NIR visibility decreased too strongly.
  \end{enumerate}

  \subsection{AS\,205\,N}\label{section:as205n}
  \begin{figure*}[!tb]
    \center
    \resizebox{0.48\textwidth}{!}{\includegraphics{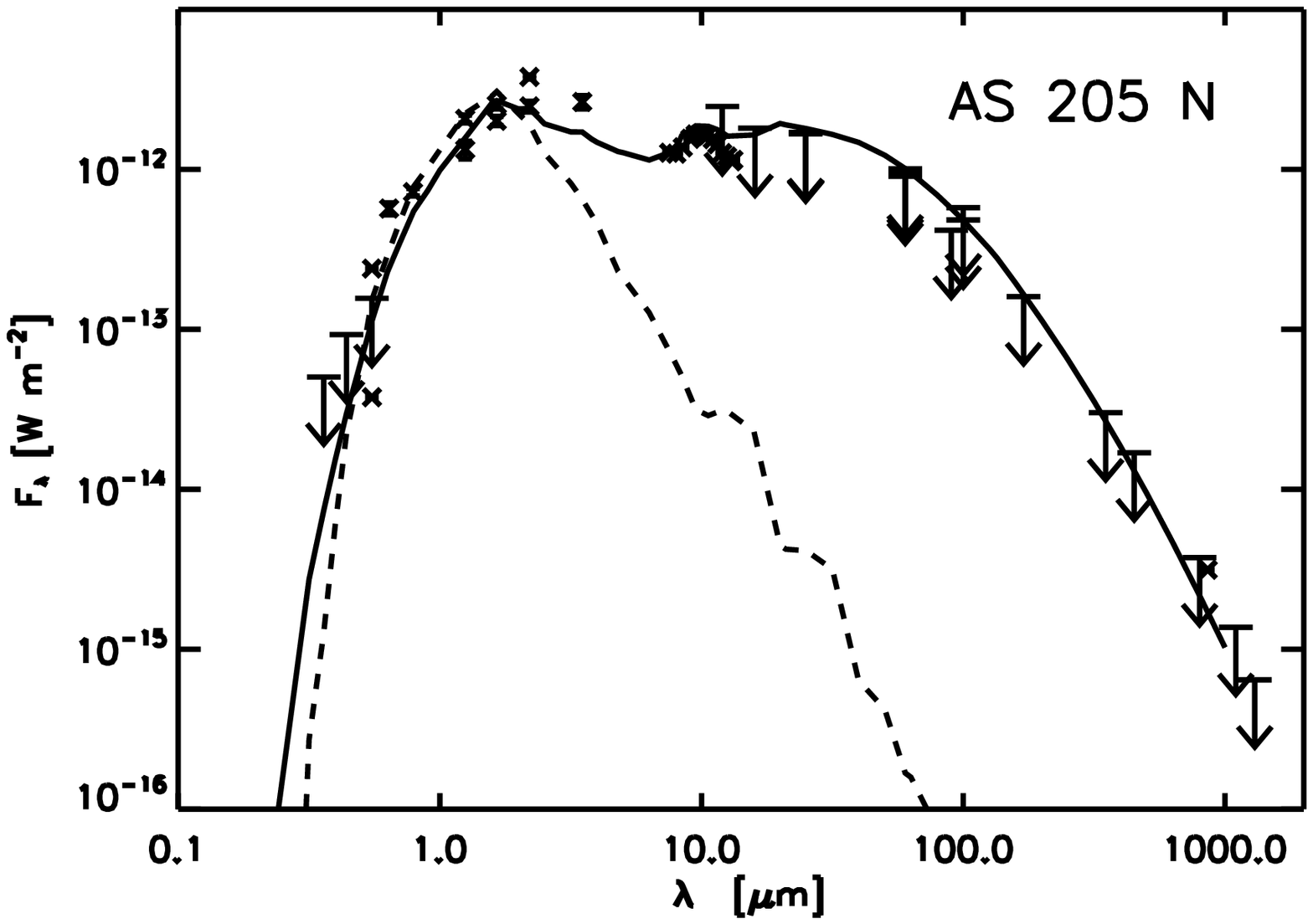}}\newline
    \resizebox{0.33\textwidth}{!}{\includegraphics{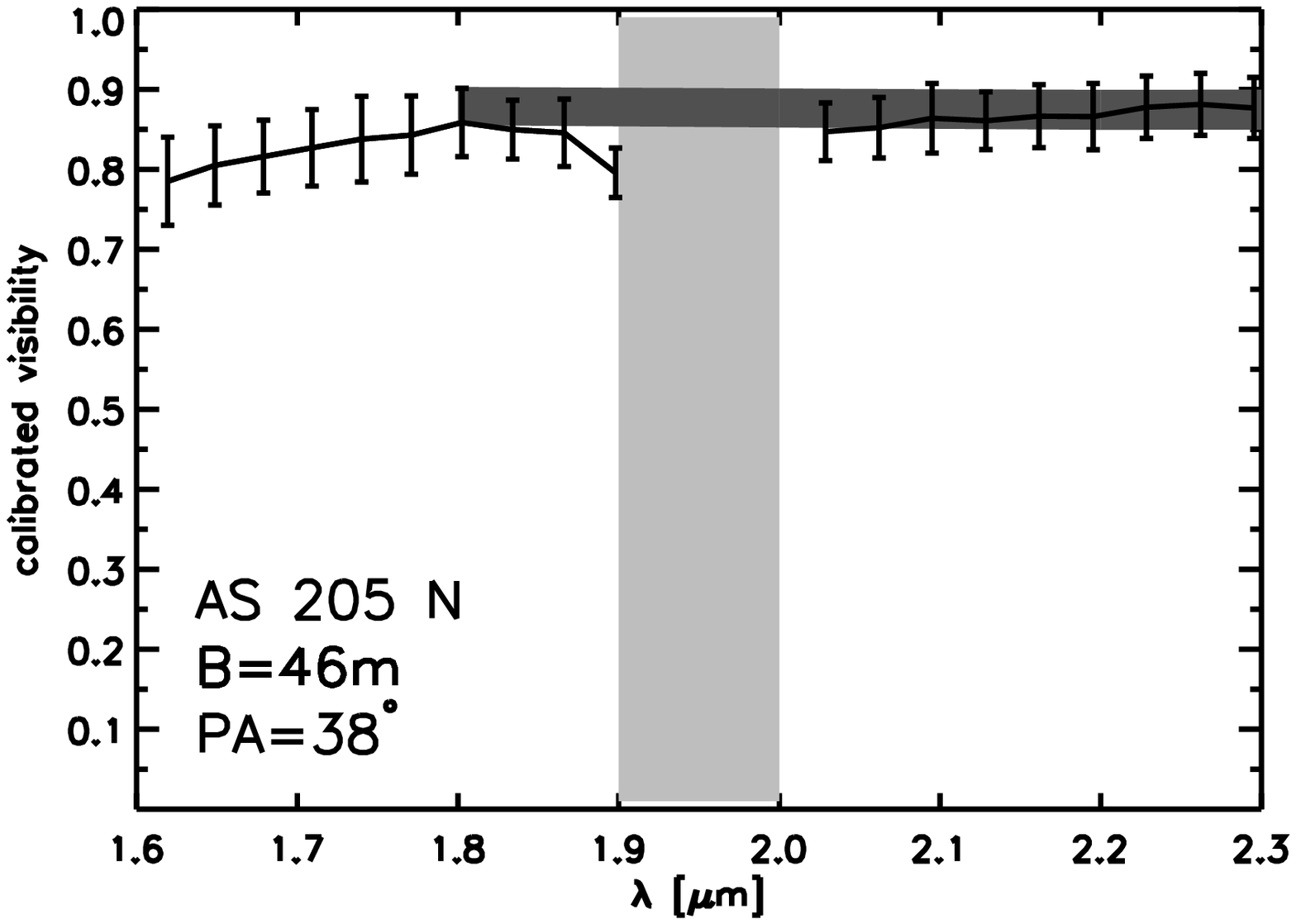}}
    \resizebox{0.33\textwidth}{!}{\includegraphics{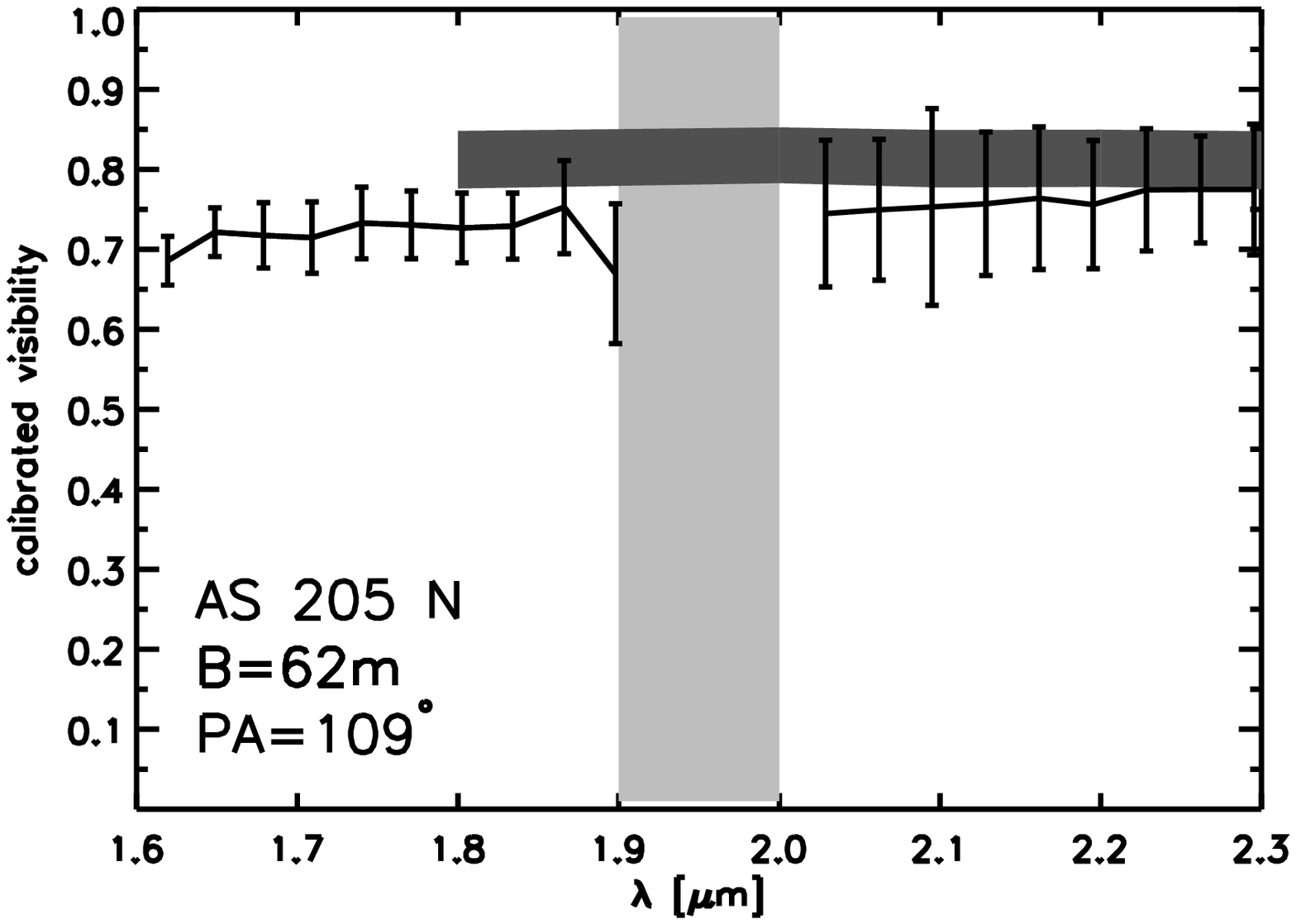}}
    \resizebox{0.33\textwidth}{!}{\includegraphics{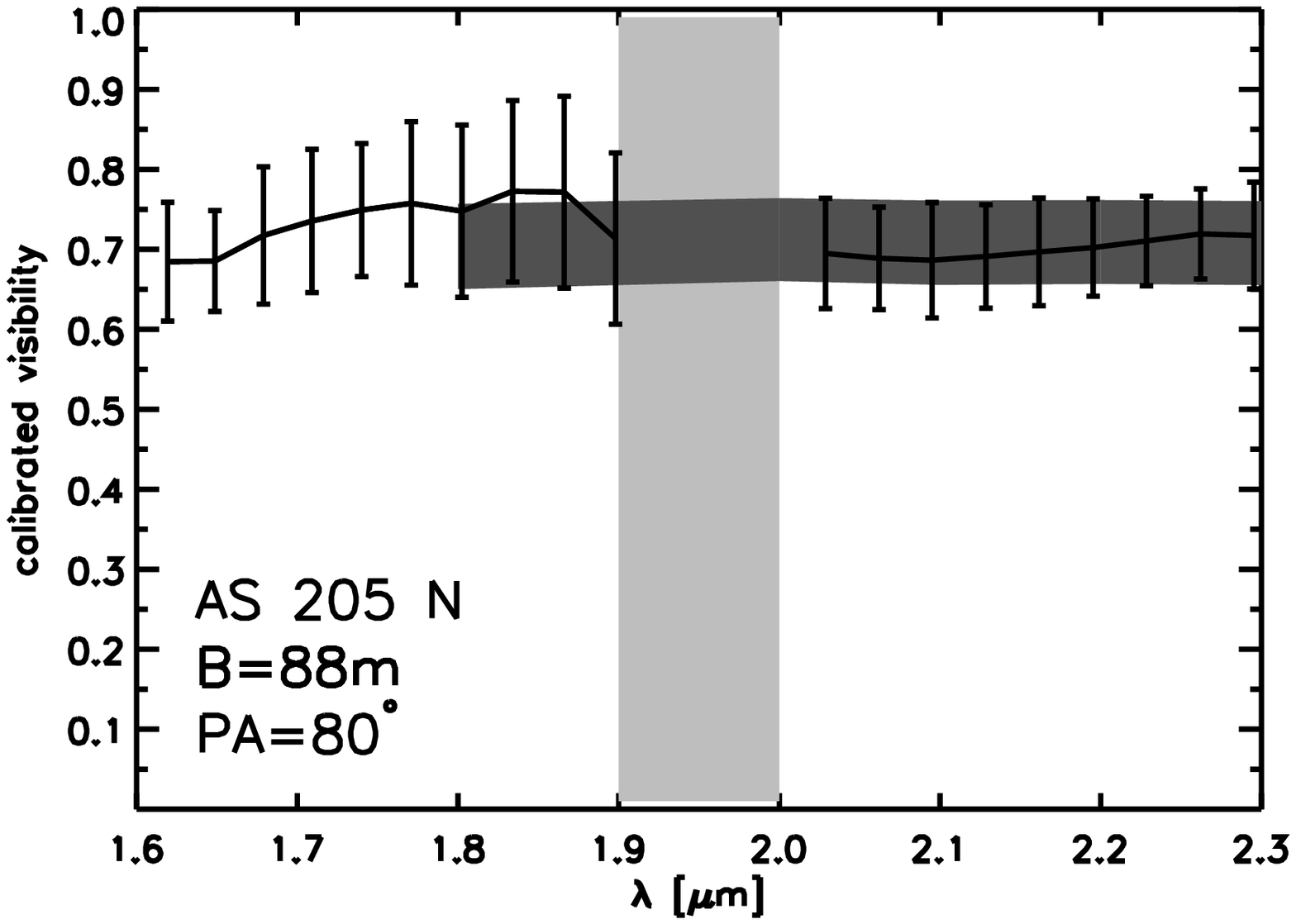}}\newline
    \resizebox{0.33\textwidth}{!}{\includegraphics{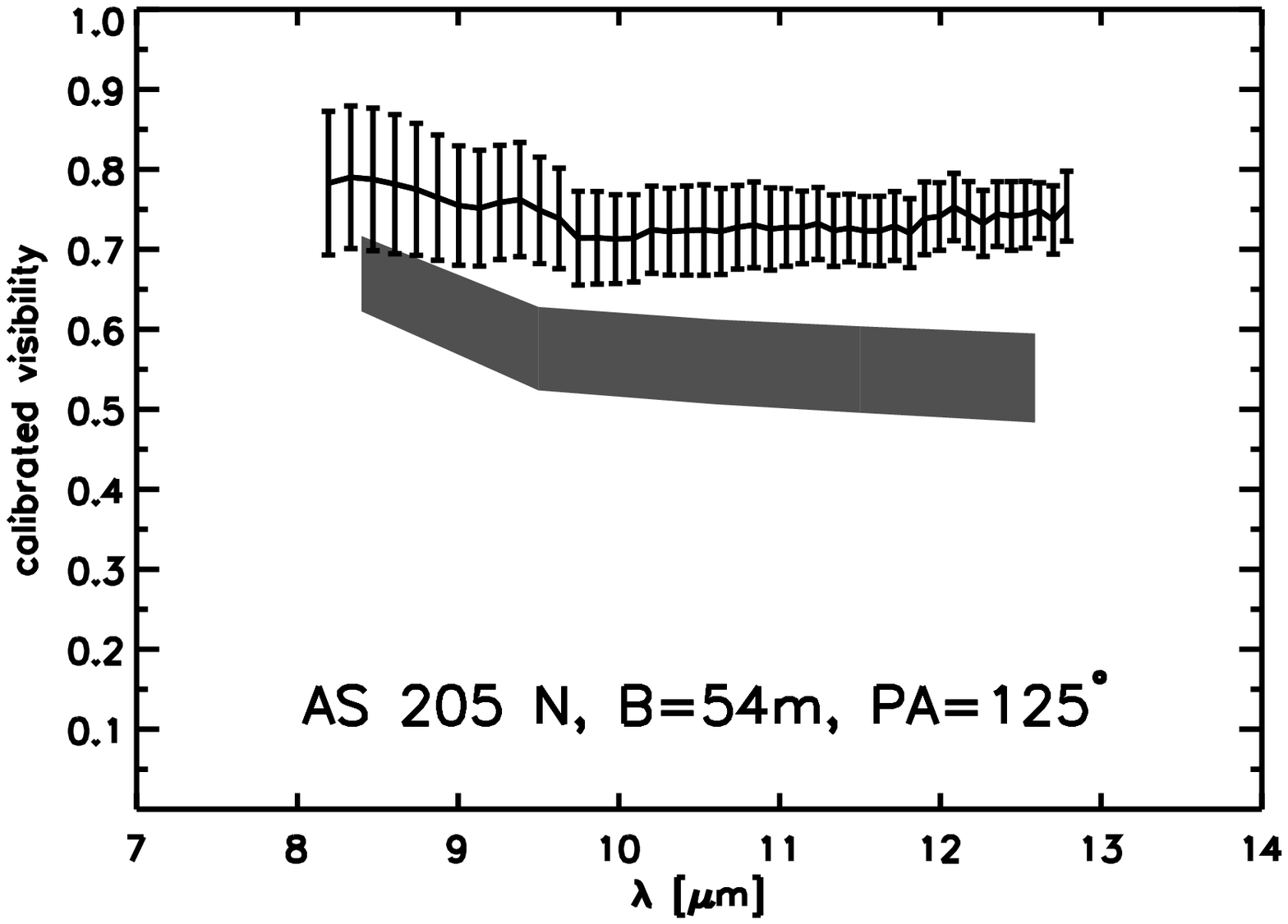}}
    \resizebox{0.33\textwidth}{!}{\includegraphics{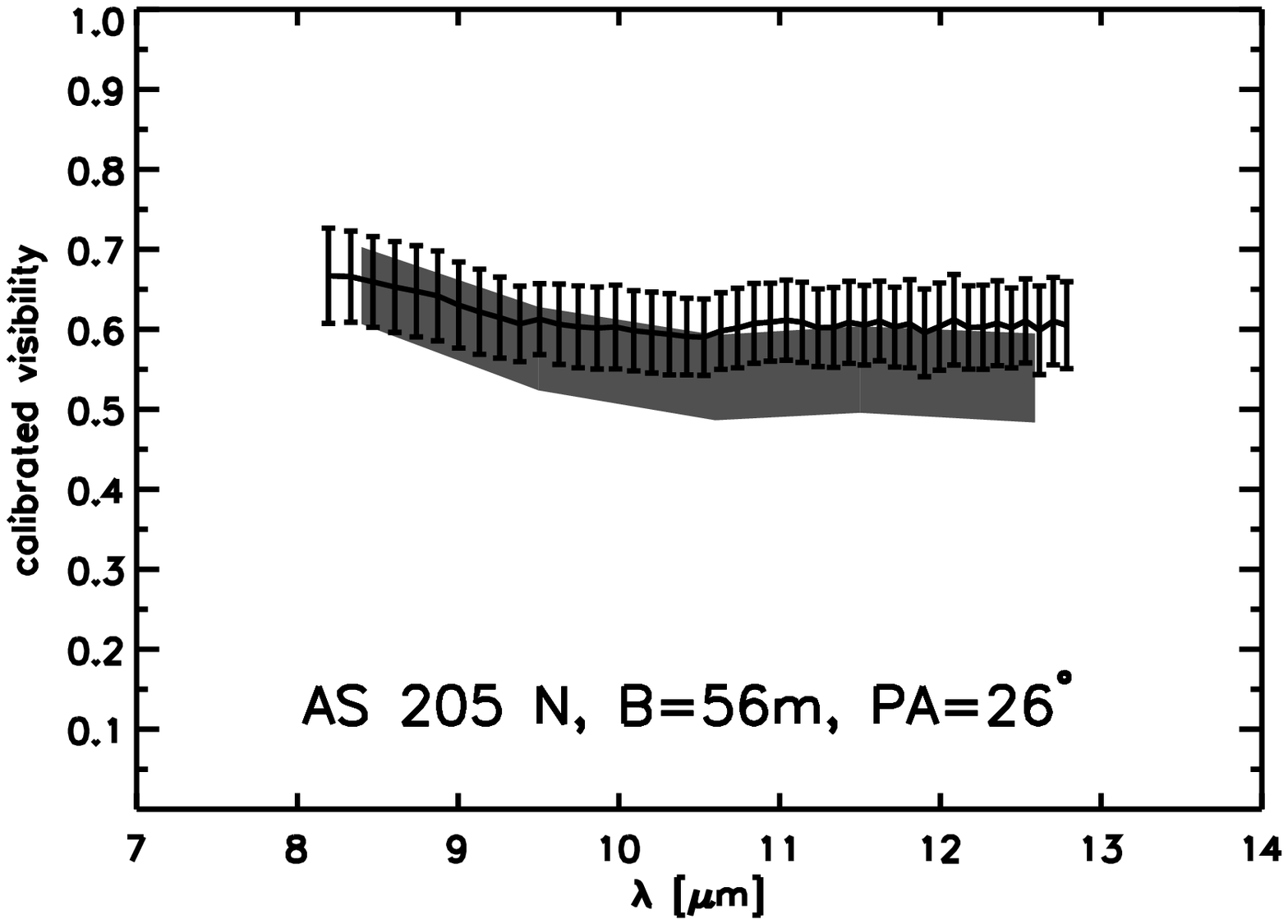}}
    \resizebox{0.33\textwidth}{!}{\includegraphics{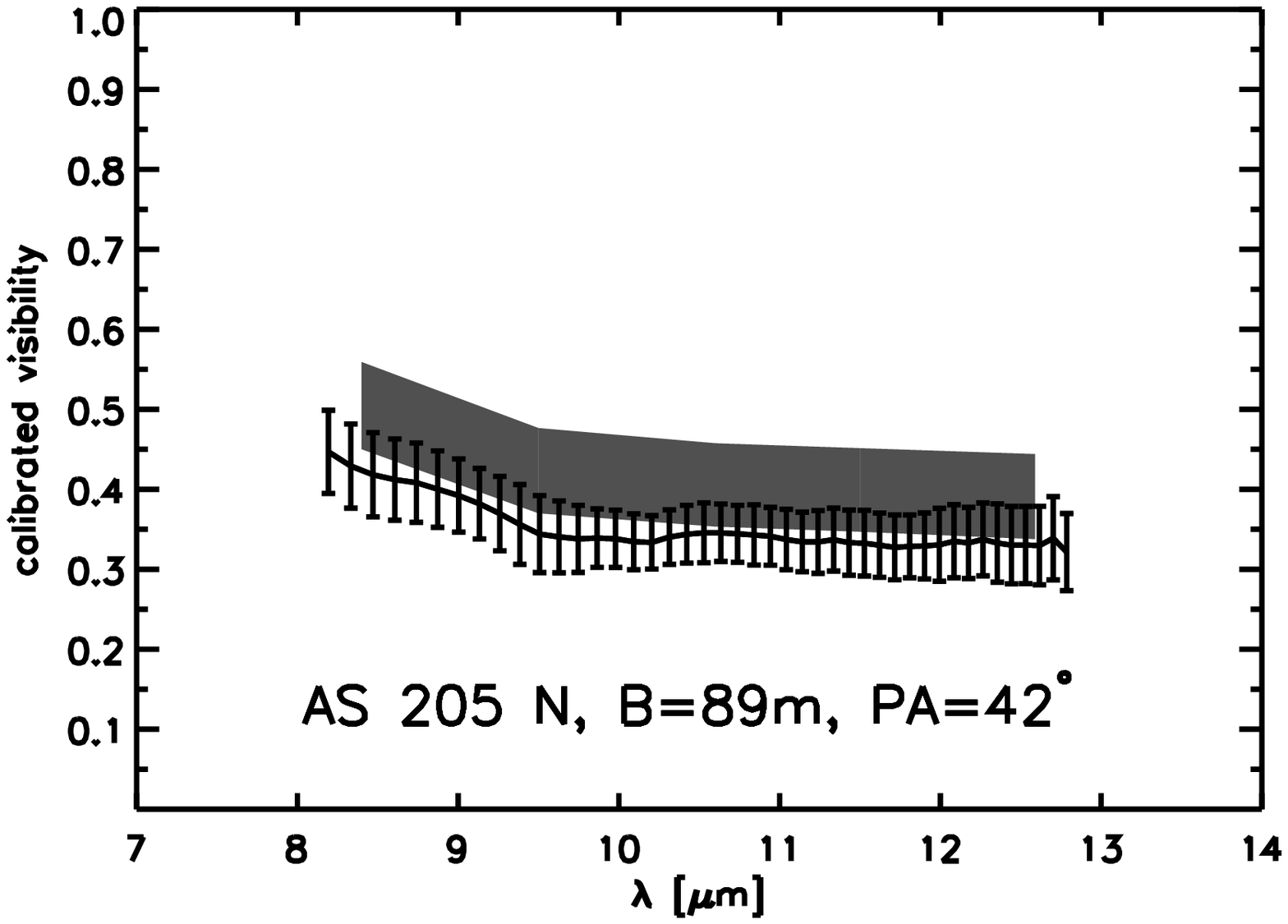}}
    \caption{SED, NIR, and MIR visibilities spectra for the different projected
      baselines that result from the measurements and our model of AS\,205\,N.} 
    \label{figure:as205n}
  \end{figure*}

  There are previous flux measurements in the NIR and MIR wavelength ranges
  that can be exclusively ascribed to the northern component (Eisner et
  al.~\cite{eisnerII}; Prato et 
  al.~\cite{prato}). Apart from a recent observation at a wavelength of
  $860\,\mathrm{\mu m}$ (Andrews et
  al.~(\cite{andrews}), we assumed that
  all other previous photometric measurements could not spatially resolve each
  single component of the system. Cohen \& Kuhi~(\cite{cohen}) could measure
  the visual flux of each 
  component, but they missed ascribing the photometric
  values to each single component. Since AS\,205\,N is an active and variable
  source (Johns-Krull et al.~\cite{johns-krullII}; Eisner et
  al.~\cite{eisnerII}), it depends on the variability of the visual
  fluxes if the measurements of Cohen \& Kuhi can be correctly
  assigned. The spectral flux in this wavelength regime and the visibilities
  have to be measured simultaneously. 

  The canonical disk model was used for modeling this source,  We found
  that an advanced disk model with a larger inner disk radius and lower
  dust density in the inner disk region, respectively, cannot reproduce the
  observations, in particular the photometric measurements. The successful
  data reproduction using the 
  canonical modeling approach confirms the primordial, unevolved character of
  the disk, which is consistent with the young age of the source of  
  $1\,\mathrm{Myrs}$ (Prato et al.~\cite{prato}). However, there is an
  ambiguity in the profile parameters $\beta$ and $h_\mathrm{100}$. Models with values $\beta \in
  [1.0,1.1]$ and $h_\mathrm{100} \in [19,21]$ reproduce the
  measurements just as well. 

  The visibility that was
  measured with the $89\,\mathrm{m}$-baseline is $\sim$$30$\,\% lower 
  than the modeled value. The SED that results from our best model for
  AS\,205\,N differs by about $40$\,\% from the photometric measurements of
  Prato et al.~(\cite{prato}) in the NIR wavelength range
  (Fig.~\ref{figure:as205n}). 
  Considering a deviation by $6$\,\%, the modeled SED agrees with the fluxes
  that were measured by Eisner et al.~(\cite{eisnerII}). This disagreement
  with Prato et al. and the agreement with the Eisner et al. probably 
  point to an intrinsic variability of the source in the NIR. However, the
  stellar luminosity of $L_{\star}=1.3\,\mathrm{L_{\odot}}$ that was derived by
  Eisner et al.~(\cite{eisnerII}) is about a factor of $\sim$$7$ lower than
  other previous results (Prato et al.~\cite{prato}; Liu et al.~\cite{liu})
  and cannot be confirmed by our model. In contrast, the accretion rate of
  $\dot{M}=7 \times 10^{-7}\,\mathrm{M_{\odot} yr^{-1}}$ that is used in our
  model corresponds to the estimation of Eisner et al.~(\cite{eisnerII}) and
  Johns-Krull et al.~(\cite{johns-krullII}: $\dot{M}=6.7 \times
  10^{-7}\,\mathrm{M_{\odot} yr^{-1}}$). The accretion luminosity is about
  $L_\mathrm{acc} \approx 6\,\mathrm{L_{\odot}}$. Therefore, AS\,205\,N is a
  T\,Tauri object with one of the highest accretion rates in comparison to
  other T\,Tauri stars (Ratzka et al.~\cite{ratzkaIII}; Schegerer et
  al.~\cite{schegerer}, Schegerer et al.~\cite{schegererII}). 

  Although we assume an accretion rate of $\dot{M}=7 \times
  10^{-7}\,\mathrm{M_{\odot} yr^{-1}}$, we have found that a model with the
  same profile parameters as listed in
  Table~\ref{table:properties-midisurveyII} but with an accretion rate that is
  lower by a factor $3-4$ assuming a larger inner disk radius of
  $R_\mathrm{in}=0.15\,\mathrm{AU}$ at the same time can also simultaneously fit the SED, MIR,
  and NIR visibilities. A lower accretion rate
  of $\dot{M} < 7 \times 10^{-7}\,\mathrm{M_{\odot} yr^{-1}}$ results in
  less NIR and MIR radiation and a larger inner disk radius of
  $R_\mathrm{in} = 0.15\,\mathrm{AU}$ (instead of $R_\mathrm{in} =
  0.10\,\mathrm{AU}$) induces a temperature decrease of $\Delta T \approx
  150\,\mathrm{K}$ on average in the innermost disk regions and a 
  shift of the thermal emission peak to longer wavelengths. Our
  modeling approach thus allows models with different modeling parameters to
  fit the whole data set. This ambiguity can be avoided by considering the
  results of the previous, independent measurements of the accretion
  rate.

  \begin{figure}[!tb]
    \center
    \resizebox{0.33\textwidth}{!}{\includegraphics{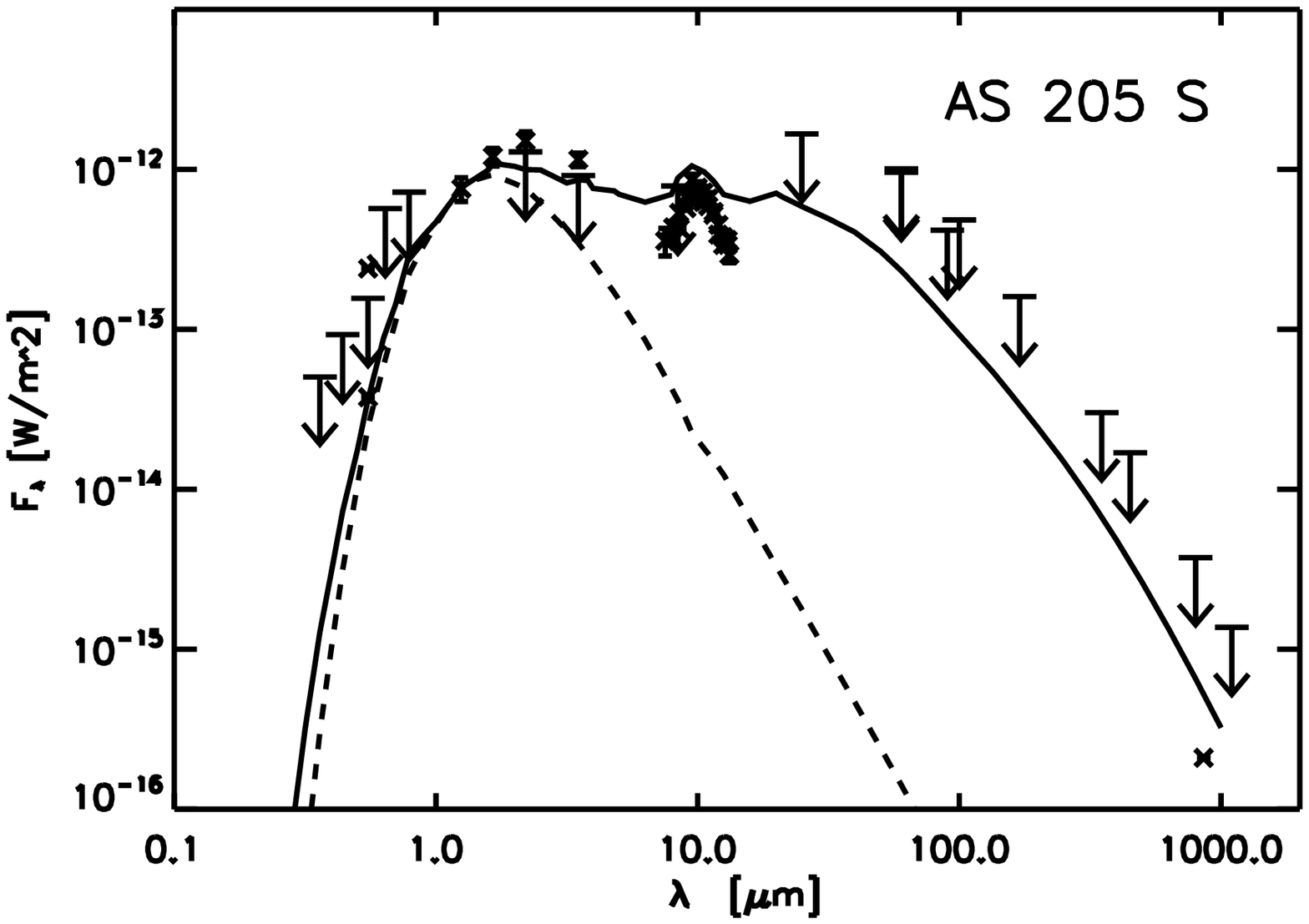}}
    \resizebox{0.33\textwidth}{!}{\includegraphics{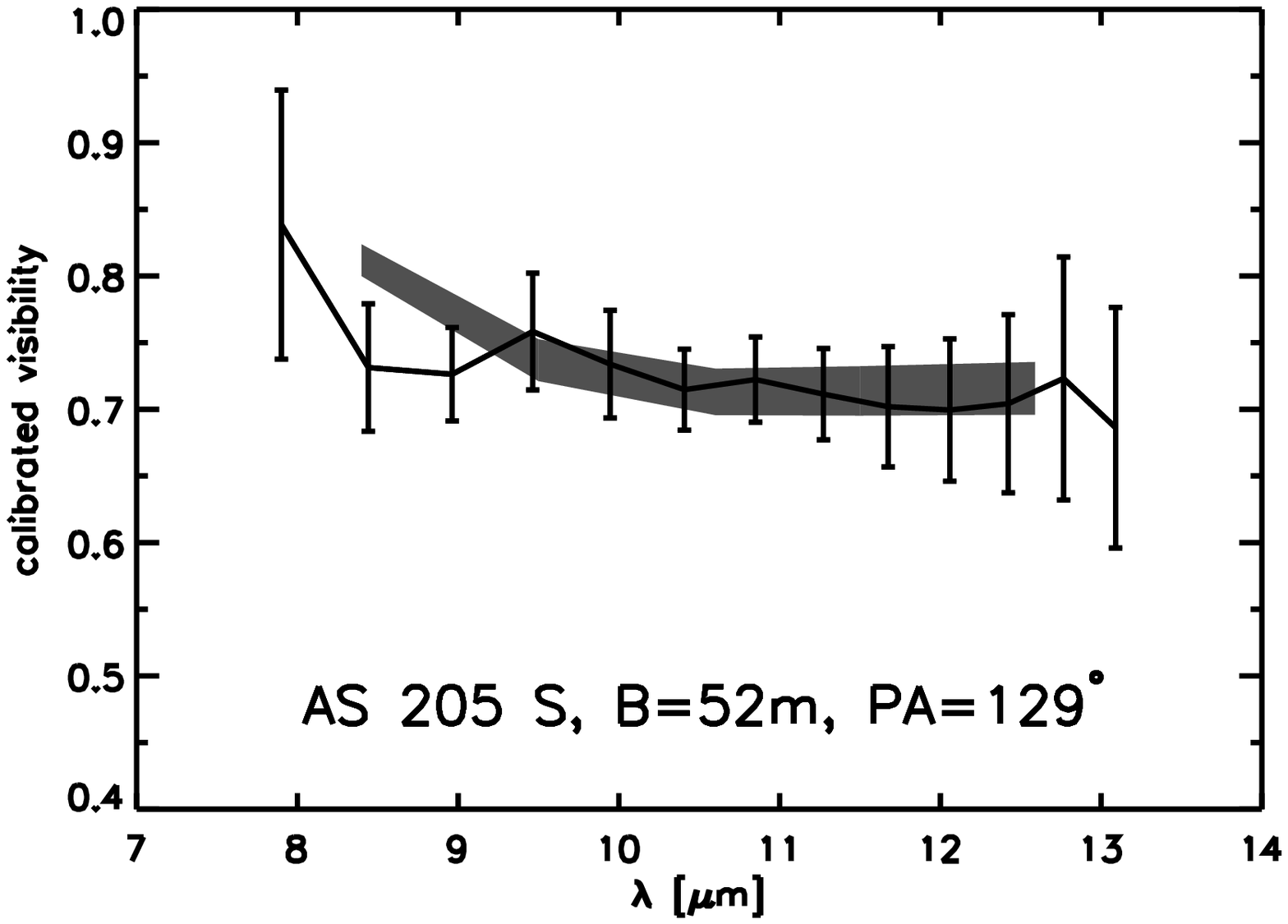}}
    \caption{SED and spectrally dispersed MIR visibility for a projected baseline length of
      B=52\,m 
      that result from the measurements and our model of AS\,205\,S.}
    \label{figure:as205s}
  \end{figure}  

  \subsection{AS\,205\,S}\label{section:as205s}  
  Apart from a recent submillimeter observation by Andrews et
  al.~(\cite{andrews}), the photometric fluxes in the visual, FIR, and millimeter
  ranges have to be ascribed 
  to AS\,205\,N and AS\,205\,S. Moreover, the visual extinction of the
  southern component, which is a binary itself, could
  not be derived unambiguously. While Prato et al.~(\cite{prato}) assume
  $A_\mathrm{V} = (2.1 \pm 1.0)\,\mathrm{mag}$, Eisner et
  al.~(\cite{eisnerII}) and McCabe et al.~(\cite{cabeII}) derived higher
  values of $A_\mathrm{V} = (3.6 \pm 1.0)\,\mathrm{mag}$ and $A_\mathrm{V} =
  2.41\,\mathrm{mag}$, respectively. Although the measurements approximately
  agree with each other within the error bars, a standard
  deviation of $\sigma=1.0\,\mathrm{mag}$ makes a subsequent allocation of the
  photometric, spatially resolved fluxes in the visual range impossible. 

  The flux in L band, which results from the model of AS\,205\,S differs by about
  $26\%$ from the fluxes measured by Prato et al.~(\cite{prato};
  Fig.~\ref{figure:as205s}). It is possible that photometric variability in the
  NIR is also responsible for this difference between the measurements and the
  model of the southern component. The accretion luminosity of this component
  is $L_\mathrm{acc}=0.32\,\mathrm{L_{\odot}}$. The interferometric measurements
  on the baselines of $B=56\,\mathrm{m}$ and $B=85\,\mathrm{m}$
  (Table~\ref{table:journal_as205}) were discarded because of the bad weather
  conditions during the observations. According to Eisner et
  al.~(\cite{eisnerII}), even AS\,205\,S is a binary. The separation of both
  components is $1.3\,\mathrm{AU}$ at a position angle of
  $PA=101^{\circ}$. Such a binary system was not considered in our models
  for AS\,205\,S because a sinusoidal visibility curve that is
  characteristic of a close binary system could not be identified (Schegerer
  et al.~\cite{schegerer}). 
  \begin{figure}[!tb]
    \center
    \resizebox{0.48\textwidth}{!}{\includegraphics{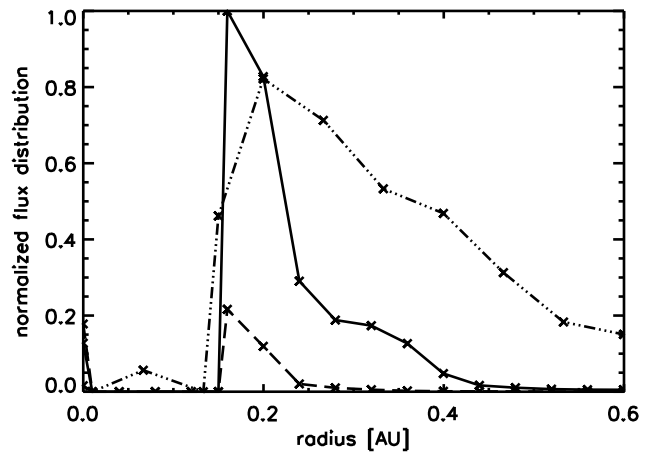}}
    \caption{Flux distributions at wavelengths of $1.65\,\mathrm{\mu m}$,
      $3.5\,\mathrm{\mu m}$, and $12.5\,\mathrm{\mu m}$ (dashed, solid,
      and dotted-dashed line, respectively) that are derived from
      a model that fits the measurements best. These
      distributions are normalized to the maximum of the $3.5\,\mathrm{\mu m}$ flux
      distribution.}  
    \label{figure:rad-int}
  \end{figure}         
 
  \section{Discussion}\label{section:discussion}
  Our comprehensive modeling approach is complex. There are some deficiencies
  (d-i -- d-vi) that should be listed.  
  \renewcommand{\labelenumi}{d-\roman{enumi}}
  \begin{enumerate}
  \item The main limitation of our observations is the limited
    uv coverage, which makes the data analysis model dependent. The
    physical quantity that can be best constrained is the radial extent of
    the NIR and MIR disk emission. The goal of this paper was to
    constrain this quantity for the sources under investigation. Studying
    SED and visibility, the inclination and the position angle of the
    disks cannot be derived precisely. Only a few visibility points are not
    sufficient to derive these values. With respect to the SED only an upper
    limit of the inclination can be determined. This limit corresponds to the
    angle where the disk allows the observations of inner disk
    regions (and the central star).
  \item There are small, local deviations between measured and modeled
    data. The modeling approach of Shakura \& Sunyaev~(\cite{shakura}) 
    assumes a rotationally symmetric structure. Deviations between modeled and
    measured visibilities at fixed baseline lengths but different PAs, for
    instance, could be caused by local asymmetric structures. In fact, such
    structural asymmetries could be evoked, e.\/g., by large-scale eddies (Klahr et
    al.~\cite{klahr}), which have already been observed in large-scale images of YSOs
    (Fukagawa et al.~\cite{fukagawaI}; Lagage et al.~\cite{lagage}). 
  \item For HD\,142666, we found a model for this putative TO by assuming a smooth
    transition to the canonical disk approach. Although this 
    approach was confirmed by our interferometric observations, there 
    are different possible approaches that are supposed to reproduce the
    measurements as well (e.\/g., Bouwman et al.~\cite{bouwmanIII}). 
  \item A latent ambiguity in our modeling approach cannot be ruled out
    (Sect.~\ref{section:as205n}).
  \item The discontinuous boundaries at the inner disk edge, as well as the edges of
    a potential ring gap, that can be implemented in our advanced disk 
    model are a simplification. 
  \item A wide variety of outer disk radii $80\,\mathrm{AU} < R_\mathrm{out}
    < 250\,\mathrm{AU}$ can reproduce the measured data of 
    all objects of this study. Because of the intrinsic ambiguity of the SED
    (Men'shchikov \& Henning~\cite{menshchikov}; Thamm et al.~\cite{thamm}),
    large photometric errors, and missing spatially resolved observations in
    the FIR range, an unambiguous determination of the model parameter
    $R_\mathrm{out}$ is not possible. 
  \item  The modeled SED can deviate from the measured SED. We
    assume a dust mixture of "astronomical silicate" and graphite with the
    canonical MRN grain-size distribution. But previous studies of the
    silicate feature and the water ice feature have already shown strong 
    modifications of the composition in circumstellar disks
    (Bouwman et al.~\cite{bouwman}; Przygodda et al.~\cite{przygodda}; Bouwman
    et al.~\cite{bouwmanII}; Schegerer \&
    Wolf~\cite{schegererIII}). The modification of the dust composition
    depends on properties of the central star and the evolutionary status of
    the source (van Boekel et al.~\cite{boekelII}; Schegerer et
    al.~\cite{schegererI}), as well as on the distance from the star (van Boekel et 
    al.~\cite{boekel}; Schegerer et al.~\cite{schegerer}) and from the
    midplane (Rodmann et al.~\cite{rodmann}). However, the analysis of dust
    features only allows investigations of the relative occurrence
    of silicate grains with sizes between
    $\sim$$0.1\,\mathrm{\mu m}$ and $\sim$$2\,\mathrm{\mu m}$ assuming
    nonporous grains. Therefore, an individual dust composition
    is not considered in our modeling approach, although the dust composition affects the disk
    structure as shown by Monnier \&
    Millan-Gabet~(\cite{monnier}).  
  \end{enumerate}

  Different modeling approaches have been used for the objects of this
  study. This is discussed in detail in the following.
  \subsection{HD\,142666}\label{section:dis-hd142666}
  Extensive modifications of the parameter of the canonical model, i.\/e., the
  inner disk radius $R_\mathrm{in}$, the
  profile parameters $\beta$ and $h_\mathrm{100}$, and the accretion rate
  $\dot{M}$, respectively, failed to reproduce all the observations.
  Therefore, the modeling of the source HD\,142666 required an extension of
  the canonical approach of Shakura \& Sunyaev~(\cite{shakura};
  Eq.~\ref{eq:shakura}). 
  In our approach for HD\,142666, we reduced the dust density at the inner
  disk edge by multiplying of the canonical disk with the factor
  $f_{\rho}=0.01$. This factor smoothly increased up to
  unity at $R_\mathrm{out}$. Additionally, we increased the inner disk radius
  from $0.1\,\mathrm{AU}$ to $0.3\,\mathrm{AU}$ and cut a dust-free ring from
  the disk between $R_\mathrm{gapin}=0.35\mathrm{AU}$ and
  $R_\mathrm{gapout}=0.8\,\mathrm{AU}$. There are several
  different approaches in advanced disk models that potentially reproduce the whole
  data set that we obtained. However, an increased inner disk radius and 
  a disk gap is putatively necessary for a successful fit. While the increased inner disk
  radius decreases the infrared flux and the NIR visibility, a low MIR
  visibility results from the disk gap. For a constant disk mass, the NIR flux
  even increases after cutting the gap from the disk. 
  A larger sublimation radius or a disk gap, alone, cannot explain the
  infrared visibilities and the SED of this source. Higher accretion rates,
  an inner optical and geometrically thin disk, a more flared disk, or a
  puffed-up inner rim can all be excluded.
  
  The disk
  gaps in HD\,142666 classify this source as TO. This result contrasts with
  previous studies where TOs were classified as sources 
  with specific color relations (McCabe et al.~\cite{cabe}; Alexander et
  al.~\cite{alexanderII}) or a significant lack of infrared excess at $\lambda
  < 6\,\mathrm{\mu m}$, solely (Table~\ref{table:colors}). The last two
  characteristics are not even fulfilled by HD\,142666. Recently, several
  such ``unexpected'' transition disks have been 
  discovered through direct imaging on the millimeter wave dust
  emission in more outer disk regions (e.\/g., Andrews et
  al.~\cite{andrews}). All these objects have  
  normal SED but cavities larger than $10$\,AU, as measured in the millimeter
  wavelength range.

  We assume that this clearing of the
  disk can be a consequence of planet formation where adjacent regions
  are strongly affected by tidal forces (Calvet et al.~\cite{calvet}; Rice et
  al.~\cite{rice}; Quillen et al.~\cite{quillen}). However, this modeling
  result also confirms the outcome of the theoretical study of Takeuchi et
  al.~(\cite{takeuchi}), in principle, where disk evolution has been modeled
  combining viscous evolution, photoevaporation, and the differential radial
  motion of dust grains and gas. In this study, the formation of a dust ring
  in the disk of HAeBe stars is
  predicted without assuming planet formation but as a natural consequence of
  photoevaporation and grain migration. HAeBe stars strongly suffer from a 
  photoevaporative mass loss from the disk. The outer disk regions are
  slowly eroded, the inner hole grows, and the dust becomes concentrated
  close to the inner disk edge. We have to note, however, that Takeuchi et
  al.~(\cite{takeuchi}) studied more massive
  ($M_\star=2.5\,$M$_\mathrm{\odot}$) and more luminous
  ($L_\star=30\,$L$_\mathrm{\odot}$) HAeBe objects where a
  dust ring appears at radii on the order of $10\,$AU. 
  
  For the case that a young planet has formed the disk gap, we studied its theoretical
  effects on our interferometric observations. In fact, a planetary companion
  of the main source evokes a specific sinusoidal visibility curve whose
  amplitude and frequency depend on the separation $a_\mathrm{sep}$, the
  position angle of the companion with respect to the position angle of the
  interferometric baseline $a_\mathrm{PA}$, and the brightness ratio
  $a_\mathrm{rat}$ 
  (Ratzka et al.~\cite{ratzka}). We assume a brightness ratio of
  $100:1$ between the main source and a potential planetary companion, 
  the primordial
  visibility curve without companion would change by $\Delta V =
  = 0.02$,
  only. Considering a standard deviation of $0.07$ on average
  (Sect.~\ref{section:hd142666}), such a modification would be too weak to be
  noticed using the interferometers available in this study. The brightness
  ratio should be $a_\mathrm{rat}< 25:1$, at least, to be directly found by our
  interferometric measurements. 

  \subsection{AS\,205}\label{section:dis-as205}
  The SED and the MIR visibilities of AS\,205 
  have been investigated and modeled in Schegerer~(\cite{buch}) and Schegerer
  et al.~(\cite{schegererII}). Only a few,
  small modifications of these previous disk models, which are canonical
  models, were enough to fit the NIR visibility spectra of these sources, as well. We
  conclude that the inner disk structure
  ($>$$0.1\,\mathrm{AU}$ up to several $1\,\mathrm{AU}$s) is 
  strongly determined by SED and MIR visibility spectra. In fact,
  considering Fig.~\ref{figure:rad-int}, the regions where MIR and NIR flux
  arise overlap. However, the spatial resolution of
  MIDI is probably too low to uniquely determine the parameter
  $R_\mathrm{in}$, in particular for less evolved objects. For the northern
  and the southern components of AS\,205, the inner disk radius is assumed 
  to be the sublimation radius that was determined by the properties of the
  star and the dust. 
  
  \begin{table}
    \centering
    \caption{K-L
      color index ranges of the objects of our sample. The indices
      were measured in previous studies that are mentioned in the right column
      (ref).} 
    \begin{tabular}{lcr}\hline\hline
      object & $K-L$ & ref.\\   
      & (mag) & \\ \hline  
      HD\,142666 & $0.98-1.07$ & (4), (5)\\[1.0ex]
      AS\,205\,N & $0.8-1.05$ & (1), (2), (3) \\[1.0ex]
      AS\,205\,S & $1.02-1.1$ & (1), (2), (3) \\[1.0ex]
      \hline
    \end{tabular}
        {\newline \scriptsize {References:} {\bf 1}: Prato et
          al.~(\cite{prato}); {\bf 2}: Koresko~(\cite{koreskoII}); {\bf 3}:
          McCabe et al.~(\cite{cabeII}); {\bf 4}: Sylvester et
          al.~(\cite{sylvester});  {\bf 5}: Meeus et al.~(\cite{meeusII})}
        \label{table:colors}
  \end{table}

  \section{Summary and outlook}\label{section:conclusion}
  In this study, we observed three well-known YSOs using the interferometers IOTA and
  AMBER in the NIR wavelength range and MIDI in the MIR wavelength range. 
  A maximum spatial resolution of $0.3\mathrm{AU}$ and $0.9\mathrm{AU}$ could
  be reached, respectively, allowing study of the
  innermost disk regions and of the inner disk edge in particular. The visibility
  $V$, i.\/e., the outcome from our interferometric observations, is a complex
  function of the dust density distribution, dust properties,
  and stellar properties of the source. Thus, a proper quantitative analysis
  was only possible by
  comparison with radiative transfer simulations for circumstellar disk
  models, which we performed with our radiative transfer code MC3D. We created 
  large samples of models of YSOs and mimicked the interferometric
  observational data to compare the results with our observations. We
  considered photometric data and object properties from the literature to obtain
  a unique disk model that
  reproduces all the data. AS\,205\,N and AS\,205\,S could be modeled using
  the canonical approach of Shakura \& Sunyaev~(\cite{shakura}) for viscous
  disks. 

  Here, we considered accretion effects, as well as a
  two-layer disk model with the classical MRN dust composition in the surface
  layer. Several model parameters were constrained by literature values,
  such as stellar properties, the distance of the object, and its accretion
  rate. Five (free) parameters were modified in the canonical
  approach. Since no parameter set could be found that model all the
  observations that are available for HD\,142666, this modeling approach was
  extended. The disk density at the inner 
  disk edge was multiplied with an exponential function of radius $r$ that
  reduces the mass density at the inner disk edge $R_\mathrm{in}$ by a
  constant $f_\mathrm{\rho}<1$ and 
  converges to unity at the outer disk radius $R_\mathrm{out}$. Additionally, a
  dust-free ring was cut from the disk. Thus, the advanced approach where three extra
  parameters ($f_{\rho}$, $R_\mathrm{gapin}$, $R_\mathrm{gapout}$) are
  available allows the putative properties of a TO to be modeled, 
  i.\/e., disk gaps and a low dust density. 

  Considering the simplifications in our modeling approach
  (Sect.~\ref{section:discussion}), the models cannot reproduce all the data
  perfectly. However, the following basic results (r-i -- r-vii) can be
  derived from our modeling approach:  
  \renewcommand{\labelenumi}{r-\roman{enumi}}
  \begin{enumerate}
  \item The canonical modeling approach failed to reproduce all the data
    obtained from high spatially resolved observations of HD\,142666, which is
    a YSO at an advanced
    evolutionary stage. An advanced modeling approach was a necessary criterion for a TO.
  \item The HAeBe star HD\,142666 could be successfully modeled by implementing an
    inner disk hole where the inner disk radius
    ($R_\mathrm{in}=0.20\,\mathrm{AU}$) is equal to the initial sublimation
    radius $R_\mathrm{sub}$. The initial sublimation radius was
      approximated after 
    assuming the stellar temperature and an optically thin disk with 
    MRN dust composition. A dust-free ring
    between $R_\mathrm{gapin}=0.35\mathrm{AU}$ and $R_\mathrm{gapout} =
    0.80\,\mathrm{AU}$ had to be cut from the disk, leaving a dust ring between
    $R_\mathrm{in}=0.3\,\mathrm{AU}$ and $R_\mathrm{gapin}=0.35\mathrm{AU}$
    with reduced mass density.
  \item If we classify a TO as a YSO with a disk gap, the object HD\,142666
    is a 
    TO. This finding is
    based on our interferometric observations, while the SED or the color relations
    are not appropriate for a final classification. Table~\ref{table:colors}
    lists the color $K-L$ for the objects of this study, but a criterion that
    separates CTTS from TOs could not be found.
  \item The canonical disk model satisfied the
    observations for the binary system AS\,205 where the components AS\,205\,N and
    AS\,205\,S were separately modeled. 
    The implementation of accretion
    effects were necessary for a successful reproduction of the
    observations. The presence of accretion could be a necessary hint that the
    transmit to a TO has not happened yet. A disk gap or lower
    dust densities than the canonical model provides can be excluded as the
    sources of this system. 
  \item This study has shown that the SED and the MIR visibilities are
    generally not sufficient to determine the inner disk radius
    of {\it more evolved} YSOs, although the inner disk edge close to the
    sublimation radius emits MIR radiation (Fig.~\ref{figure:rad-int}). NIR
    visibilities obtained from long enough interferometric baselines
    can avoid such an ambiguity. 
  \item For both components of the CTTS-system AS\,205, the NIR visibility data could confirm
    the model that we previously derived for both sources considering their SED
    and MIR visibilities alone. For these sources, the inner disk radius
    $R_\mathrm{in}$ is the sublimation radius $R_\mathrm{sub}$ that we
    formerly approximated. 
  \end{enumerate}
  Apart from our findings, the 
  following questions (q-i -- q--ii) arise from this study:  
  \renewcommand{\labelenumi}{q-\roman{enumi}}
  \begin{enumerate}
  \item Is the small dust ring between $R_\mathrm{in}=0.30\mathrm{AU}$ and
    $R_\mathrm{gapin} = 0.35\,\mathrm{AU}$ in HD\,142666 a region that favors
    planet formation?
  \item Do disk gaps (inner disk hole, dust-free ring) result from
    photoevaporation 
    or can they be attributed to the action of planet formation/motion of
    young planets? In fact, inner disk gaps such as an inner disk hole in
    T\,Tauri objects and an inner hole and dust-free ring in HAeBe stars 
    are an outcome of the theoretical study of Takeuchi et
    al.~(\cite{takeuchi}), which only assumes viscous motion and
    photoevaporation. However, Takeuchi et al.~(\cite{takeuchi})
    studied more massive sources and predicted
    dust gaps at larger disk radii ($r>1\,\mathrm{AU}$) than we
    found. An inner disk hole has also been found and confirmed for
    the nearby TO TW\,Hya with an age of $8-10$ million years (Calvet et
    al.~\cite{calvet}; Ratzka et al.~\cite{ratzkaIII}). But Setiawan et
    al.~(\cite{setiawan}) report the detection of a planetary companion with
    a mass of $9.8\pm3.3\,\mathrm{M_{\mathrm{Jupiter}}}$ around TW\,Hya in an
    orbit with radius $r=0.04\,\mathrm{AU}$. The authors of that
    study supposed that this planet is responsible for clearing the
    inner disk through the accretion of gas and dust, and they conclude that disk
    evolution and planet formation are probably directly connected. However,
    the actual existence of the planet orbiting TW\,Hya is still being discussed.
  \end{enumerate}
  
  \begin{acknowledgements}
    A.~A.~Schegerer and S.~Wolf were supported by the German Research
    Foundation (DFG) through the Emmy-Noether grant WO 857/2 ({\it ``The
      evolution of circumstellar dust disks to planetary
      systems''}). Financial support from the Hungarian OTKA grant NN102014
    and K101393 is acknowledged.
  \end{acknowledgements}

  \onecolumn
  \appendix
  \section{Previous measurements}\label{appendix}
  {\bf HD\,142666}, that is  
  also known as V\,1026\,Sco, was classified as Herbig\,Ae/Be-star
  by Gregorio-Hetem et al.~(\cite{gregorio-hetem}). It has a spectral type of A\,7\,III
  (e.\/g., Blondel et al.~\cite{blondel}). The source belongs to the star formation
  region R\,1 in Scorpius (Vieira et al.~\cite{vieira}). According to
  Meeus et al.~(\cite{meeus}), HD\,142666 is a group-II object whose disk is
  flat. The object is a photometrically variable UX\,Ori star (Natta et
  al.~\cite{nattaVI}), i.\/e., dust clouds with a size in the range of the
  stellar diameter move into the line of sight of the observer from time to
  time. The extinction 
  then increases and the stellar light can be reddened depending on the
  absorbing material. Considering polarimetric
  measurements (Hales et al.~\cite{hales}) and the high variability of the flux
  in the NIR and visual wavelength range
  ($\Delta V = 1.2$; Meeus et al.~\cite{meeusII}), the disk is assumed to be
  inclined. The approximate age 
  of the object is $10\,$million years (Natta et
  al.~\cite{nattaVI}). Millimeter measurements indicate this age, as well. The low
  decrease in the millimeter flux towards longer wavelengths could be a hint
  of 
  cm-sized dust grains (Natta et al.~\cite{nattaII}). Br$\gamma$ measurements
  showed that the circumstellar disk around HD\,142666 is still active
  with an accretion rate of
  $\dot{M} = 1 \times 10^{-8}\,\mathrm{M_{\odot}yr^{-1}}$ (Garcia-Lopez et al.~\cite{garcia-lopez}).

  \begin{table}[!b]
    \centering
    \begin{minipage}{0.29\textwidth}
      \caption{Photometric fluxes of HD\,142666. The 1.3\,mm flux is an upper
        limit. 
      }
      \label{table:photo-hd142666}
        \begin{tabular}{llr} \hline\hline
          wavelength & flux & refs.\\ 
          ($\mathrm{\mu m}$) & (Jy) & \\ \hline
          $0.36$ & $0.33$ & { 1}\\
          $0.55$ & $1.07$ & { 1}\\
          $0.64$ & $1.23$ & { 1}\\
          $1.25$ & $1.80\pm0.04$ &  { 2}\\
          $1.65$ & $2.07\pm0.05$ &  { 2}\\
          $2.20$ & $2.36\pm0.04$ & { 2}\\
          $3.50$ & $2.71$ & { 3}\\
          $3.80$ & $2.46$ & { 3}\\
          $4.80$ & $1.75$ &  { 3}\\
          $25$ & $11.5\pm0.6$ & { 4}\\
          $60$ & $7.5\pm0.4$ & { 4} \\
          $100$ & $5.1\pm1.0$ & { 4}\\
          $450$ & $1.09 \pm 0.060$ & { 3}\\
          $450$ & $1.140 \pm 0.035$ & { 5}\\
          $800$ & $0.35 \pm 0.023$ & { 3}\\
          $850$ & $0.313 \pm 0.005$ & { 5} \\
          $1100$ & $0.18 \pm 0.01$ &  { 3}\\ 
          $1200$ & $0.079\pm0.004 $ & { 6} \\
          $1300$ & $<0.064$ & { 3} \\ 
          $3100$ & $0.013\pm0.001$ & { 6} \\
          $3300$ & $0.011\pm0.001$ & { 6} \\ 
          $7000$ & $0.0017\pm0.0002$ & { 6} \\ \hline
      \end{tabular}
               {\newline \scriptsize {\it References}~-- {\bf 1}: Zacharias et  
                 al.~(\cite{zacharias}); {\bf 2}: 2\,MASS catalogue~(Cutri et
                 al.~\cite{cutri}); {\bf 3}: Sylvester et
                 al.~(\cite{sylvester}); {\bf 4}: IRAS catalogue~(\cite{iras});
                 {\bf 5}: Sandell et al.~(\cite{sandell}); {\bf 6}:~Natta et al.~(\cite{nattaII})}
    \end{minipage}
  \end{table} 

  {\bf AS\,205},   
  also known as V866\,Scorpii, is a member of the Upper Scorpius association,
  West from the $\rho$ Ophiuc star forming region (Reipurth \& Zinnecker~1993). As formerly
  observed by Herbig \& Rao~(1974), AS\,205 has an infrared companion 
  at a position angle of $PA=211^{\circ}$ and at an angular distance of
  $1.32$'', i.\/e., at a projected distance of $210\,\mathrm{AU}$ (Prato et
  al.~2003). Observations in the I and NIR bands could spatially resolve both
  components with the northern component as the brighter source (Reipurth \&
  Zinnecker~1993; Liu et al.~1996). Visual and NIR Speckle-interferometric
  observations (Koresko~\cite{koreskoII}) as well as spectrally highly resolving
  measurements (Eisner et al.~2005), have shown that the southern component is
  a close 
  binary system (angular distance $8.5$'', position angle $PA=101^{\circ} \pm
  1^{\circ}$), as well. Both components of this system have similar
  brightnesses in the R and  
  I bands and similar stellar properties ($L_{\star}=0.44\,\mathrm{L_{\odot}}$
  and $0.44\,\mathrm{L_{\odot}}$; $M_{\star}=0.74\,\mathrm{M_{\odot}}$
  and $0.54\,\mathrm{M_{\odot}}$; $A_\mathrm{V}=3.9\,\mathrm{mag}$ and
  $3.4\,\mathrm{mag}$, respectively; Eisner et al.~2005). Although the age
  of the entire system is still discussed, it
  is assumed that all the components have simultaneously formed by
  fragmentation from 
  the same region of the molecular cloud. The northern and southern components
  show hints of
  accretion (Cohen \& Kuhi~1979; Prato et al.~1997; Eisner et
  al.~2005). Johns-Krull et al.~(2000) have derived an accretion rate of
  $\dot{M}=6.7 
  \times 10^{-7}\,\mathrm{M_{\odot} yr^{-1}}$ from the profile of the C\,IV line
  of the northern component. Considering the high activity and variability of
  the source
  which has been known for $30$ years already, Welin~(1976) assumed that AS\,205 is
  a good candidate for future FU\,Ori variability outbursts.   

\begin{table}[!b]
  \centering
  \begin{minipage}{0.36\textwidth}
  \caption{Photometric measurements of AS\,205. 
    The two components could only be spatially 
    resolved in the NIR and MIR range. Cohen \& Kuhi~(1979) were able to spatially
    resolve both components, but they missed declaring which flux belongs to
    which component. The southern
    system is a binary system, as well. }
  \label{photo-as205}
  \begin{tabular}{lllr} \hline\hline
    & AS\,205\,N & AS\,205\,S & \\
    wavelength & flux & flux & refs.\\ 
    ($\mathrm{\mu m}$) & (Jy) & (Jy) & \\ \hline
    $0.36$ & \multicolumn{2}{c}{$0.0060 \pm 0.0033$} & { 1}\\
    $0.44$& \multicolumn{2}{c}{$0.014 \pm 0.001$} & { 1}\\
    $0.55$& \multicolumn{2}{c}{$0.029 \pm 0.001$} & { 1}\\
    $0.55$& \multicolumn{2}{c}{$0.044$/$0.0069$} & { 2} \\
    $0.64$& \multicolumn{2}{c}{$0.12 \pm 0.01$} & { 1}\\
    $0.79$& \multicolumn{2}{c}{$0.19 \pm 0.01$} & { 1}\\
    $1.25$ & $0.55 \pm 0.07$ & -- & { 1}\\
    $1.65$ & $1.1 \pm 0.07$ & -- & { 1}\\
    $2.2$ & $1.80 \pm 0.14$ & -- & { 1}\\
    $1.25$ & $0.86 \pm 0.05$ & $0.32 \pm 0.05$ & { 3}\\
    $1.65$ & $1.40 \pm 0.08$ & $0.66 \pm 0.08$ & { 3}\\
    $2.2$ & $2.80 \pm 0.08$ & $ 1.11 \pm 0.15$ & { 3}\\
    $3.5$ & $3.10 \pm 0.30$ & $1.34 \pm 0.13$ & { 3}\\
    $25$ & \multicolumn{2}{c}{$14 \pm 2$} & { 4}\\
    $60$ & \multicolumn{2}{c}{$19 \pm 2$} & { 4}\\
    $60$ & \multicolumn{2}{c}{$20 \pm 1$} & { 5} \\
    $90$ & \multicolumn{2}{c}{$13 \pm 1$} & { 5} \\ 
    $100$ & \multicolumn{2}{c}{$16 \pm 3.2$} & { 4}\\ 
    $170$ & \multicolumn{2}{c}{$9.1 \pm 1.6$} & { 5} \\
    $350$ & \multicolumn{2}{c}{$3.5 \pm 0.5$} & { 6} \\
    $450$ & \multicolumn{2}{c}{$2.5 \pm 0.2$} & { 6} \\
    $800$ & \multicolumn{2}{c}{$1.00 \pm 0.04$}  & { 6} \\
    $859$ & $\sim$$0.9 \pm 0.006$ & $\sim$$ 0.060 \pm 0.001$ & {7 } \\
    $1100$ & \multicolumn{2}{c}{$0.50 \pm 0.02$} & { 6} \\
    \hline
  \end{tabular}
  {\newline \scriptsize {\it References}~-- {\bf 1}: Eisner et
    al.~(\cite{eisnerII});  {\bf 2}: Cohen \& Kuhi~(\cite{cohen}); {\bf 3}:
    Prato et al.~(\cite{prato}); {\bf 4}: IRAS catalogue~(\cite{iras}); {\bf 5}:
    ISO data archive; {\bf 6}:~Jensen et al.~(\cite{jensen}); {\bf 7}:
    Andrews et al.~(\cite{andrews})}
  \end{minipage}
\end{table}

\end{document}